\documentclass{aa} 

\usepackage{graphicx}
\usepackage{multirow}
\usepackage{amsmath}
\usepackage{txfonts}
\usepackage{longtable}
\usepackage{subfig}

\usepackage{appendix}
\usepackage{enumerate}
\usepackage[usenames]{color}
\usepackage{bigdelim}
\usepackage{soul,ulem}
\usepackage[comma,authoryear]{natbib}
\def\cm#1{\ifmmode {\,{\rm cm^{-#1}}}                  
        \else \hbox{$\,${\rm cm$^{\rm -#1}$}}\fi}
\def\raw {\ifmmode\rightarrow\else$\rightarrow$\fi}
\def\ex#1{\ifmmode {\times 10^{#1}}         
        \else \hbox{{$\times 10^{\rm #1}$}}\fi}

\newcommand{\iram}{IRAM-30\,m}

\newcommand{\kms}{\mbox{km~s$^{-1}$}}

\newcommand{\mloss}{\mbox{$\dot{M}$}}
\newcommand{\my}{\mbox{$M_{\odot}$~yr$^{-1}$}}

\newcommand{\rstar}{\mbox{$R_{*}$}}

\newcommand{\vexp}{\mbox{$\varv_{\mathrm{exp}}$}} 
\newcommand{\vinf}{\mbox{$\varv_{\infty}$}}

\newcommand{\vsys}{\mbox{$V_{\mathrm{sys}}^{\mathrm{LSR}}$}}

\newcommand{\eu}{\mbox{$E_{\mathrm{u}}$}}

\newcommand{\trot}{\mbox{$T_{\mathrm{rot}}$}}

\newcommand{\tkin}{\mbox{$T_{\mathrm{kin}}$}}

\newcommand{\doce}{$^{12}$CO\,($J$=1$-$0)}

\newcommand{\oh}{\mbox{OH\,231.8$+$4.2}}
\newcommand{\microns}{\mbox{$\mu$m}}
\newcommand{\arcsecp}{\mbox{\rlap{.}$''$}} 
\newcommand{\secp}{\mbox{\rlap{.}$^{\mathrm{s}}$}} 

\newcommand{\tant}{\mbox{T$_{\mathrm{A}}$$^*$}}
\newcommand{\tmb}{\mbox{T$_{\mathrm{mb}}$}}
\newcommand{\tb}{\mbox{T$_{\mathrm{B}}$}}

\begin{document}

   \title{The millimeter IRAM-30\,m\thanks{Based on observations carried out with the \iram\, Telescope. The Institut de Radioastronomie Millim\'{e}trique (IRAM) is supported by INSU/CNRS (France), MPG (Germany) and IGN (Spain).} 
   line survey toward \object{IK Tau}\thanks{Tables\,\ref{tab:so2ener} and \ref{tab:so2coef} are only available in electronic format at the CDS. 
   Additionally, FITS files of the line survey are available at the CDS. }}
   \author{L.~Velilla Prieto\inst{1,2}
          \and
          C.~S\'anchez Contreras\inst{2}
	  \and
	  J.~Cernicharo\inst{1}
	  \and
	  M.~Ag\'undez\inst{1}
	  \and
	  G.~Quintana-Lacaci\inst{1}
          \and
          V.~Bujarrabal\inst{3}
          \and
          J.~Alcolea\inst{4}
	  \and 
	  C.~Balan\c ca\inst{5}
	  \and
	  F.~Herpin\inst{6,7}
	  \and
	  K.~M.~Menten\inst{8}
	  \and
	  F.~Wyrowski\inst{8}
          }
   \institute{Instituto de Ciencia de Materiales de Madrid, CSIC, c/ Sor Juana In\'es de la Cruz 3, 28049 Cantoblanco, Madrid, Spain\\
              \email{lvelilla@icmm.csic.es}
         \and
         Centro de Astrobiolog\'ia, INTA-CSIC, E-28691 Villanueva de la Ca\~nada, Madrid, Spain
         \and
         Observatorio Astron\'omico Nacional (IGN), Ap 112, 28803 Alcal\'a de Henares, Madrid, Spain
         \and
         Observatorio Astron\'omico Nacional (IGN), Alfonso XII No 3, 28014 Madrid, Spain
         \and
         LERMA, Observatoire de Paris, Sorbonne Universit\'e, UPMC, UMR 8112, F-92195 Meudon, France
         \and
         Universit\'e de Bordeaux, LAB, UMR 5804, F-33270 Floirac, France
         \and
         CNRS, LAB, UMR 5804, F-33270, Floirac, France
         \and
         Max-Planck-Institut f\"{u}r Radioastronomie, Auf dem H\"{u}gel 69, 53121 Bonn, Germany
             }
   \date{Arxiv VERSION. Accepted for A\&A 24/08/2016.}

   \abstract
  {} 
   {We aim to investigate the physical and chemical properties of the molecular envelope of the oxygen-rich
   AGB star \object{IK Tau}.}
   {We carried out a millimeter wavelength line survey between $\sim$79 and 356\,GHz with the IRAM-30\,m telescope.
   We analysed the molecular lines detected in \object{IK Tau} using the population diagram technique to derive rotational temperatures
   and column densities. We conducted a radiative transfer analysis of the SO$_2$ lines, which also helped us 
   to verify the validity of the approximated method of the population diagram for the rest of the molecules.}
  {For the first time in this source we detected rotational lines in the ground vibrational state of
  HCO$^{+}$, NS, NO, and H$_2$CO, as well as several isotopologues of molecules previously identified, namely,
  C$^{18}$O, Si$^{17}$O, Si$^{18}$O, $^{29}$SiS, $^{30}$SiS, Si$^{34}$S, H$^{13}$CN, $^{13}$CS, C$^{34}$S, 
  H$_2$$^{34}$S, $^{34}$SO, and $^{34}$SO$_2$. 
  We also detected several rotational lines in vibrationally excited states of SiS and SiO isotopologues, as well
  as rotational lines of H$_2$O in the vibrationally excited state $\nu_{\mathrm{2}}$=2.
  We have also increased the number of rotational lines detected of molecules
  that were previously identified toward \object{IK Tau}, including vibrationally excited states, 
  enabling a detailed study of the molecular abundances and excitation temperatures.
  In particular, we highlight the detection of NS and H$_2$CO with fractional abundances of
  $f$(NS)$\sim$10$^{-8}$ and $f$(H$_2$CO)$\sim$[10$^{-7}$--10$^{-8}$].
  Most of the molecules display rotational temperatures between 15 and 40\,K. NaCl and  SiS
  isotopologues display rotational temperatures higher than the average ($\sim$65\,K).
  In the case of SO$_2$ a warm component with \trot$\sim$290\,K is also detected.}
   {With a total of $\sim$350 lines detected of 34 different molecular species (including different isotopologues), \object{IK Tau} displays a rich chemistry 
   for an oxygen-rich circumstellar envelope.
   The detection of carbon bearing molecules like H$_2$CO, as well as the discrepancies found between our derived abundances
   and the predictions from chemical models for some molecules, highlight the need for a revision of standard chemical models.
   We were able to identify at least two different emission components in terms of rotational temperatures. 
   The warm component, which is mainly traced out by SO$_2$, is probably arising from the inner regions 
   of the envelope (at $\lesssim$8\rstar) where SO$_2$ has a fractional abundance of $f$(SO$_2$)$\sim$10$^{-6}$. 
   This result should be considered for future investigation of the main formation channels of this, and other, parent species 
   in the inner winds of O-rich AGB stars, which at present are not well reproduced by current chemistry models.}


   \keywords{Astrochemistry -- 
             Circumstellar matter --
             Line: identification -- 
             Stars: abundances --
             Stars: AGB and post-AGB --
             Stars: individual: \object{IK Tau}}

   \maketitle
\section{Introduction}\label{sec:intro}
Asymptotic giant branch (AGB) stars are the main contributors to the interstellar medium (ISM) chemical enrichment.
The physical conditions, that is, the high densities ($\gtrsim$10$^{12}$\,cm$^{-3}$) 
and temperatures ($\sim$2000--3000\,K),  in their atmospheres allow the formation of stable molecules.
All this molecular material is driven by the slow AGB wind creating a circumstellar envelope (CSE) that surrounds the star,
up to regions where the interstellar UV field photodissociates the molecules. 
Carbon and oxygen are the two most abundant and reactive elements in the atmospheres and winds of AGB stars, after hydrogen.
All the possible carbon monoxide, which is a very stable molecule, is formed and then depending on which element (carbon or oxygen) 
is in excess, other molecules will be formed.
Hence, the chemistry in these objects mainly depends on the elemental carbon to oxygen ratio, being O-rich ([C]/[O]$<$1), 
C-rich ([C]/[O]$>$1) or S-type stars ([C]/[O]$\sim$1) \citep[e.g.][]{olo96Rev}.

Since the first detection of CO in the millimeter wavelength range toward the CSE of an AGB star \citep{sol71}, the observations 
of the molecular emission of CSEs in that wavelength domain have increased \citep[e.g.][]{mor75,buj91,cer15}. 
These studies have been mostly focused on C-rich CSEs given that carbon is more chemically active than oxygen 
and, therefore, C-rich envelopes are expected and observed to display a large variety of different 
molecular species \citep[e.g.][]{cer00,smi15}.
However, the number of studies in the millimeter wavelength range has increased in the last years, and these studies have 
evidenced that O-rich CSEs do also host a rich variety of molecules \citep[][and references therein]{ziu07,kim10,deb13,san15}.

Our motivation is to observe and to study the molecular content of the oxygen-rich CSE \object{IK Tauri} which
is one of the most studied O-rich CSEs, and is considered a reference of its class.

\subsection{\object{IK Tauri}}\label{sec:iktau}
\object{IK Tauri} (hereafter \object{IK Tau}), also known as \object{NML Tau}, is a Mira-type variable star with a period of 
470\,days and a spectral type $\sim$M9 \citep{pes67,win73,alc99}.
This star was discovered by \cite{neu65} and it is located at 
$\alpha$(J2000)=3$^{\mathrm{h}}$53$^{\mathrm{m}}$28\secp87 and $\delta$(J2000)=11$^{\circ}$24$'$21\arcsecp7
\citep{cut03}.
The distance to \object{IK Tau} was estimated to be 250--265\,pc \citep{hal97,olo98}.
The systemic velocity of the star with respect to the local standard of rest is \vsys$\sim$34\,\kms\ 
\citep[][and references therein]{kim10}.
Its effective temperature is $T_{\mathrm{eff}}$$\sim$2200\,K and the stellar radius is \rstar$\sim$2.5$\times$10$^{13}$\,cm 
\citep{dec10a}.

The star is surrounded by a an O-rich CSE, composed of dust and gas, 
which is the result of mass loss at a rate that has been estimated by different methods.
\cite{ner98} estimated \mloss$\sim$3.8$\times$10$^{-6}$\,\my\ 
from the model of $^{12}$CO $J$=1--0 and $J$=2--1 lines.
\cite{gon03} estimated \mloss$\sim$3.0$\times$10$^{-5}$\,\my\ 
from the model of $^{12}$CO $J$=1--0, $J$=2--1, $J$=3--2, and $J$=4--3 lines.
Recent modelling of the $J$=3--2, $J$=4--3, and $J$=7--6 lines of $^{12}$CO yielded \mloss$\sim$4.7$\times$10$^{-6}$\,\my\ \citep{kim10}.
The terminal expansion velocity of the CSE is \vinf$\sim$18.5\,\kms\ as measured from
\cite{deb13} and references therein.
The size of the CSE depends on the molecule used to trace it out.  
\cite{buj91} gave a half-intensity diameter of $\theta_{\mathrm{1/2}}$$\sim$16-17\arcsec\
measured for $^{12}$CO $J$=1--0 and $^{12}$CO $J$=2--1 emission detected with the IRAM-30\,m telescope.
\cite{kim10} measured a $\theta_{\mathrm{1/2}}$$\sim$20\arcsec\ for $^{12}$CO $J$=3--2 with the Atacama Pathfinder 
EXperiment (APEX) telescope.
Recent observations with the Plateau de Bure Interferometer (PdBI), showed that \doce\ displays a
$\theta_{\mathrm{1/2}}$$\sim$18\arcsec\ \citep{cas10}. 
Also, HCN $J$=1$-$0 was observed with the Owens Valley millimeter-array by \cite{mar05},
displaying a size of $\theta_{\mathrm{1/2}}$$\sim$3\arcsecp85.
The size of the SiO v=0 $J$=2$-$1 emission is $\theta_{\mathrm{1/2}}$=2\arcsecp2$\pm$0\arcsecp1\ as determined with
PdBI \citep{luc92}.
Finally, the emission of several lines of PN and PO has been mapped with the SubMillimeter Array (SMA), 
with $\theta_{\mathrm{1/2}}$$\lesssim$0\arcsecp65\ \citep{deb13}. 

The O-rich CSE around \object{IK Tau} displays maser emission of OH, H$_2$O, and SiO \citep[][and references therein]{lan87,bow89,alc92,kim10}. 
It also shows thermal molecular emission of
$^{12}$CO, $^{13}$CO, SiO, $^{29}$SiO, $^{30}$SiO, OH, SiS, 
HCN, HC$_3$N, H$_2$S, SO, SO$_2$, NaCl, H$_2$O, H$_{2}$$^{17}$O, H$_{2}$$^{18}$O, NH$_3$, CS, CN, PO, PN, 
AlO, and tentatively, HNC 
\citep[][and references therein]{lin88,omo93,buj94,mil07,kim10,dec10b,men10,jus12,deb13,deb15}.
Some of the molecules observed up to date in \object{IK Tau} have been compared with chemical models
and their line emission analysed with radiative transfer models
\citep{wil97,dua99,che06,kim10,dec10a,dan16,gob16,lix16}.
Nevertheless, most of the reported abundances in previous studies were derived from the analysis of a moderated 
number of lines \citep[e.g.][]{mil07,deb13}. 
Additionally, discrepancies remain between the predicted abundances and the observations, for example for
SO$_2$ \citep{dec10a,gob16}.

\subsection{This paper}\label{sec:thispaper}
In this article we report the millimeter wavelength survey between $\sim$79 and $\sim$356\,GHz 
carried out with the IRAM-30\,m telescope toward \object{IK Tau}, which allowed us to detect 
rotational lines of 34 different species (including isotopologues).
We detected for the first time in this source HCO$^+$, NS, H$_2$CO, and NS, as well as
several isotopologues of previously identified molecules, such as C$^{18}$O, Si$^{17}$O, Si$^{18}$O, $^{29}$SiS, $^{30}$SiS, 
Si$^{34}$S, H$^{13}$CN, $^{13}$CS, C$^{34}$S, H$_2$$^{34}$S, $^{34}$SO, and $^{34}$SO$_2$.
For molecules with previous detections reported in the literature, we increased significantly the number of transitions observed, 
which is needed for a robust estimate of the excitation conditions and molecular abundances.
We report the results of our analysis based on population diagrams, used to derive rotational temperatures 
and column densities.
We also estimated fractional abundances which have been compared with 
values derived from previous observations and with predictions by chemical models.
In the particular case of SO$_2$, we performed a radiative transfer calculation to study the excitation conditions of the $\sim$90 
lines detected with more detail. This suggests that this molecule is rather abundant 
not only in the intermediate and outer envelope but also in the inner, hotter and more dense regions of the CSE, 
where we estimate an average fractional abundance of $\sim$10$^{-6}$ (with respect to H$_2$).
This radiative transfer model also allowed us to verify the validity of the population diagram analysis 
for the rest of the species detected. 

\section{Observations}\label{sec:obs}
The observations presented in this paper correspond to a sensitive millimeter-wavelength ($\sim$79-356\,GHz) survey carried out
with the \iram\ telescope toward the CSEs of two O-rich evolved stars, \object{IK Tau} and \object{OH231.8+4.2}, during several observing runs from 2009 to 2014.
Partial results from the survey toward \oh\ are reported in \citet{san14,san15} and \citet{vel13,vel15}.

We used the heterodyne Eight MIxer Receiver (EMIR) working at four different wavelengths bands: 
E090=3\,mm, E150=2\,mm, E230=1.3\,mm, and E330=0.9\,mm \citep[][]{car12}.
This receiver system was operated in single-sideband (SSB) mode for the E150 band and in dual sideband (2SB) mode
for bands E090, E230, and E330.
Two polarizations -- horizontal and vertical -- were available per sideband.
The typical EMIR value image band rejection is  approximately -14\,dB or better,
this implies that the peak intensity of a line entering through the image band is only $\lesssim$4\,\%\ of its real value.
We verified this value for the image band rejection measuring the relative intensities of strong lines that appear both in 
the signal and image bands.

Four different backends or spectrometers were connected to each receiver depending on their availability:
the WIde Lineband Multiple Autocorrelator (WILMA), the fast Fourier Transform Spectrometer (FTS),
the VErsatile SPectrometer Array (VESPA) and the 4\,MHz spectrometer.
The capabilities of these spectrometers and their usage are summarised in Table\,\ref{tab:spectrometers}.

\begin{table}[hbt!]
\caption{Specifications for the spectrometers used.}
\label{tab:spectrometers}
\begin{center}
\begin{tabular}{l c c c c}
\hline\hline
Name & SR & VR & IBWEP & Usage \\
-            & (MHz)               & (\kms)              & (GHz)                     & - \\
\hline
WILMA	& 2.000	& 1.7$-$7.6	& 3.7	& E090, E150, E230, E330 \\
FTS	& 0.195	& 0.2$-$0.7	& 4.0	& E230, E330		 \\
VESPA	& 1.250	& 1.0$-$4.7	& 0.4	& E090, E150, E230, E330 \\
4\,MHz	& 4.000	& 3.4$-$15.2	& 4.0	& E150, E230, E330  	 \\
\hline
\hline
\end{tabular}
\end{center}
\tablefoot{
(Col. 2) Spectral resolution; 
(Col. 3) spectral resolution in velocity units, the highest velocity resolution (the first number of the range) corresponds to 356\,GHz and the lowest resolution to 79\,GHz; 
(Col. 4) instantaneous bandwidth in each polarization;
(Col. 5) EMIR bands observed with the corresponding spectrometer. 
}
\end{table}

The observational technique used was the wobbler switching with a single pointing toward the position of the 
source (see Sect.\,\ref{sec:iktau}) and a wobbler throw of 120\arcsec\ in azimuth.
We configured different setups (tuning steps) to observe both polarizations simultaneously until we covered the total frequency
range available for each EMIR band. 
We selected a small overlap between adjacent setups to ensure a uniform calibration across the bands.
During the observations we checked regularly the pointing and focus of the antenna every $\sim$1.5 and $\sim$4\,h, respectively,
on strong nearby sources. On-source integration times per setup were $\sim$1\,h.
Calibration scans on the standard two loads + sky system were taken every $\sim$18\,min
using the atmospheric transmission model (ATM) adopted by \iram\ \citep{cer85,par01}.
Errors in the absolute flux calibration are expected to be $\la$25\%.

\begin{figure}[hbtp!] 
\centering
\includegraphics{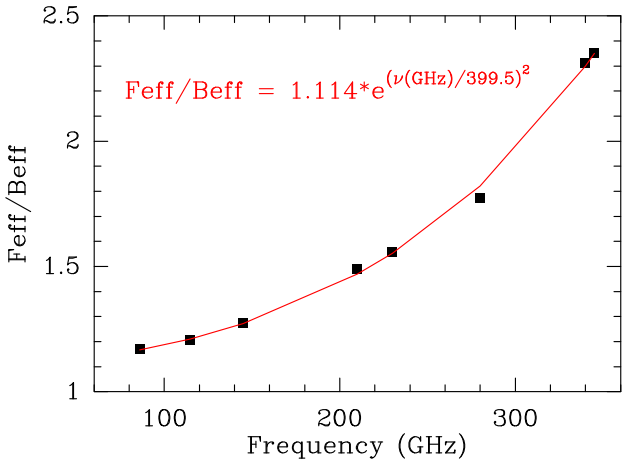}
\caption{Fit of the inverse of the main beam efficiency for the \iram\ with EMIR.}\label{fig:eff_fits}
\end{figure}

The data reduction, analysis and also most of the graphic representation were done using the
GILDAS\footnote{GILDAS is a world-wide software to process, reduce and analyse astronomical single-dish and interferometric 
observations mainly. It is maintained by the Institut de Radioastronomie Millim\'{e}trique (IRAM). 
See \tt{\,http://www.iram.fr/IRAMFR/GILDAS}}
software package.
The standard procedure followed to reduce the data and obtain the final spectra consists of
the flagging of bad channels, the flagging of low-quality scans, the baseline substracting, and finally, 
the averaging of individual scans.

The output spectra obtained from the antenna are calibrated in antenna temperature (\tant),
which can be converted to main beam temperature (\tmb) and brightness temperature (\tb) according to:
\begin{alignat}{3}
T_{\rm B} &= T^{*}_{\rm A} \; \eta^{-1} \; b\!f\!f^{-1},\label{eq:1} \\
\eta^{-1} &= F_{\rm eff}/B_{\rm eff},\label{eq:2}\\
b\!f\!f^{-1} &= (\theta^2_{\mathrm{b}}+\theta^2_{\mathrm{s}})/\theta^2_{\mathrm{s}},\label{eq:3}
\end{alignat}
where $\eta$ is the main beam efficiency,
$b\!f\!f$ is the beam-filling factor,
$B_{\mathrm{eff}}$ is the main-beam efficiency of the antenna,
$F_{\mathrm{eff}}$ is the forward efficiency of the antenna,
$\theta_{\mathrm{s}}$ is the size (diameter) of the emitting region of the source, and 
$\theta_{\mathrm{b}}$ is the half power beamwidth (HPBW) of the main beam of the antenna. 
See Table\,\ref{tab:iram_param} for a summary of the relevant telescope parameters.

\begin{table}[hbt!] 
\caption{Main parameters of the \iram\ antenna measured with EMIR between 2009 and 2013 at representative 
frequencies.}
\label{tab:iram_param}
\begin{center}
\begin{tabular}{c c c c c}
\hline\hline
Frequency & B$_{\mathrm{eff}}$ & F$_{\mathrm{eff}}$ & HPBW & S/T$_A^*$ \\
(GHz)     &    (\%)         &   (\%)             & (\arcsec)  & (Jy/K) \\
\hline
86      & 81 & 95 & 29\tablefootmark{\dag}     & 5.9\tablefootmark{\dag}  \\ 
115     & 78 & 94 & -                          & -                        \\
145     & 73 & 93 & 16\tablefootmark{\dag}     & 6.4\tablefootmark{\dag}  \\ 
210     & 63 & 94 & 11\tablefootmark{\dag}     & 7.5\tablefootmark{\dag}  \\
230     & 59 & 92 & 10.7\tablefootmark{\ddag}  & -                        \\
280     & 49 & 87 & -                          & -                        \\
340     & 35 & 81 & 7.5\tablefootmark{\dag}    & 10.9\tablefootmark{\dag} \\
345     & 34 & 80 & -                          & -                        \\ 
\hline 
\hline
\end{tabular}
\end{center}
\tablefoot{
(Col. 1) Representative frequency;
(Col. 2) beam efficiency; 
(Col. 3) forward efficiency; 
(Col. 4) half power beam width; 
(Col. 5) Flux density to antenna temperature conversion factor in Jansky per Kelvin for a point-like source.
\tablefoottext{\dag}{2009 values.}
\tablefoottext{\ddag}{2012 value.}\\
The values without symbols were reported on 2013.
Source:\ {\tiny{\tt{\,http://www.iram.es/IRAMES/mainWiki/Iram30mEfficiencies\label{foot}}}}
}
\end{table}

The half power width of the main beam can be approximated, using the values in Table\,\ref{tab:iram_param},
to a good accuracy by the expression:
\begin{equation}\label{eq:hpbw}
\theta_{b}\,(\arcsec) = 2460/\nu\,(GHz),
\end{equation}
and the inverse of the main beam efficiency can be fitted (see Fig.\ref{fig:eff_fits})
using the parameters in Table\,\ref{tab:iram_param}, to obtain:
\begin{equation}\label{eq:mbeff}
\eta^{-1}\equiv F_{\mathrm{eff}}/B_{\mathrm{eff}} = 1.114\,\exp\{(\nu\,(GHz)/399.5)^{2}\}.
\end{equation}

\begin{figure*}[hbtp!] 
\centering
\includegraphics{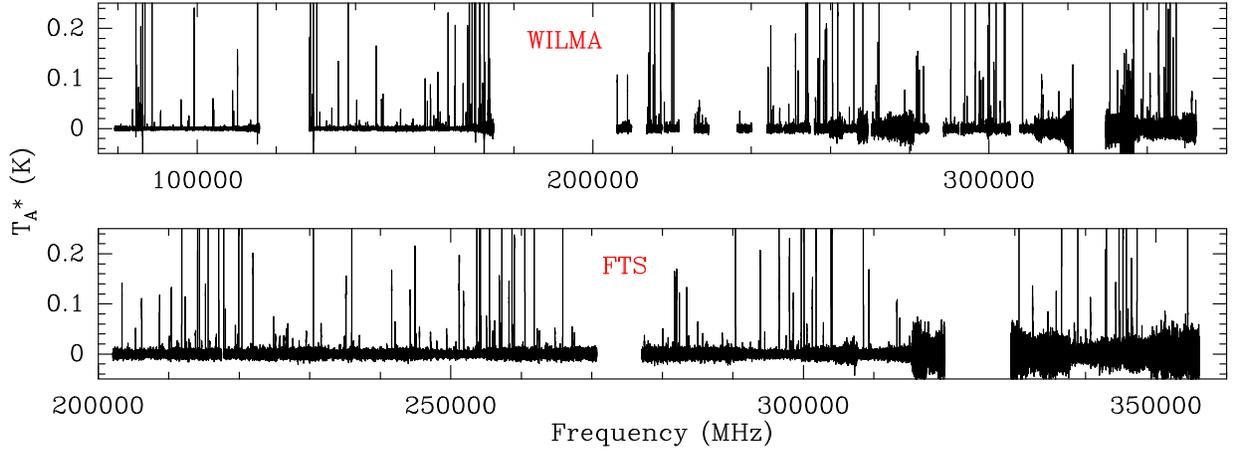} 
\caption{Overall view of the survey observed with WILMA (top) and FTS (bottom).
The spectral resolution is $\Delta\nu$=2\,MHz.}\label{fig:allsurvey}
\end{figure*}

\subsection{Observational results}\label{sec:obsres}
Our results are based mainly on the spectra obtained with WILMA and FTS given their better spectral resolution and bandwidth. 
The data obtained with the 4\,MHz spectrometer were used to check the edges of the setups that were only
observed with WILMA, given that the 4\,MHz bandwidth is slightly larger than the bandwidth of WILMA, but it has
the same bandwidth than that of the FTS (see Table\,\ref{tab:spectrometers}).
VESPA data were used only to check certain line profiles. 
However, only WILMA was available to cover the full EMIR wavelength range. For example, FTS was not avaliable
in the E150 band at the epoch of the observations and also some technical issues prevented us from use the FTS with 
the E090 band.

\begin{figure}[] 
\centering
\includegraphics{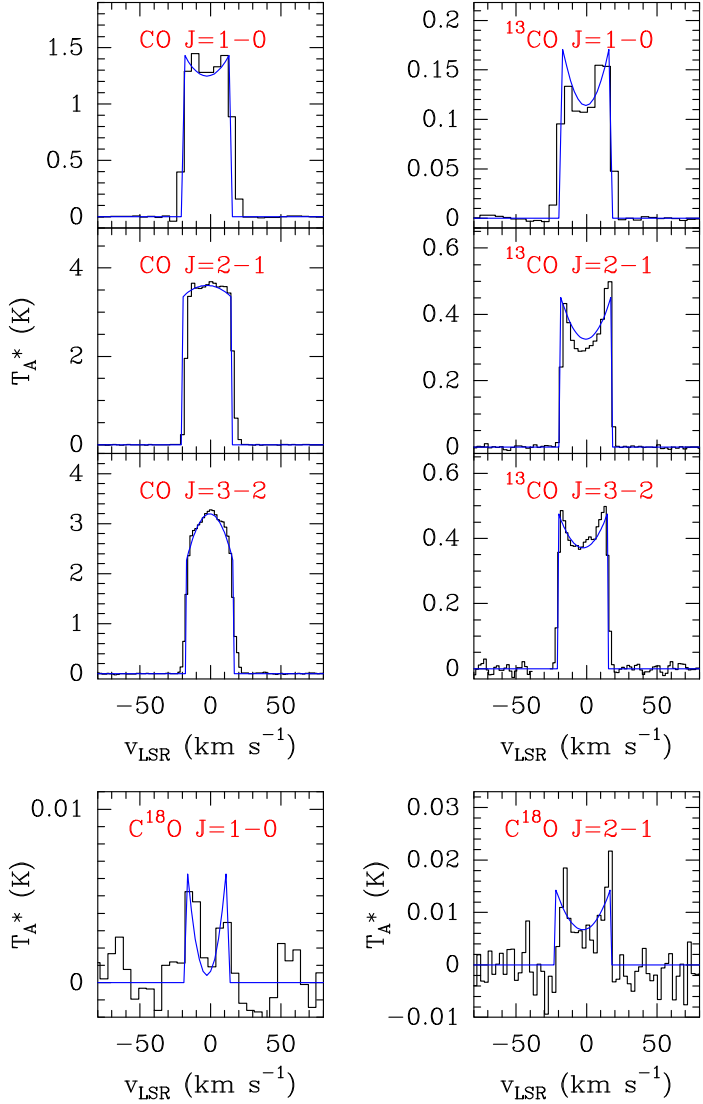}
\caption{$^{12}$CO, $^{13}$CO, and C$^{18}$O lines detected with our millimeter wavelength survey with a spectral resolution
of $\Delta\nu$=2\,MHz. The fit of each line to a function of the type given by Eq.(\ref{eq:shell}) is shown in blue.
}
\label{fig:coprofs}
\end{figure}

The spectra of the full mm-wavelength survey carried out with \iram\ toward \object{IK Tau} with the WILMA and the FTS spectrometers 
can be seen in Fig.\,\ref{fig:allsurvey} and also in more detail in Fig.\,\ref{fig:survey1}.
The $^{12}$CO, $^{13}$CO, and C$^{18}$O line profiles are shown in Fig.\,\ref{fig:coprofs}.
In Table\,\ref{tab:survey_results} we show a summary of the observational results.

\begin{table}[hbt!] 
\caption{Summary of the observational results.}
\label{tab:survey_results}
\begin{center}
\begin{tabular}{c c c c c}
\hline\hline
Band	& SP	& $\nu_{obs}$	& rms	& Opacity \\
-	& -	& (GHz)		& (mK)	& -	  \\
\hline
E090	& WILMA	& 79.3--115.7	&  1.5\,(0.5) & 0.12\,(0.05)  \\
E150	& WILMA	& 128.4--174.8	&  2.7\,(1.3) & 0.19\,(0.11)  \\ 
E230	& FTS 	& 202.1--270.7\tablefootmark{\dag} & 4.7\,(0.6)   & 0.18\,(0.03) \\
E330	& FTS 	& 277.1--356.2\tablefootmark{\ddag} & 12.8\,(10.4) & 0.26\,(0.16) \\ 
\hline 
\hline
\end{tabular}
\end{center}
\tablefoot{
Values given between parentheses represent 1\,$\sigma$\ of the value.
(Col. 1) EMIR band;
(Col. 2) spectrometer; 
(Col. 3) observed frequency windows in GHz; 
(Col. 4) root mean square (rms) noise in units of T$_a^*$ for a spectral resolution of 2\,MHz;
(Col. 5) zenith atmospheric opacity at the observed frequency.
\tablefoottext{\dag}{217.5--217.8 gap.}
\tablefoottext{\ddag}{325.0--329.5 gap.}
}
\end{table}

\section{Line identification}\label{sec:id}
For the line identification, we used the public line catalogues from the Cologne
Database for Molecular Spectroscopy \citep[CDMS,][]{mul05} and the Jet
Propulsion Laboratory \citep[JPL,][]{pic98}, together with 
a private spectroscopic catalogue that assembles information for
more than five thousand spectral entries (atoms and molecules, 
including vibrationally excited states), compiled
from extensive laboratory and theoretical works \citep[the MADEX code,][]{madex}.

Given the wavelength range covered in our observations, we detected, mainly, rotational transitions 
in the ground vibrational state, but also, rotational
transitions in higher vibrational states (i.e. SiO v=1, SiO v=2, SiO v=3, $^{29}$SiO v=1, $^{29}$SiO v=2, $^{29}$SiO v=3, 
$^{30}$SiO v=1, $^{30}$SiO v=2, H$_2$O $\nu_{\mathrm{2}}$=1, and H$_2$O $\nu_{\mathrm{2}}$=2).
We established the detection limit at $\geq$5$\sigma$ with respect to the integrated intensity
for well detected features, and between 3$\sigma$ and 5$\sigma$ for tentative detections, although, these tentative 
detections may be more reliable when they belong to well-known species that have been identified through other stronger lines.

We detected $\sim$450 spectral features in the spectra of \object{IK Tau}, $\sim$90\,\% of which are lines 
in the signal side band of the receivers.
Of the $\sim$400 signal band features, $\sim$350 lines have been unambiguously identified with
rotational transitions of 34 different species 
(including vibrationally excited states), which are reported in Table\,\ref{tab:measures}
and Table\,\ref{tab:variability},
along with some of their spectroscopic parameters and their fitted parameters. 
The rest of the lines ($\sim$35) remain unidentified, although, we proposed a tentative identification for some of them 
(see Table\,\ref{tab:uis}). 

There were several lines that we assigned to spurious features produced in the receivers in the range 167475--174800\,MHz
(Table\,\ref{tab:loproblems}). These spurious features are symmetrical replicas at both sides of certain strong real 
emission lines, with the intensities of the replicas decreasing with the frequency distance to the real feature. 
We show in Fig.\,\ref{fig:ikprobs} an example of this problem. 
The features of the image band and the spurious features were blanked in the final data.

We identified for the first time toward \object{IK Tau}, rotational lines of: 
HCO$^+$, NO, H$_2$CO, NS,
C$^{18}$O, SiO v=2, $^{29}$SiO v=2, $^{29}$SiO v=3, $^{30}$SiO v=2, Si$^{17}$O, Si$^{18}$O, 
$^{29}$SiS, $^{30}$SiS, Si$^{34}$S, SiS v=1, 
H$^{13}$CN, $^{13}$CS, C$^{34}$S, H$_2$$^{34}$S, $^{34}$SO, 
$^{34}$SO$_2$, and H$_2$O $\nu_{\mathrm{2}}$=2, .
We also detected rotational lines of:
$^{12}$CO, $^{13}$CO, SiO, SiO v=1, SiO v=3, $^{29}$SiO, $^{29}$SiO v=1, $^{30}$SiO v=1, SiS, HCN, HNC, CS, NaCl, 
H$_2$S, SO, SO$_2$, PN, PO, CN, H$_2$O, and H$_2$O $\nu_{\mathrm{2}}$=1.
The result of the identification can be seen in Fig.\,\ref{fig:survey1}.

\begin{figure}[!hbtp]
\centering
\includegraphics{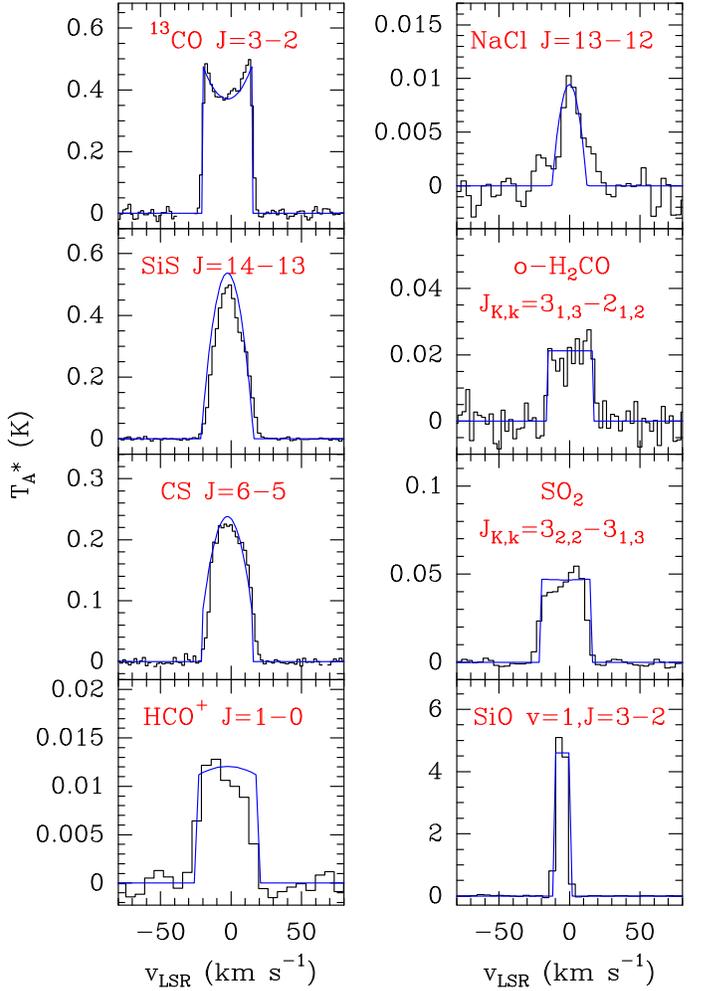}
\caption{Representative profiles of the lines observed. From top-left to bottom-right: U-shaped, 
gaussian-like, triangular, flat-topped, parabolic, complex, flat-topped, and a narrow maser-like line.
The spectral resolution is $\Delta\nu$=2\,MHz. The fit of each line to a function of the type
given by Eq.(\ref{eq:shell}) is shown in blue.}                                                                                               
\label{fig:profiles}
\end{figure}

\section{Data analysis}\label{sec:analysis}
\subsection{Line profiles}\label{sec:linepro}
Most of the lines show profiles that can be reasonably well fitted with the so-called shell profile provided by the software 
CLASS\footnote{See \tt{\,http://www.iram.fr/IRAMFR/GILDAS}}:
\begin{equation}\label{eq:shell}
f(\nu)=\frac{A}{\Delta\nu}\frac{1+4H[(\nu - \nu_o)/\Delta\nu]^{2}}{1+H/3},
\end{equation}
\noindent
where $A$ is the area under the profile, $\nu_{o}$ is the central frequency, $\Delta\nu$ is the full width at zero 
intensity, and $H$ is the horn to centre ratio.
The expansion velocity (\vexp) can be related to the $\Delta\nu$ of a line through the expression:
\begin{equation}\label{eq:vexp}
\varv_{exp}=c\frac{\Delta\nu/2}{\nu_{o}},
\end{equation}
\noindent
where $c$ is the speed of light. 

In the case of a spherical CSE, there are several typical line profiles which are commonly found: U-shaped, parabolic, flat-topped, 
gaussian-like or triangular.
Each type of profile has a particular interpretation in terms of the size of the emitting region compared to the 
$\theta_{\mathrm{b}}$ of the telescope, the optical thickness of the line, and the kinematical properties of the gas responsible 
of the spectroscopic feature \citep[e.g.][]{zuc87,hab04}.
All these profiles are described to a good accuracy by the shell function described before in Eq.\,(\ref{eq:shell}).

In Fig.\,\ref{fig:profiles} we show a sample of the line profiles observed.
Most of the lines display profiles that match one of the types mentioned before, although, some profiles are 
complex and have to be considered carefully in the analysis. 
We also observed lines that are significatively narrower and 
more intense than the average, that is, they show maser like spectral profiles (see Fig.\,\ref{fig:profiles} bottom-right panel). 
For the sake of consistency we used the shell profile described in Eq.\,(\ref{eq:shell}) to fit all the lines detected,
even for the narrow lines since that should give us an approximated idea of the characteristic velocity in the regions of the 
wind acceleration.
The aim of the fit is to estimate the centroids and linewidths, however, the velocity integrated intensities given in 
Table\,\ref{tab:measures}, Table\,\ref{tab:variability}, and Table\,\ref{tab:uis} were obtained integrating the whole line profiles.
The line profiles observed are discussed in detail for each detected molecule in Sect\,\ref{sec:discussion}.

\subsection{Population diagrams and fractional abundances}\label{sec:anapop}
Using the population diagram technique \citep{gol99}, we derived 
rotational temperatures and column densities averaged in the emitting region of the molecules detected.
These values were derived using the following equation:
\begin{equation} 
\label{eq:rotdiag}
\ln\,\left(\frac{N_{u}}{g_{u}}\right)=\ln\,\left(\frac{3k_{B}W}{8\pi^{3}\, \nu \, S_{ul}\,\mu^{2}}\right)=\ln\,\left(\frac{N}{Z}\right)-\frac{E_{u}}{k_B T_{rot}},
\end{equation}
\noindent
where $N_{\mathrm{u}}$ is the column density of the upper level,
$g_{\mathrm{u}}$ is the degeneracy of the upper level, $W$ is the
velocity-integrated intensity of the line, $k_{\mathrm{B}}$ is
the Boltzmann constant, $\nu$ is the rest frequency of the transition,
$S_{\mathrm{ul}}$ is the line strength, $\mu$ is the
dipole moment of the molecule, $N$ is the total column density, $Z$ is the partition
function, \eu\ is the upper level energy of the transition, and \trot\ is the rotational temperature. 
The results of the populations diagrams are reported in Sect.\,\ref{sec:respop} and Appendix\,\ref{sec:app_rd}.

This method relies on local thermodynamical equilibrium (LTE), optically thin emission, 
as well as an uniform rotational temperature for the gas shell, and it permits the analysis of all the molecules using a homogeneus criteria.
Also, the size of the emitting region for each molecule has to be known to account for the proper dilution
of the emission. 
We have adopted approximate sizes of the emitting regions of 
$\theta_{\mathrm{s}}$($^{12}$CO,$^{13}$CO)=18\arcsec\ which is a representative value for the 
sizes measured in \cite{buj91,cas10}, and \cite{kim10}.
We also adopted the sizes
$\theta_{\mathrm{s}}$(HCN,H$^{13}$CN)=4\arcsec\ \citep{mar05},
$\theta_{\mathrm{s}}$=2\arcsec\ for SiO isotopologues \citep{luc92}, and 
$\theta_{\mathrm{s}}$$\lesssim$0\arcsecp7 for PO and PN \citep{deb13}.
Similar values have been also used for these molecules by \cite{kim10}.
For the rest of the molecules, which have not been mapped, we assumed emitting region sizes according to observational 
constraints, previous estimations, or predictions by radiative transfer and chemical models. 
These values are given in Table\,\ref{tab:rdresult} and discussed in Sect.\,\ref{sec:discussion}.
We note that the adopted sizes are uncertain for some molecules and also that the emitting region size
may vary for each transition of a given molecule.
In general, an underestimation of the size of any emitting region 
will cause an underestimation of the beam filling factor (see Eq.\,(\ref{eq:3})) and, thus, 
an overestimation of the abundance, and vice versa.

There are additional sources of uncertainty in the values derived from the population 
diagrams. In particular, the kinetic temperature (\tkin) throughout the CSE is not expected to be constant.
This effect is less important for the molecules whose emission arises from regions with very uniform physical conditions 
(inner layers, photodissociation layers, etc.), and it must be more important for species whose emission extends through the whole 
envelope, such as probably SO$_2$, for which we did a more detailed radiative transfer analysis (see Sect.\,\ref{sec:madex}).

In the population diagrams, we have included lines (unambigously identified and unblended) with fluxes above 5$\sigma$. 
For those molecules well known to be present in the envelope of \object{IK Tau}, we have also included lines with 3--5$\sigma$ detections. 
In the case of molecules with a hyperfine structure that cannot be spectrally resolved with the spectral resolution achieved,
that is, CN, NS, and NO, we measured the total integrated intensity of the blend of the hyperfine components and we 
calculated the sum of the strength of the hyperfine components to compute their population diagrams.
In the case of H$_2$S we fitted together the lines of the ortho and para species due to the 
scarce number of lines detected for each species individually.

We have not calculated population diagrams for the molecules SiO v=1, SiO v=2, SiO v=3, $^{29}$SiO v=1, 
$^{29}$SiO v=2, $^{29}$SiO v=3, $^{30}$SiO v=1, $^{30}$SiO v=2, H$_2$O, H$_2$O $\nu_{\mathrm{2}}$=1 or H$_2$O $\nu_{\mathrm{2}}$=2. 
Some of these lines are masers and their populations are expected to strongly deviate from a Boltzmann distribution.
Also, the intensity of the lines of these molecules may vary with the stellar pulsation phase and, given 
that the observations spread over a period of time of approximately five years, the excitation conditions 
may have changed along the observational runs (see Sect.\,\ref{sec:variability}).

The fractional abundances averaged in the emitting region for the molecules detected are calculated using 
the following equation:
\begin{equation}\label{eq:abundances}
f(X)=N(X)\frac{f(^{13}CO)}{N(^{13}CO)},
\end{equation}
\noindent
where $f$(X) and $N$(X) represent the fractional abundance (with respect to H$_2$) and column density of the molecule analysed,
respectively.
$N$($^{13}$CO) is the column density that we derived from the $^{13}$CO population diagram, 
and 
$f$($^{13}$CO)=1.4$\times$10$^{-5}$ \citep[][]{dec10a}.

\subsection{Radiative transfer model: MADEX}\label{sec:madex}
MADEX \citep{madex} is a radiative transfer code which is able to operate under LTE and large velocity gradient (LVG) 
approximation \citep{gol74}. 
It solves the radiative transfer problem coupled with statistical equilibrium equations to derive the radiation field and 
populations of the levels on each point of a gas cloud. Then, the emergent profile of the lines is obtained through ray-tracing. 

We used MADEX to calculate line opacities, excitation temperatures, and critical densities for the physical conditions
expected at a given distance from the star, that is, $n(r)$ and \tkin$(r)$ (see Table\,\ref{tab:iktau}), along with the 
column densities derived from the population diagrams (given in Table\,\ref{tab:rdresult}), and the \vexp\ derived from
the line fitting (given in Table\,\ref{tab:measures}) for all the molecules detected.
We systematically used this procedure to compare the line opacities obtained from the interpretation of the line profiles and to verify 
if a given molecule could be sub-thermally excited or not. These calculations are detailed in each subsection of the Sect\,\ref{sec:discussion}.
The sets of the collisional coefficients used for these calculations (when available) are:
CO \citep{yan10},
SiO \citep{day06},
H$_2$O \citep{dan11},
HCN \citep{ben12},
CS \citep{liq07cs},
SiS \citep{tob08},
NaCl \citep[][]{qui16},
SO \citep{liq07so},
PN \citep{tob07},
CN \citep{liq11},
HCO$^+$ \citep{flo99}, and 
H$_2$CO \citep[][and references therein]{sch05}. 
We also obtained synthetic spectra that helped us to correctly identify the spectral features.

As it is shown in Sect\,\ref{sec:respop}, the detected SO$_2$ transitions have a large span in energies 
(i.e. from 7.7 up to 733.4\,K). The population diagram analysis of this molecule indicates that it is possibly 
tracing out an inner region of the CSE, with a \tkin$\gtrsim$290\,K which is not observed for other molecules, and 
also a more external region with \trot$\sim$40\,K from where most of the emission of the rest of the molecules arises.
To investigate more precisely the excitation conditions of SO$_2$ and to confirm or not the temperature stratification 
inferred from the population diagrams, we have performed a detailed radiative transfer calculation for SO$_2$.
Our main goal is to investigate the presence of SO$_2$ in the innermost regions of the CSE.

\begin{figure}
\centering
\includegraphics{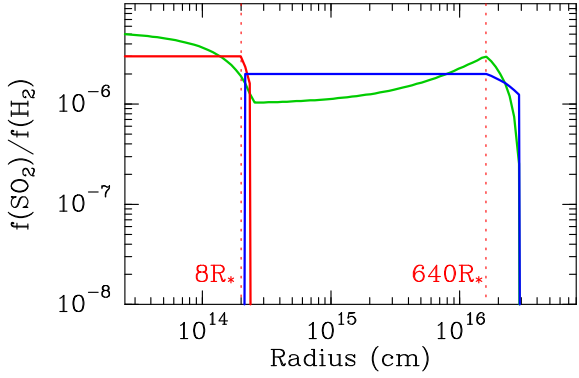}
\caption{Radial profiles of the abundance adopted for the radiative transfer model of the lines of SO$_2$.
We tested three different radial profiles: 
$i$) an inner distribution of SO$_2$ with $f$(SO$_2$)=3$\times$10$^{-6}$ between 1-8\rstar\ (red line), 
$ii$) a constant radial profile with $f$(SO$_2$)=2$\times$10$^{-6}$ between 8-640\rstar\ (blue line), and
$iii$) a two component profile (green line) which includes an inner warm component plus an extended component of SO$_2$.
The results of the different models are shown in Fig.\,\ref{fig:so2rt}.
}                                                                                               
\label{fig:rpso2}
\end{figure}

The physical model consists on a spherical expanding envelope of dust and gas with a constant mass loss rate,
similar to the model presented in \cite{agu12} but corrected for \object{IK Tau} (see Table\,\ref{tab:iktau}).
The radius of the star is 2.5$\times$10$^{13}$\,cm \citep{dec10a}.
The density of particles is calculated with the law of conservation of mass, that is, $n\propto r^{-2}$ (valid for a CSE expanding
at constant velocity). 
In terms of \vexp, the envelope is 
divided in three regions: 
($i$) from 1 to 5\,\rstar\ where \vexp=5\,\kms\ as an average value in this region \citep{dec10b},
($ii$) from 5 to 8\,\rstar\ where \vexp=10\,\kms\ according to the dust condensation radius \citep{gob16}, and 
($iii$) from 8\,\rstar\ to the end of the CSE where \vexp=18.5\,\kms (see Sect.\,\ref{sec:vexp}).
Concerning dust, we adopted a dust-to-gas ratio of 1.3$\times$10$^{-4}$ for the region ($i$), a 
value of 1.2$\times$10$^{-3}$ for region ($ii$) and a value of 2.3$\times$10$^{-3}$ for the region ($iii$),
according to the values given by \cite{gob16} for silicates. 
The dust temperature has been taken from \cite{dec06} (see Table\,\ref{tab:iktau}).
The optical properties of the silicate dust and the size of the grains (0.1\,\microns) have been adopted from \cite{suh99}.
For the microturbulence velocity we adopted the values given in \cite{agu12} for the C-rich CSE IRC+10216.
We used different SO$_2$ abundance profiles (see Fig.\,\ref{fig:rpso2}) in order to investigate the presence of 
warm SO$_2$ in the inner regions of the CSE.
For the calculations we used a set of collisional coefficients which are described in Appendix\,\ref{sec:app_so2}.
The results of the radiative transfer model are shown in Fig.\,\ref{fig:so2rt} and discussed in Sect.\,\ref{sec:so2}.

\begin{table} 
\caption{Parameters of the central Mira-type star and the CSE of \object{IK Tau}.}
\label{tab:iktau}
\centering    
\begin{tabular}{l c c}
\hline\hline
Parameter & Value & Reference      \\ 
\hline   
Distance ($d$)                                 &  265\,pc                                   & a \\
Stellar radius (\rstar)                        &  2.5x10$^{13}$\,cm                         & c \\ 
Stellar effective temperature ($T_\mathrm{*}$) &  2200\,K                                   & c \\
Terminal expansion velocity (\vinf)            &  18.5\,\kms                                & e \\
AGB mass loss rate (\mloss)                    &  8$\times$10$^{-6}$\,\my                   & c \\
Gas kinetic temperature (\tkin)                &  T$_*(r/R_{*})^{-0.6}$                     & c \\
Dust temperature ($T_\mathrm{d}$)              &  T$_*(2r/R_{*})^{-0.4}$                    & b \\
Dust condensation radius ($R_{\mathrm{c}}$)    &  8\,\rstar                                 & d \\
Density of particles (n) & $\dot{M} / (4 \pi r^{2} \langle m_{g} \rangle v_{exp})$ & - \\
\hline                        
\hline                      
\end{tabular}
\tablefoot{
The density of particles is calculated with the law of conservation of mass, where 
$\langle m_{g} \rangle$ is the mean mass of gas particles (2.3\,amu,
after considering H$_2$, He, and CO).
(a): \cite{hal97}; 
(b): \cite{dec06}; 
(c): \cite{dec10a}; 
(d): \cite{gob16}; 
(e): this work (see Sect.\,\ref{sec:vexp}).
}
\end{table}

\begin{figure*}[hbtp!] 
\centering
\includegraphics[scale=1]{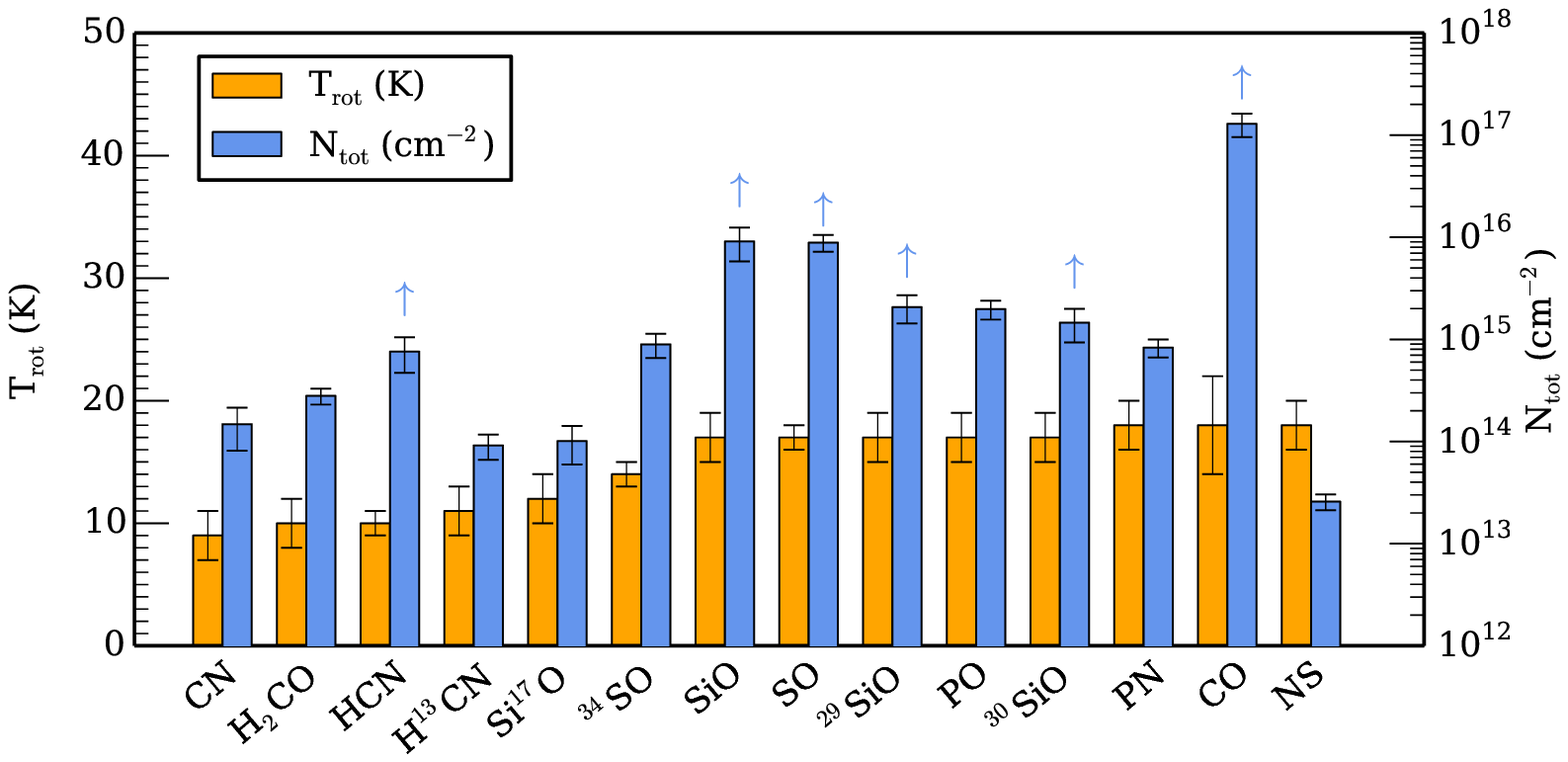}
\includegraphics[scale=1]{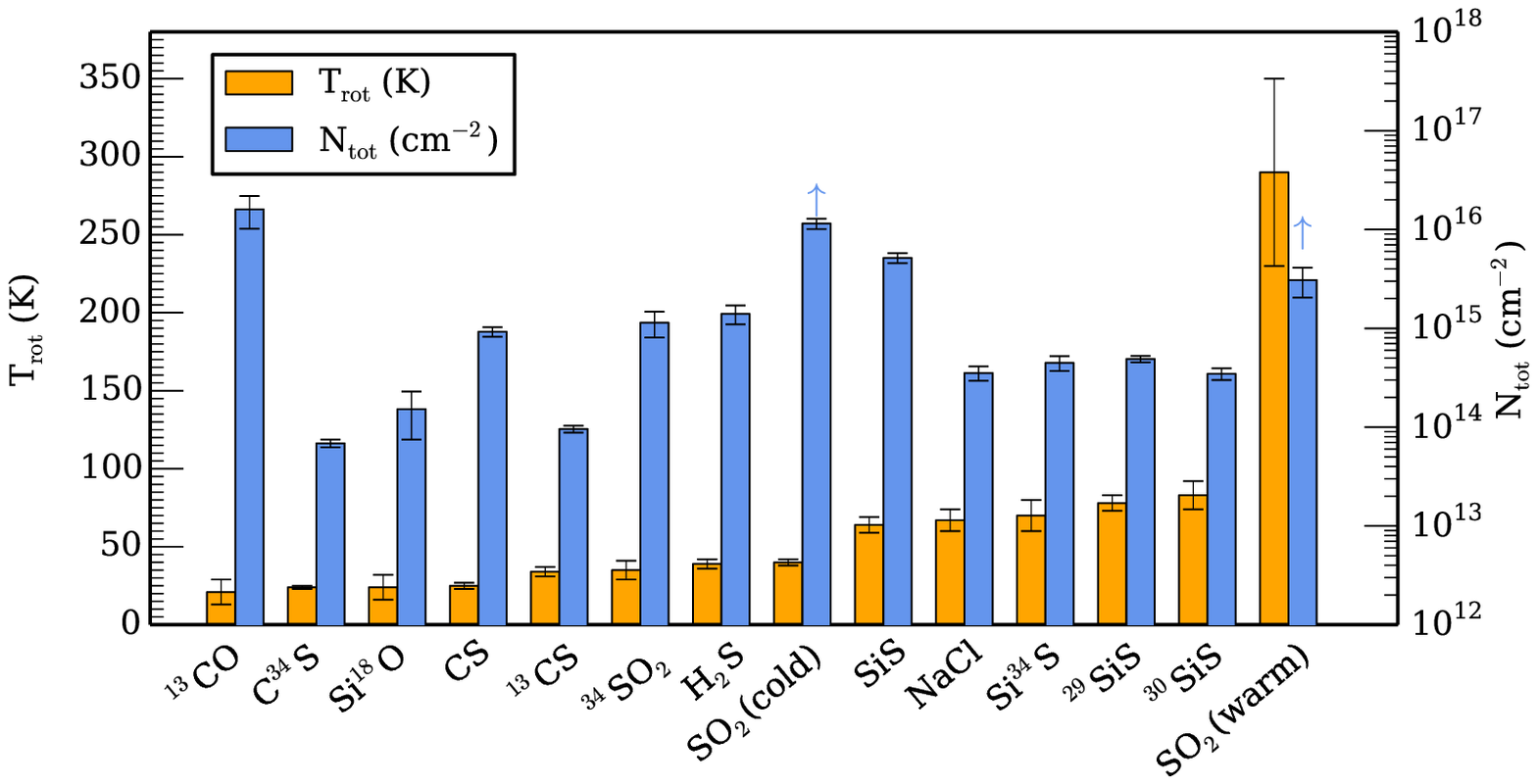}
\caption{Graph of the rotational temperatures (orange bars) and column densities (blue bars), with their 
formal uncertainties (black lines), derived from the population diagrams.
Vertical blue arrows are plotted to indicate column density lower limits over each corresponding molecule.
The temperature scale is represented on the left vertical axis, and the column density logarithmic scale is represented 
on the right vertical axis.
The limits of the temperature scale are different in both boxes to improve the visual aspect of the figures.
The values represented are tabulated in Table\,\ref{tab:rdresult}.
}
\label{fig:bargraph}
\end{figure*}

\section{Results}\label{sec:results}
Here we present the general results from the analysis, which are summarised in 
Table\,\ref{tab:rdresult} and Fig.\,\ref{fig:bargraph}.

\begin{table*}[hbtp!] %
\caption{Results from the population diagrams, sorted according to the molecular fractional abundance (relative to H$_2$)
in descending order.}
\label{tab:rdresult}
\centering    
\begin{tabular}{l l l l l l l l l}
\hline\hline
Molecule & T$_{\mathrm{rot}}$ & Column density & Abundance & $\theta_{\mathrm{s}}$ & Lit. observational $f$(X) & Ref. & Lit. chemical model $f$(X) & Ref. \\
 & (K) & (cm$^{-2}$) & - & \arcsec\ & - & - & - & - \\
\hline
\noalign{\vskip 1mm}    
CO                   & 18\,(4)   &  $\gtrsim$1.3\,(0.3)$\times$10$^{17}$                & $\gtrsim$1.1$\times$10$^{-4}$                & 18                     & [2--3]$\times$10$^{-4}$                          & d, f  & 7$\times$10$^{-4}$--1$\times$10$^{-3}$  & $\gamma$             \\
$^{13}$CO            & 21\,(8)   &  1.6\,(0.6)$\times$10$^{16}$                         & 1.4$\times$10$^{-5}$\tablefootmark{\dag}     & 18                     & [1--3]$\times$10$^{-5}$                          & d, f  & -                                       & -                    \\
SO$_2$ cold          & 40\,(2)   &  $>$1.1\,(0.1)$\times$10$^{16}$\tablefootmark{\ddag} & $>$9.6$\times$10$^{-6}$\tablefootmark{\ddag} & 2\tablefootmark{\ddag} & 2$\times$10$^{-6}$--1$\times$10$^{-5}$           & b, f  & 2$\times$10$^{-7}$                      & $\epsilon$           \\
SiO                  & 17\,(2)   &  $>$9.1\,(3.3)$\times$10$^{15}$                      & $>$8.0$\times$10$^{-6}$                      & 2                      & 4$\times$10$^{-7}$--2$\times$10$^{-5}$           & e, d  & 2$\times$10$^{-5}$--9$\times$10$^{-5}$  & $\gamma$             \\                
SO                   & 17\,(1)   &  $\gtrsim$8.9\,(1.6)$\times$10$^{15}$                & $\gtrsim$7.8$\times$10$^{-6}$                & 2                      & 3$\times$10$^{-7}$--3$\times$10$^{-6}$           & f, a  & 2$\times$10$^{-12}$--9$\times$10$^{-7}$ & $\gamma$, $\epsilon$ \\                        
SiS                  & 64\,(5)   &  5.2\,(0.6)$\times$10$^{15}$                         & 4.6$\times$10$^{-6}$                         & 2                      & 8$\times$10$^{-9}$--1$\times$10$^{-5}$           & d     & 8$\times$10$^{-10}$--3$\times$10$^{-6}$ & $\beta$, $\epsilon$  \\                       
SO$_2$ warm          & 290\,(60) &  $>$3.1\,(1.0)$\times$10$^{15}$\tablefootmark{\ddag} & $>$2.7$\times$10$^{-6}$\tablefootmark{\ddag} & 2\tablefootmark{\ddag} & -                                                & -     & 2$\times$10$^{-14}$--4$\times$10$^{-9}$ & $\gamma$             \\               
$^{29}$SiO           & 17\,(2)   &  $\gtrsim$2.1\,(0.6)$\times$10$^{15}$                & $\gtrsim$1.8$\times$10$^{-6}$                & 2                      & -                                                & -     & -                                       & -                    \\                       
PO                   & 17\,(2)   &  2.0\,(0.4)$\times$10$^{15}$                         & 1.7$\times$10$^{-6}$                         & 0.7                    & 5$\times$10$^{-8}$--6$\times$10$^{-7}$           & c     & 2$\times$10$^{-10}$--1$\times$10$^{-7}$ & $\gamma$             \\               
$^{30}$SiO           & 17\,(2)   &  $\gtrsim$1.5\,(0.5)$\times$10$^{15}$                & $\gtrsim$1.3$\times$10$^{-6}$                & 2                      & -                                                & -     & -                                       & -                    \\                       
H$_2$S               & 39\,(3)   &  1.4\,\,(0.3)$\times$10$^{15}$                       & 1.2$\times$10$^{-6}$                         & 2                      & -                                                & -     & 6$\times$10$^{-13}$--3$\times$10$^{-5}$ & $\beta$, $\gamma$    \\                     
$^{34}$SO$_2$        & 35\,(6)   &  1.1\,(0.3)$\times$10$^{15}$                         & 9.6$\times$10$^{-7}$                         & 2                      & -                                                & -     & -                                       & -                    \\                       
CS                   & 25\,(2)   &  9.2\,(1.0)$\times$10$^{14}$                         & 8.0$\times$10$^{-7}$                         & 2                      & 8$\times$10$^{-8}$--3$\times$10$^{-7}$           & f, g  & 2$\times$10$^{-11}$--2$\times$10$^{-5}$ & $\gamma$, $\beta$    \\                     
$^{34}$SO            & 14\,(1)   &  8.9\,(2.4)$\times$10$^{14}$                         & 7.8$\times$10$^{-7}$                         & 2                      & -                                                & -     & -                                       & -                    \\                       
PN                   & 18\,(2)   &  8.3\,(1.7)$\times$10$^{14}$                         & 7.3$\times$10$^{-7}$                         & 0.7                    & 3$\times$10$^{-7}$                               & c     & 4$\times$10$^{-10}$--6$\times$10$^{-7}$ & $\gamma$             \\                
HCN                  & 10\,(1)   &  $>$7.6\,(2.9)$\times$10$^{14}$                      & $>$6.6$\times$10$^{-7}$                      & 4                      & 4$\times$10$^{-7}$--1$\times$10$^{-6}$           & i, f  & 6$\times$10$^{-12}$--3$\times$10$^{-4}$ & $\gamma$, $\beta$    \\                     
$^{29}$SiS           & 78\,(5)   &  4.9\,(0.4)$\times$10$^{14}$                         & 4.3$\times$10$^{-7}$                         & 2                      & -                                                & -     & -                                       & -                    \\                       
Si$^{34}$S           & 70\,(10)  &  4.5\,(0.8)$\times$10$^{14}$                         & 3.9$\times$10$^{-7}$                         & 2                      & -                                                & -     & -                                       & -                    \\                       
NaCl                 & 67\,(7)   &  3.5\,(0.6)$\times$10$^{14}$                         & 3.1$\times$10$^{-7}$                         & 0.3                    & 4$\times$10$^{-9}$                               & h     & 4$\times$10$^{-12}$--1$\times$10$^{-8}$ & $\gamma$             \\                
$^{30}$SiS           & 83\,(9)   &  3.5\,(0.5)$\times$10$^{14}$                         & 3.1$\times$10$^{-7}$                         & 2                      & -                                                & -     & -                                       & -                    \\                       
H$_2$CO              & 10\,(2)   &  2.8\,(0.5)$\times$10$^{14}$                         & 2.4$\times$10$^{-7}$                         & 2                      & -                                                & -     & 1$\times$10$^{-9}$--5$\times$10$^{-7}$  & $\alpha$, $\epsilon$ \\                        
CN                   & 9\,(2)    &  1.5\,(0.7)$\times$10$^{14}$                         & 1.3$\times$10$^{-7}$                         & 6                      & 2$\times$10$^{-10}$--2$\times$10$^{-7}$          & d, f  & 3$\times$10$^{-13}$--3$\times$10$^{-7}$ & $\beta$, $\delta$    \\                     
Si$^{18}$O           & 24\,(8)   &  1.5\,(0.8)$\times$10$^{14}$                         & 1.3$\times$10$^{-7}$                         & 2                      & -                                                & -     & -                                       & -                    \\                       
Si$^{17}$O           & 12\,(2)   &  1.0\,(0.4)$\times$10$^{14}$                         & 8.7$\times$10$^{-8}$                         & 2                      & -                                                & -     & -                                       & -                    \\                       
$^{13}$CS            & 34\,(3)   &  9.6\,(0.8)$\times$10$^{13}$                         & 8.4$\times$10$^{-8}$                         & 2                      & -                                                & -     & -                                       & -                    \\                       
H$^{13}$CN           & 11\,(2)   &  9.2\,(2.5)$\times$10$^{13}$                         & 8.0$\times$10$^{-8}$                         & 4                      & -                                                & -     & -                                       & -                    \\                       
C$^{34}$S            & 24\,(1)   &  6.9\,(0.6)$\times$10$^{13}$                         & 6.0$\times$10$^{-8}$                         & 2                      & -                                                & -     & -                                       & -                    \\                       
NS                   & 18\,(2)   &  2.6\,(0.5)$\times$10$^{13}$                         & 2.3$\times$10$^{-8}$                         & 6                      & -                                                & -     & 7$\times$10$^{-13}$--8$\times$10$^{-9}$ & $\beta$, $\delta$    \\                      
\hline
\hline
\end{tabular}
\tablefoot{We give the formal uncertainties derived from the population diagram fits within parentheses. 
(Col. 5) Size adopted for the emitting region. (Col. 6\,\&\,7) Range of the fractional abundances given in the literature, 
derived from the following works based on observations: 
(a): \cite{buj94};
(b): \cite{dan16};
(c): \cite{deb13};
(d): \cite{dec10a};
(e): \cite{gon03};
(f): \cite{kim10};
(g): \cite{lin88}:
(h): \cite{mil07};
(i): \cite{sch13}.
(Col. 8\,\&\,9) Range of the fractional abundances given in the literature, which were derived from the following chemical and shocks models:
($\alpha$): \cite{agu10};
($\beta$): \cite{dua99};
($\gamma$): \cite{gob16};
($\delta$): \cite{lix16};
($\epsilon$): \cite{wil97}.\\
\tablefoottext{\dag}{Value given in Table\,6 in \cite{dec10a}.}\\
\tablefoottext{\ddag}{See Sect.\,\ref{sec:so2}.}\\
}
\end{table*}


\subsection{Expansion velocity}\label{sec:vexp}
The terminal expansion velocity of the CSE (\vinf) can be estimated from the linewidths of the spectral features 
that arise from the outer \citep[$r$$>$8\rstar,][]{gob16} envelope regions where the gas has been fully accelerated to 
this maximal velocity.
We estimated the \vinf\ from the $^{13}$CO linewidths (see Table\,\ref{tab:measures}) given that $^{13}$CO emission certainly
extends beyond the wind acceleration region, and the line profiles display a clear U-shaped profile (see Fig.\,\ref{fig:coprofs})
and, thus, no significant opacity broadening is expected \citep{phi79}.
Assuming a mean density of $n$(H$_2$)=10$^{5}$\,cm$^{-3}$, and the temperature and column density given 
in Table\,\ref{tab:rdresult}, the highest opacity measured with MADEX is $\tau$=0.13 for the $^{13}$CO $J$=3--2 line.
We derived \vinf=18.6$\pm$1.2\,\kms\ which is in good agreement with previous measurements \citep[e.g.][]{deb13}.

\begin{figure}[hbtp!]
\centering
\includegraphics{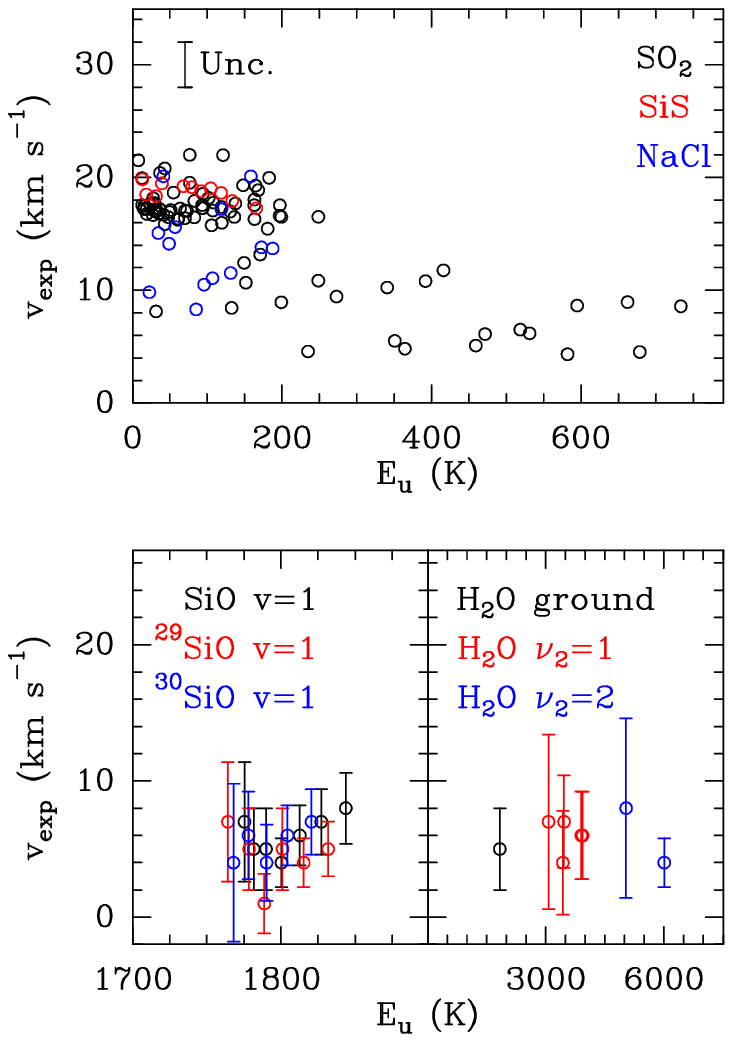}
\caption{Plot of the \vexp\ derived from the fit of the lines as a function of the \eu\ of the 
corresponding transition for several molecules. In the top figure, we did not represent each error bar to improve the 
visualization of the figure. A typical error bar of 4\,\kms\ is plotted in the top left corner of the box.
The values and uncertainties of the bottom figure correspond to the values given in Table\,\ref{tab:variability}.
For those lines that were observed in different epochs, we adopted an average value of the different measurements for the 
\vexp\ and its uncertainty.}                                                                                               
\label{fig:allvexp}
\end{figure}

\begin{figure}[hbtp!]
\centering
\includegraphics{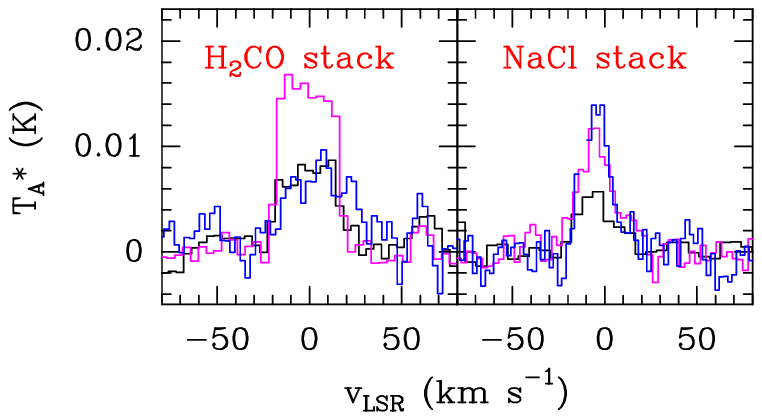}
\caption{Line stacking of the H$_2$CO (left) and the NaCl (right) lines detected, with a spectral resolution of 2\,MHz. 
Black spectrum correspond to the H$_2$CO $K_{\mathrm{a}}$=0 lines and the NaCl lines with \eu=[22.5-56.9]\,K,
pink spectrum correspond to the H$_2$CO $K_{\mathrm{a}}$=1 lines and the NaCl lines with \eu=[85.0-118.7]\,K, and
blue spectrum correspond to the H$_2$CO $K_{\mathrm{a}}$=2 lines and the NaCl lines with \eu=[131.2-187.4]\,K.
}                                                                                               
\label{fig:stack}
\end{figure}

Most of the rest of the lines detected display linewidths consistent with \vinf\ like for example SiS  
(see Table\,\ref{tab:measures}, Table\,\ref{tab:variability} and Fig.\,\ref{fig:allvexp}).
In the particular case of H$_2$CO, we stacked the lines with $K$=0, $K$=1, and $K$=2 to confirm that H$_2$CO linewidths are 
consistent with \vexp=\vinf\ (Fig.\,\ref{fig:stack}).
There are also several lines with linewidths larger than \vinf, due to
a blend of several hyperfine components (as occurs for NO and NS), or a poor fitting for lines 
detected below 5\,$\sigma$.

Additionally, the lines with \eu$\gtrsim$160\,K, that is, the lines of vibrationally excited states (Table\,\ref{tab:variability}),
H$_2$O, PO, NaCl, and several high-\eu\ SO$_2$lines, have line profiles indicative of \vexp$\lesssim$10\,\kms, consistent with emission 
from the inner regions of the CSE where the gas is still being accelerated 
\citep[$r$$<$8\rstar, i.e. $\sim$2$\times$10$^{14}$\,cm$^{-2}$,][]{dec10b,gob16}.
In order to obtain a more reliable estimate of the NaCl linewidths,
we stacked the lines with \eu=[22.5-56.9]\,K (group 1), \eu=[85.0-118.7]\,K (group 2), and \eu=[131.2-187.4]\,K (group 3) 
(see Fig.\,\ref{fig:stack}). 
The linewidths measured are consistent with 
\vexp=14.8$\pm$2.2\,\kms\ (group 1), \vexp=20.1$\pm$1.1\,\kms\ (group 2), and \vexp=11.8$\pm$1.1\,\kms\ (group 3). 
Therefore, it seems that \vexp$<$\vinf\ at least for the lines of the group 3, that is to say, the high-\eu\ lines.

There are also some lines with low \eu\ ($<$200\,K) which appear to have expansion velocities of $\sim$10\,\kms\ (e.g. SO$_2$, see 
Fig.\,\ref{fig:allvexp}). Such low values of the \vexp\ were measured for weak SO$_2$ lines with low $S/N$, thus, the linewidths
observed in their profiles are uncertain.

\subsection{Rotational temperatures, column densities and fractional abundances}\label{sec:respop}
The population diagrams for all the molecules detected are shown in Appendix\,\ref{sec:app_rd}, 
in Fig.\,\ref{fig:rdallco}--\ref{fig:rdallh2co}.
Most diagrams display a linear trend, however,
the population diagrams of, for example, SiO, $^{29}$SiO, or SO,
display departures from a linear behaviour, which are more notable for low $J$ transitions 
(see e.g. Fig.\,\ref{fig:rdallsio}). These departures are discussed in detail in Sect.\,\ref{sec:discussion}
and they reflect the effect of optically thick emission and/or sub-thermal excitation.
For SO$_2$ we see two different trends for lines below and above \eu=160\,K.
We fitted both separately and their implications are explained in Sect.\,\ref{sec:so2}.

The rotational temperatures derived range from 9\,K (for CN) to 290\,K (for the warm component of SO$_2$),
with most of the molecules displaying rotational temperatures between 15\,K and 40\,K (see Fig.\,\ref{fig:bargraph}).
The column densities range from $\gtrsim$1.3$\times$10$^{17}$\,cm$^{-2}$ for $^{12}$CO, down to 2.6$\times$10$^{13}$\,cm$^{-2}$
for NS. 
We obtained averaged fractional abundances using the Eq.\,(\ref{eq:abundances}),
which range between $>$1.1$\times$10$^{-4}$ for $^{12}$CO, down to 2.3$\times$10$^{-8}$ for NS.
We calculated the isotopic ratios of the molecules for which several isotopologues were detected. 
Results are presented in Table\,\ref{tab:isorat} and discussed in Sect.\,\ref{sec:isot}.

\begin{table}[hbtp!] %
\caption{Isotopic ratios obtained from the abundances derived with the population diagrams.
}
\label{tab:isorat}
\centering    
\begin{tabular}{l l l}
\hline\hline
Ratio 	& Value & From \\
\hline
$^{12}$C/$^{13}$C	& $>$8\,($\sim$10)	& CO    \\ 
$^{12}$C/$^{13}$C	& 10	                & CS	\\
$^{12}$C/$^{13}$C	& $>$8\,($\sim$15)	& HCN	\\
$^{16}$O/$^{18}$O	& $\gg$61		& SiO	\\ 
$^{16}$O/$^{17}$O	& $\gg$91		& SiO	\\ 
$^{28}$Si/$^{29}$Si	& $>$4\,($\sim$18)	& SiO	\\ 
$^{28}$Si/$^{29}$Si	& 11	                & SiS	\\ 
$^{28}$Si/$^{30}$Si	& $>$6\,($\sim$34)	& SiO	\\ 
$^{28}$Si/$^{30}$Si	& 16	                & SiS	\\
$^{32}$S/$^{34}$S	& $>$8\,($\sim$15)	& SO	\\ 
$^{32}$S/$^{34}$S	& 13	                & CS	\\
$^{32}$S/$^{34}$S	& 12	                & SiS	\\
$^{32}$S/$^{34}$S	& $>$10\,($\sim$13)\tablefootmark{\dag} & SO$_2$  \\ 
\hline
\hline
\end{tabular}
\tablefoot{
Opacity corrected values are given between parentheses (see Sect.\,\ref{sec:isot})
\tablefoottext{\dag}{This value was computed using only the cold SO$_2$ component in Table\,\ref{tab:rdresult}.}\\
}
\end{table}


\section{Discussion: an overall picture of the whole envelope}\label{sec:discussion}
\subsection{Variability}\label{sec:variability}
The excitation mechanisms of the lines are a mix of collisional and radiative procceses, where 
the radiation emitted by the central star has an impact on the population of the rotational levels \citep[e.g. H$_2$O][]{agu06}.
Since AGB stars are variable with periods of one to two years, 
the net excitation mechanism for those molecules is variable and the results obtained from the analysis of the molecular 
lines of AGB CSEs could be affected by this variability. In the case of O-rich stars, the impact of radiative pumping effects  
on masers is well-known \citep[e.g.][]{nak07}. 
Thermal line observations of CSEs have also to be considered 
carefully in the sub-millimeter and far-IR domain, while for the millimeter wavelength range, specially for low-$J$
lines, the variation of the stellar light has not a major impact \citep{cer14}.
Given that our observations spread over a period of approximately five years, we observed a few spectral ranges at different 
epochs and here we discuss the results observed.

\begin{figure*}[hbtp!]
\centering
\includegraphics[width=\textwidth]{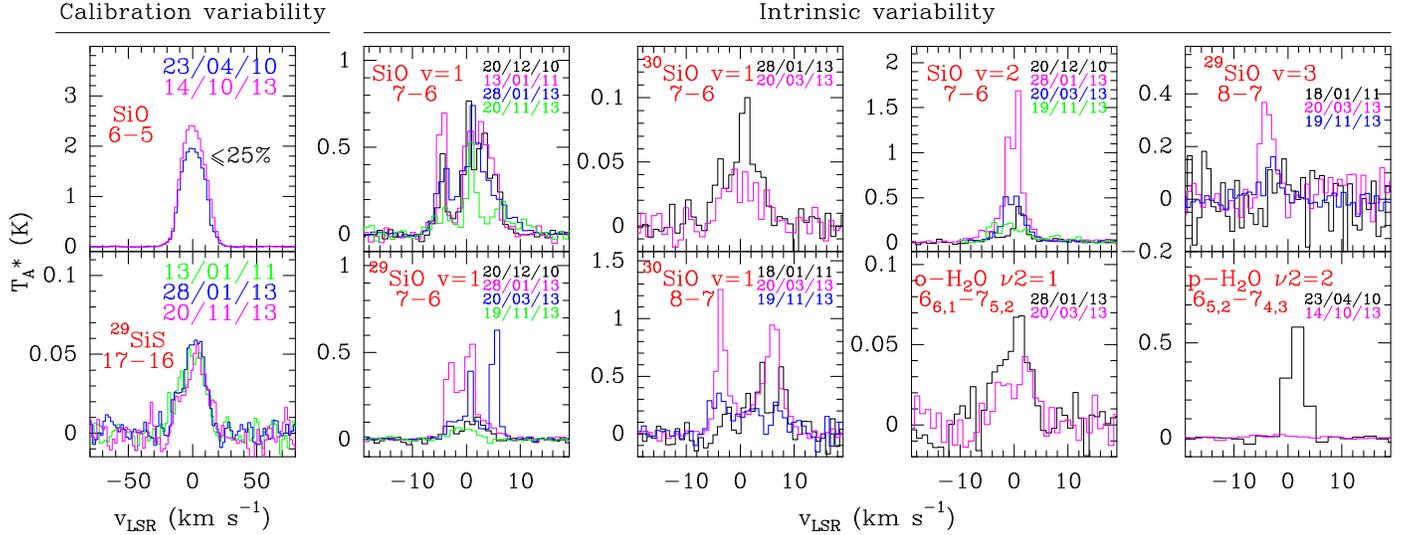}
\caption{Plot of the variability of some of the lines observed. The observation dates which correspond to the spectra 
are shown in the top right corner of each box with its corresponding colour. The spectral resolution is
2\,MHz. As discussed in Sect.\,\ref{sec:variability}, variations of up to a 25\% are within the uncertainties of the calibration,
pointing and baseline substraction (i.e. left boxes). 
Variations of intensity higher than a 25\% are observed in vibrationally excited lines or masers
(i.e. right boxes).}                                                                                               
\label{fig:var}
\end{figure*}

For the spectra observed at different epochs we observed intensity variations of less than 25\% for the rotational lines of the ground 
vibrational levels.
This value (i.e. 25\%) was taken as the calibration uncertainty (see Sect.\,\ref{sec:obs}), and considered in the results presented 
throughout this work (Fig.\,\ref{fig:var} left). 
However, we observed strong intensity variations ($>$25\%) for several lines, like for example the SiO $v$=1 $J$=7--6, 
(see Table\,\ref{tab:variability} and Fig.\,\ref{fig:var}).
These strongly variable lines correspond to rotational transitions of vibrationally excited states with \vexp$\lesssim$10\,\kms, which 
arise from very inner regions of the CSE.
The excitation mechanism of these lines may be correlated with the stellar phase, thus, 
it is not possible to extract any information of their abundances.
In Table\,\ref{tab:variability} we present the fit parameters for these vibrationally excited and maser lines, where we included 
the mean julian date of the observation and, in case that the feature was observed in different epochs, multiple measurements 
of that spectral feature.

\subsection{O-bearing molecules}\label{sec:orich}  
CO is the most abundant molecule in \object{IK Tau} (after H$_2$). 
It is distributed along the whole CSE with 
a size of $\sim$[7--8]$\times$10$^{16}$\,cm (see Sect.\,\ref{sec:iktau}).
The $^{12}$CO $J$=2--1 and $^{12}$CO $J$=3--2 lines display parabolic profiles typical of optically thick lines,
while the $^{12}$CO $J$=1--0 line, all the $^{13}$CO lines, and the C$^{18}$O lines
show U-shaped profiles typical of optically thin lines (Fig.\,\ref{fig:coprofs}). The emission of all the CO 
isotopologues is probably spatially resolved considering the typical beam size (Table\,\ref{tab:iram_param}) and the shape of the line 
profiles.
The population diagrams of $^{12}$CO and $^{13}$CO hint small departures from a linear trend, owing to 
high optical depth and/or sub-thermal excitation (Fig.\,\ref{fig:rdallco}). 
The results can be seen in Table\,\ref{tab:rdresult} and Fig.\,\ref{fig:bargraph}.
We derived a fractional abundance $f$($^{12}$CO)$\gtrsim$1.1$\times$10$^{-4}$ (with respect to H$_2$) in agreement with previous 
estimates.
For C$^{18}$O, we estimated $f$(C$^{18}$O)$\sim$4$\times$10$^{-8}$, assuming the same excitation temperature (20\,K)
and emitting size as for $^{13}$CO.

We used MADEX to estimate the opacities of the CO lines 
(using the \tkin, the \vexp\ derived from the linewidhts, the column density derived from the population diagram, and the $n$(H$_2$) 
density at the outer radius of the shell according to the equation given in Table\,\ref{tab:iktau}). 
With these input parameters, MADEX predicts that the $^{12}$CO $J$=2--1 and the $J$=3--2 lines are 
moderately thick ($\tau$$\lesssim$1.4), while the $J$=1--0 line is optically thin ($\tau$$\sim$0.2).
This is consistent with CO tracing the coolest, outermost layers of the CSE and $^{13}$CO probing also regions deeper 
inside.

For SiO isotopologues we have adopted an emitting size equivalent to 7.9$\times$10$^{15}$\,cm (2\arcsec) at a distance of 265\,pc 
(Sect.\,\ref{sec:iktau}). All the line profiles of the different SiO isotopologues are parabolic, consistent with optically 
thick emission.
The rotational diagrams of all the SiO isotopologues display small departures from a linear trend, except 
perhaps for Si$^{17}$O (Fig.\,\ref{fig:rdallsio}).
We found similar \trot\ for all the isotopologues, that is, \trot$\sim$20\,K, which are much lower than the \tkin\
expected at a distance of 1\arcsec\ from the star (i.e. \tkin$\sim$105\,K). Hence, sub-thermal excitation 
may have an impact on the values derived from the population diagrams of SiO isotopologues. 

MADEX (see Sect.\,\ref{sec:madex}) predicted $\tau$$>$1 for all of the SiO, $^{29}$SiO, and $^{30}$SiO lines
(except for the $J$=2--1 lines of $^{29}$SiO and $^{30}$SiO, for which MADEX predicted $\tau$$\lesssim$0.6.)
These values are consistent with the observed line profiles.
For Si$^{18}$O and Si$^{17}$O lines, MADEX predicted $\tau$$<$0.5.
Therefore, except for Si$^{18}$O and Si$^{17}$O, the column densities and the abundances derived for the SiO isotopologues,
given in Table\,\ref{tab:rdresult}, should be considered lower limits.
The lower limit obtained, $f$(SiO)$>$8.0$\times$10$^{-6}$, is in good agreement with previous measures 
\citep[][and references therein]{dec10a},
and, in principle, is also consistent with the low SiO abundances predicted by the model by \citep{gob16} that proposes the 
formation of SiO in abundance under thermodynamical equilibrium (TE) in the stellar photosphere, and a significant abundace decay (to 1.5$\times$10$^{-5}$) already 
at 6\rstar\ mainly due to dust condensation.

\subsection{C-bearing molecules}\label{sec:crich}
We detected emission of molecules, like CS, HCN and HNC, that are typically found in C-rich CSEs 
\citep[e.g.][]{buj94,cer00,zha09}. 

According to \cite{mar05}, the HCN emission arises from a compact region with $\theta_{\mathrm{s}}$$\sim$4\arcsec, 
which we adopted in this work. For CS, there are no observational constraints on the size of the emission, therefore, 
we adopted a size of $\theta_{\mathrm{s}}$=2\arcsec\ which is the same size used for SiO and 
it is consistent with the extent of the CSE emission predicted by chemical models \citep{lix16}. 
Adopting these sizes we obtained \trot(HCN)$\sim$10\,K and \trot(CS)$\sim$25\,K, and 
$N_{\mathrm{tot}}$(HCN)$\gtrsim$8$\times$10$^{14}$\,cm$^{-2}$ and 
$N_{\mathrm{tot}}$(CS)$\sim$9$\times$10$^{14}$\,cm$^{-2}$ (Fig.\,\ref{fig:rdallhcn} and Fig.\,\ref{fig:rdallcs}).

Using MADEX, we found optically thick lines ($\tau$$\gtrsim$1.5) for HCN, and moderately thick lines ($\tau$$\lesssim$1.2)
for H$^{13}$CN. Therefore, the column density and the abundance of HCN should be considered as lower limits.
In the case of CS isotopologues, with the physical conditions expected at r$\sim$1\arcsec, MADEX predicted optically 
thin lines ($\tau$$\lesssim$1.0). 
Finally, given that \tkin($r$$\lesssim$1\arcsec)$\gtrsim$100\,K, the lines of HCN and CS isotopologues are most likely sub-thermally 
excited.

The rotational temperatures and abundances derived for HCN and CS isotopologues (see Table\,\ref{tab:rdresult})
are consistent with previous measurements in \object{IK Tau} \citep{lin88,kim10}. 
The HCN abundance is in the high end of the abundance range deduced by \cite{sch13} in a sample of M-type AGB stars.
The TE models for O-rich CSEs do not account for the HCN and CS abundances observed in O-rich CSEs, predicting 
values of $f$(HCN)$\sim$10$^{-11}$ and $f$(CS)$\sim$10$^{-10}$ \citep{dua99,gob16}.
The inclusion of shocks can contribute to enhance the formation of HCN in O-rich stars \citep{dua99,che06,gob16},
but it also brings up a theoretical homogeinity on the expected HCN abundances 
among different chemical types of stars that it is not observed, as noted by \cite{sch13}.
Other authors invoked the photochemistry to try to explain the abundances of C-bearing molecules observed in O-rich 
CSEs \citep{ner89,ziu09}.
In particular, the chemical model of \cite{wil97} predicts peak abundances of 
$f$(HCN)$\sim$1$\times$10$^{-7}$ and $f$(CS)$\sim$3$\times$10$^{-7}$, although the carbon source proposed by these 
authors is CH$_4$ which has been later on refuted \citep{mar05}. 
Additionally, it has been also proposed that CSEs could be clumpy, hence, photochemistry could be important also in the inner 
layers of the envelopes \citep{agu10}.
Our results do not conclude clearly which is the most likely scenario, although, our derived abundances are more 
similar to those predicted by the models of \cite{gob16}.

Finally, we detected two HNC lines, in particular, the $J$=1--0, and the $J$=3--2 line, which is blended with the image of the 
SiO $J$=6--5 line.
We estimated $f$$\sim$8$\times$10$^{-9}$, assuming an excitation temperature of $\sim$30\,K and a size of 2\arcsec\ for the emitting 
region, which are average values for these parameters.

\subsection{Refractory species}\label{sec:refra}
We confirmed the presence of two important refractory molecules which are mainly found in C-rich envelopes: SiS and NaCl.
The emission of these molecules has not been mapped in previous studies and, therefore, the size of the emitting 
region is unknown. 
The NaCl line profiles, with \vexp$\sim$14\,\kms, are narrower than those of SiS, consistent with \vexp$\sim$18.5\,\kms,
which suggests a more inner distribution of NaCl around the star (see Fig.\,\ref{fig:allvexp}). 
This molecule may condense onto the dust grains beyond the dust condensation radius, as proposed by \cite{mil07}. 
The profiles of the lines of NaCl and SiS isotopologues indicate spatially unresolved emission, which is compatible with 
the gaussian-like profiles observed in the case of NaCl, and the triangular or parabolic profiles observed for the SiS 
isotopologues.
Moreover, the gaussian-like profiles observed for NaCl may support that the emission of this molecule arises from the innermost 
regions of the CSE, where the gas has not been fully accelerated.

We adopted a size of $\theta_{\mathrm{s}}$=2\arcsec\ for SiS isotopologues, as a first  
guess considering the emission size of SiO ($\theta_{\mathrm{s}}$=2\arcsec).
This size is also consistent with the size predicted by recent chemical model of \object{IK Tau} \citep{lix16}.
For NaCl we adopted a size of $\theta_{\mathrm{s}}$(NaCl)=0\arcsecp3\ \citep{mil07}.
We derived similar rotational temperatures for SiS and NaCl (i.e. \trot$\sim$65\,K) even higher for $^{29}$SiS, $^{30}$SiS, and 
Si$^{34}$S (see Fig.\,\ref{fig:rdallsis} and Fig.\,\ref{fig:rdallnacl}). 
According to the size adopted for SiS (i.e. $r$$\lesssim$4$\times$10$^{15}$\,cm), we estimated \tkin$\gtrsim$105\,K, 
and $n$(H$_2$)$\gtrsim$3.5$\times$10$^{5}$\,cm$^{-3}$.
For the size adopted for NaCl (i.e. r$\lesssim$6$\times$10$^{14}$\,cm), we estimated 
\tkin$\gtrsim$330\,K. Hence, SiS and NaCl are most likely sub-thermally excited.
We estimated critical densities for the SiS lines of $n_{\mathrm{crit}}$$\sim$[10$^{4}$--10$^{6}$]\,cm$^{-3}$ 
for a temperature of $\sim$105\,K, therefore, $n$$\lesssim$$n_{\mathrm{crit}}$ for several lines of SiS 
confirming sub-thermal excitation. 
For NaCl the critical densities expected are even higher, $n_{\mathrm{crit}}$$\gtrsim$5$\times$10$^{7}$\,cm$^{-3}$, 
due to the high dipole moment of NaCl.
MADEX predicted optically thin lines for both SiS ($\tau$$<$0.6) and NaCl ($\tau$$<$0.3). 

We derived $f$(SiS)$\sim$5$\times$10$^{-6}$.
This value is in good agreement with the estimations by \cite{kim10}.
The chemical model by \cite{gob16} predicts a SiS abundance of 4$\times$10$^{-8}$ under 
TE and up to $\sim$3$\times$10$^{-7}$ including dust condensation and shocks due to the 
pulsation of the star, which is at least one order of magnitude lower than our results. 
\cite{wil97} used SiS in their chemical models as a parent molecule with an abundance consistent with our observations 
(see Table\,\ref{tab:rdresult}).

Concerning NaCl, \cite{mil07} derived a rotational temperature and a column density consistent with our results. 
However, \cite{mil07} derived a fractional abundance $\sim$80 times lower than ours, through the population diagram of two 
low $S/N$ NaCl emission lines, and also a radiative transfer calculation using the code by \cite{bie93}, with a set of SiO-corrected 
collisional coefficients. 
We detected 13 NaCl lines which cover a wide range in \eu\ and have better $S/N$, from which we derived an average fractional abundance 
of $f$(NaCl)=3$\times$10$^{-7}$ (see Table\,\ref{tab:rdresult} and Fig.\,\ref{fig:rdallnacl}).
We cannot rule out uncertainties in our estimation due to the emitting size, and 
the $^{13}$CO column density adopted, which could not be representative in the region of NaCl emission.
Moreover, NaCl line profiles are not incompatible with $\theta_{\mathrm{s}}$(NaCl)$\gtrsim$0\arcsecp3, 
in particular, if $\theta_{\mathrm{s}}$(NaCl)=1\arcsec\ we would derive $f$(NaCl)=3$\times$10$^{-8}$. 
TE calculations predict abundances of 10$^{-11}$ up to 10$^{-7}$ \citep{tsu73,mil07}, while \cite{gob16} models 
(TE and shocks) predicts NaCl abundances between 4$\times$10$^{-12}$ up to 1$\times$10$^{-8}$. 
It would be necessary to obtain maps of the NaCl spatial distribution in order to clarify these discrepancies.

\subsection{S-bearing molecules}\label{sec:sbear} 
Here we discuss the detected emission of H$_2$S, SO, and SO$_2$. 
The first detection of these molecules toward \object{IK Tau} and their chemistry in O-rich CSEs was presented 
in \cite{omo93} and references therein. 
\cite{omo93} only detected one line of H$_2$S, and they were not able to estimate its abundance toward \object{IK Tau}.
The emission of SO and SO$_2$ molecules in O-rich CSEs, including \object{IK Tau}, has been recently reviewed and modelled by \cite{dan16}.

We detected three ortho and one para lines of H$_2$S as well as one line of o-H$_2$$^{34}$S, 
which point out \vexp$\sim$\vinf. 
The profiles of the lines indicate spatially unresolved emission.
Since there are not maps of the H$_2$S emission, we adopted a size of 
$\theta_{\mathrm{s}}$(H$_2$S)=2\arcsec\ which is consistent with the size predicted by 
chemical models \citep{lix16}.
Given that we only detected two lines of o-H$_2$S with $S/N$$>$5, we calculated the population 
diagram (Fig.\,\ref{fig:rdallsh2}) of ortho and para species together adopting an ortho-to-para ratio of 3:1 
which is the value that could be expected from the formation processes of H$_2$S.
We derived \trot$\sim$40\,K and $f$(H$_2$S)$\sim$1$\times$10$^{-6}$.
This value is consistent with the chemical models presented by \cite{gob16} at a few stellar radii.
We found that the lines of H$_2$S are likely to be
sub-thermally excited at the distances adopted for the H$_2$S emission ($r$$\lesssim$1\arcsec).
However, we did not made further non-LTE calculations due to the lack of a set of collisional coefficients for H$_2$S.
Under LTE conditions, MADEX predicts optically thin lines ($\tau$$<$0.6).

\begin{figure}
\centering
\includegraphics{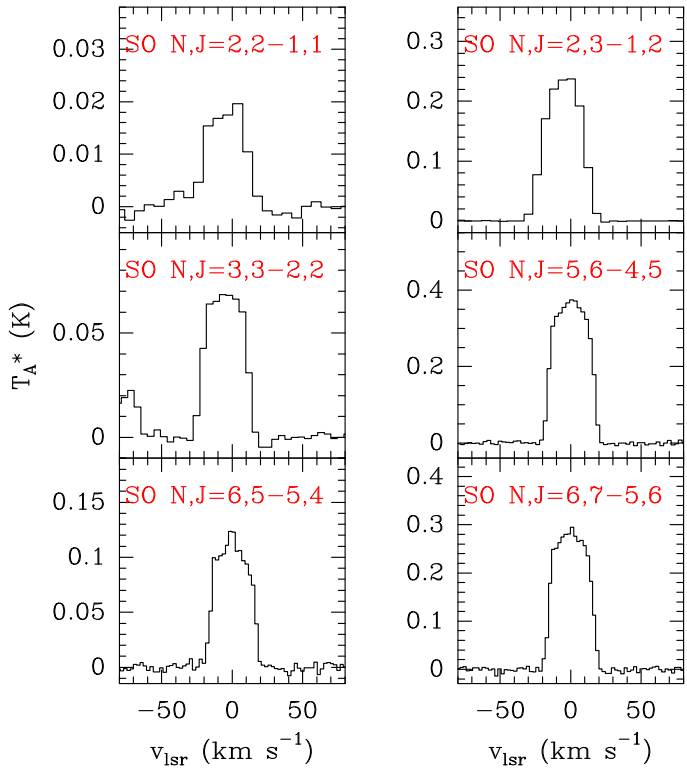}
\caption{Line profiles of some of the SO lines identified. The quantum numbers of each transition are plotted in red in each 
box.The spectral resolution is 2\,MHz for all the spectra shown.}                                                                                               
\label{fig:so_prof}
\end{figure}

\begin{figure*}
\centering
\includegraphics{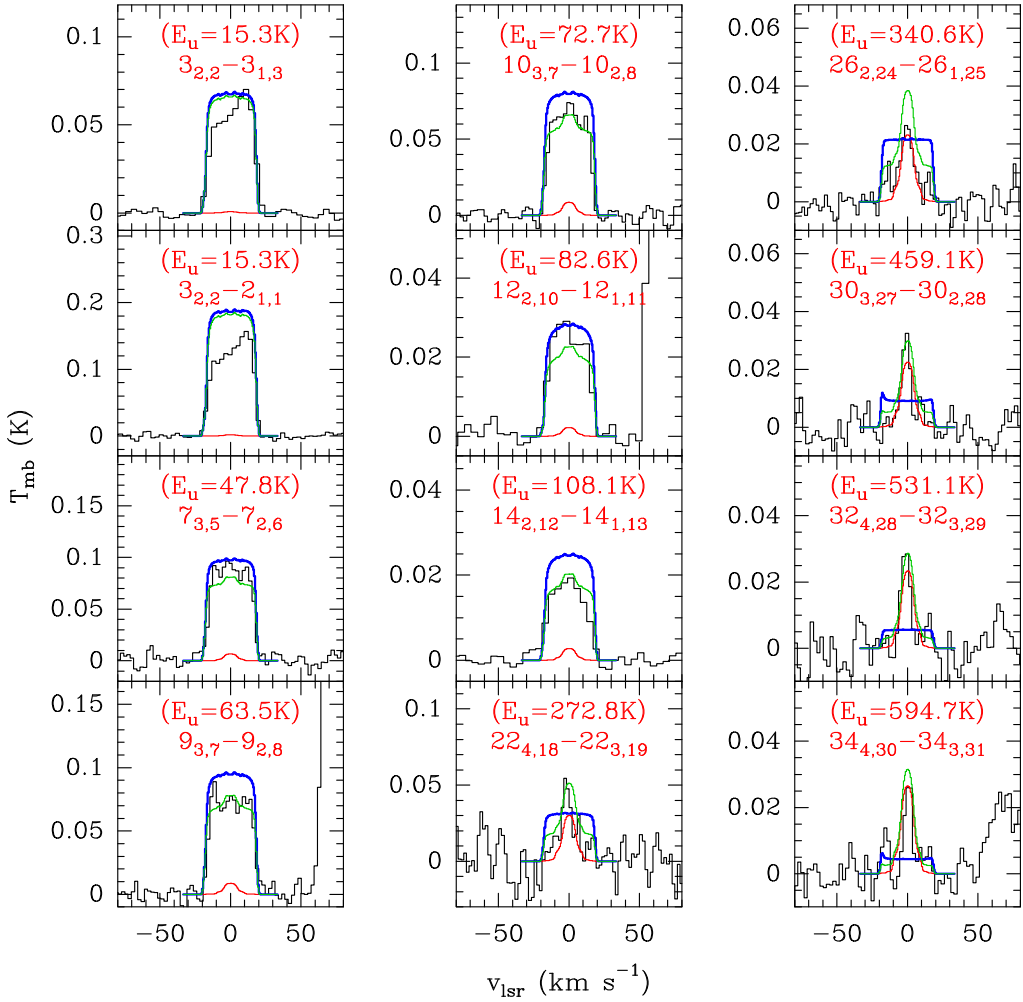}
\caption{Results of the radiative transfer models of SO$_2$ explained in Sect.\,\ref{sec:madex} and Sect.\,\ref{sec:so2}.
The coloured curves correspond to the models adopting the abundance profiles shown in Fig.\,\ref{fig:rpso2}.
The spectral resolution is 2\,MHz. The temperature scale is in main beam temperature.}                                                                                               
\label{fig:so2rt}
\end{figure*}

The profiles of the SO lines observed are varied (see Fig.\,\ref{fig:so_prof}). Most of them are flat-topped 
(optically thin and spatially unresolved emission), some of them display parabolic profiles (optically thick or moderately thick 
emission) and a few SO lines display profiles which seem to be composed of two components, at least: one dominant 
flat-topped or parabolic component with linewidths consistent with \vinf, and an additional narrow feature
which may indicate SO gas inside the gas acceleration region.
These two components are more clearly seen in several SO$_2$ lines (Fig.\,\ref{fig:so2rt}) which display a broad component plus a 
bulge-like centred narrow component, which we interpreted as SO$_2$ emission arising from r$<$8\rstar\ (see Sect.\,\ref{sec:so2}).
Nevertheless, we have not detected high-\eu\ SO narrow lines or two different trends in the SO population 
diagram (see below) which could prove a very inner component of warm SO gas (contrary to SO$_2$ as discussed in Sect.\,\ref{sec:so2}). 
Thus we have no firm evidence that could prove the presence of warm SO gas in the innermost regions of the CSE with a noticeable
abundance. Finally, the $^{34}$SO flat-topped (optically thin emission) profiles yield also \vexp$\sim$\vinf.

The brightness distribution of SO has not been mapped before, thus, we adopted a size of $\theta_{\mathrm{s}}$(SO)=2\arcsec\ 
as well as for $^{34}$SO, according to the models by \cite{lix16}.
The population diagrams (Fig.\,\ref{fig:rdallso}) display departures from a linear trend for SO.
We obtained \trot(SO)$\sim$\trot($^{34}$SO)$\sim$15\,K, 
$N_{\mathrm{tot}}$(SO)$\gtrsim$9$\times$10$^{15}$\,cm$^{-2}$ and
$N_{\mathrm{tot}}$($^{34}$SO)$\sim$9$\times$10$^{14}$\,cm$^{-2}$.
We verified with MADEX (see Sect.\,\ref{sec:madex}) that SO lines would be moderatelly thick 
($\tau$$\lesssim$1.5) with a \tkin=105\,K and n$\sim$4$\times$10$^{5}$\,cm$^{-3}$ at r=1\arcsec.  
MADEX predicted optically thin lines for $^{34}$SO ($\tau$$\lesssim$0.2).
Furthermore, we estimated $n_{\mathrm{crit}}$(SO)$\sim$[10$^5$--10$^7$]\,cm$^{-3}$ and
similar values for $^{34}$SO, which suggests sub-thermal excitation of several transitions. 

We derived abundances of $f$(SO)$\gtrsim$8$\times$10$^{-6}$ and $f$($^{34}$SO)$\sim$8$\times$10$^{-7}$.
The abundance measured of SO is at least a factor three higher compared to previous observational works toward 
\object{IK Tau} \citep{omo93,buj94,kim10}.
On the other hand, TE models predict abundances of $f$(SO)$\sim$[2--4]$\times$10$^{-8}$ \citep{dua99,gob16}.
\cite{wil97} derived peak abundances up to $f$(SO)$\sim$9$\times$10$^{-7}$ with a chemical model for an O-rich CSE 
that used only H$_2$S and SiS as parent S-bearing molecules. 
Compared to these models, our derived abundance is at least nine times higher than the highest value obtained from the models. 
Our analysis seems to overestimate the SO abundance compared to previous measurements and chemical models. These dicrepancies
may be explained given the uncertainty on the size of the SO emitting region adopted, and the $f$($^{13}$CO) adopted 
(see Eq.(\ref{eq:abundances})). In particular, a size of $\theta_{\mathrm{s}}$(SO)$\sim$5\arcsec\ would fix this discrepancy.

\subsubsection{SO$_2$ and $^{34}$SO$_2$}\label{sec:so2} 
We detected $\sim$90 lines of SO$_2$ displaying complex profiles which can be grouped according to 
their \vexp: ($i$) $\sim$60 lines with \vexp$\sim$18\,\kms\ consistent 
with the \vinf\ of the CSE, and ($ii$) $\sim$30 lines with \vexp$<$\vinf, with velocities as low as $\sim$5\,\kms 
(see Fig.\,\ref{fig:allvexp}).
The lines of SO$_2$ display parabolic profiles, flat-topped profiles, and complex profiles.
Several lines display a self-absorption in the blue side of the line (see Fig.\,\ref{fig:so2rt}).
This self-absorption may be explained considering that part of the SO$_2$ emission arising
from the inner and warm shells of the CSE, is absorbed by the external and cold shells of the CSE, which are located (within the 
line of sight) between us and the warm gas.
The narrow lines seem to be spatially unresolved, although, for the lines that have \vexp$\sim$\vinf\ it is not clear 
whether they are spatially resolved or not. 
Most of the $^{34}$SO$_2$ lines show flat-topped profiles indicative of spatially unresolved emission. 

The rotational diagram of SO$_2$ was done adopting an emission size of $\theta_{\mathrm{s}}$(SO$_2$)=2\arcsec, 
which should be considered as an educated guess derived from the chemical models 
by \cite{lix16}.
From the population diagram of SO$_2$ (Fig.\,\ref{fig:rdallso2}) we observed also two components, 
a cold component with \trot$\sim$40\,K traced out by $\sim$60 lines with 
$E_{\mathrm{u}}$$\lesssim$160\,K and a warm component with \trot$\sim$290\,K traced out by 
$\sim$30 lines with $E_{\mathrm{u}}$$>$160\,K.
The cold component displays a slight change in the trend for the lines with $E_{\mathrm{u}}$$\lesssim$50\,K
which may be explained as a result of moderate optically thick emission for those lines and/or sub-thermal 
excitation.
Using Eq.\,(\ref{eq:abundances}) we derived 
$f$(SO$_2$,cold)$\gtrsim$9.6$\times$10$^{-6}$ and
$f$(SO$_2$,warm)$\gtrsim$2.7$\times$10$^{-6}$.
For $^{34}$SO$_2$ we derived a \trot$\sim$35\,K and $f$($^{34}$SO$_2$)$\sim$9.6$\times$10$^{-7}$.

As we said in Sect.\,\ref{sec:madex}, the SO$_2$ lines detected span over a wide range of energies, thus, 
the parameters derived from the population diagram may be unreliable given that the homogenous temperature 
assumption may turn out to be a very crude approximation. 
We carried out several LVG models adopting different radial abundance profiles (see Sect.\,\ref{sec:madex} and 
Fig.\,\ref{fig:rpso2}). As it is discussed below, the different abundance profiles were adopted mainly to illustrate the need of 
the presence of warm SO$_2$ at $r$$<$8\rstar\ in order to reproduce the profiles of the high-\eu\ lines.
The results of the radiative transfer models are shown in Fig.\,\ref{fig:so2rt} for several lines of SO$_2$.

The first model (red line in Fig.\,\ref{fig:rpso2} and Fig.\,\ref{fig:so2rt}) reproduces to a good accuracy the profiles of the narrow 
high-\eu\ (i.e. \eu$>$160\,K) lines observed. However, it is unable to reproduce the line profiles of the low-\eu\ 
(i.e. \eu$<$160\,K) lines, underestimating their emission completely. 
We created a second model (blue line in Fig.\,\ref{fig:rpso2} and Fig.\,\ref{fig:so2rt})
which reproduces, within a factor two in intensity, most of the low-\eu\ SO$_2$ lines observed. 
However, this model is unable to explain the narrow profiles of the high-\eu\ SO$_2$ lines, predicting wide flat-topped lines or 
even no emission for these high energy lines. We may note, that the fractional abundance adopted for this second model is 
approximately a factor five lower than the abundance derived from the population diagram of the SO$_2$ cold component. If the abundance is increased 
up to a value consistent with the population diagram results for the cold component (i.e. $f$(SO$_2$)=9.6$\times$10$^{-6}$), the model 
highly overestimates the line profiles observed. This discrepancy may arise from the lack of precise information about the spatial 
distribution of SO$_2$ toward the CSE. 

Finally, we tested the possibility of SO$_2$ being distributed as a sum of two components, in particular, the sum of a compact inner component 
plus an extended component with a shell-like enhancement in the outermost part of the CSE, in order to reproduce the whole set of 
SO$_2$ lines observed with a single radial abundance profile.
This last model (green line in Fig.\,\ref{fig:rpso2} and Fig.\,\ref{fig:so2rt}) is able to approximately reproduce both the low- and 
high-\eu\ SO$_2$ lines observed, although, it does not reproduces perfectly all the line profiles. 
The discrepancies found between this best model and the observations are within a factor two or three in intensity for most of the lines 
observed. This disagreement emerges probably from the lack of precise information about the spatial distribution of SO$_2$. 
The radial abundance profile adopted for the model is consistent, within a factor two or three in intensity, with the results obtained from the 
population diagram for the SO$_2$ warm and cold components. 
Additionally, the best model predicts both optically thin and optically thick lines 
which is consistent with the variety of profiles observed.

Hence, we can conclude that SO$_2$ is distributed along the CSE with an average fractional abundance of $f$(SO$_2$)$\sim$10$^{-6}$.
Our models evidenced the presence of an inner (1--8\rstar) warm ($\gtrsim$290\,K) component of SO$_2$ with fractional abundances 
$\sim$10$^{-6}$, which produces most of the emission from SO$_2$ lines with \eu$\gtrsim$160\,K.
This is also consistent with the narrow profiles of the high-\eu\ SO$_2$ lines.
Nevertheless, our model is not able to reproduce the complexity observed in the profiles, which probably 
indicates the simplicity of the approximated physical model. 
In particular, the outer radius of the SO$_2$ cold component is critical 
to control the expected intensities of the low-\eu\ lines, which we adopted from the chemical model by \cite{lix16}.
Therefore, it would be necessary to map the brightness distribution of this molecule in order to constrain the outer radius of 
SO$_2$ emission. High angular resolution observations are also required to map the innermost regions of the CSE, since 
the distribution of SO$_2$ in the region between 1--20\rstar\ would improve the results obtained not only for the high-\eu\ lines 
but also for those lines that display a narrow core component (e.g. SO$_2$ 7$_{\mathrm{3,5}}$--7$_{\mathrm{2,6}}$ in 
Fig.\,\ref{fig:so2rt}).

Previous works toward \object{IK Tau} pointed out SO$_2$ abundances in the intermediate and outer envelope consistent 
with our results \citep{omo93,kim10,dec10a}.
\cite{dec10a} hinted the presence of SO$_2$ in the inner wind, since they were not able to reproduce with their radiative 
transfer model simultaneously the emission of a few lines with \eu$\sim$140\,K detected with APEX and the low-\eu\ transitions 
observed with the IRAM-30\,m telescope. Given the limited number of high-\eu\ lines detected by \cite{dec10a}, these authors 
were unable to reach conclusive results on the presence of SO$_2$ in the inner wind of \object{IK Tau} and its abundance.
Recent research was conducted to investigate one SO$_2$ line with \eu$\sim$600\,K detected with Herschel/HIFI as well as other 
SO$_2$ lines reported in the literature toward \object{IK Tau} with a radiative transfer model by \cite{dan16}. 
Their best-fit model has a peak abundance of $f$(SO$_2$)=2$\times$10$^{-6}$ and an $e$-folding radius $R_{\mathrm{e}}$=10$^{16}$\,cm,
although, it is unable to reproduce all the SO$_2$ observed line profiles as indicated by these authors.  

TE models predict the formation of SO$_2$ in the photosphere of the star with abundances of $f$(SO$_2$)$\sim$10$^{-11}$
\citep{tsu73,gob16}, which is approximately five orders of magnitude lower than our results. 
\cite{wil97} predicted a peak abundance of $f$(SO$_2$)$\sim$2$\times$10$^{-7}$ in the intermediate and outer parts 
of the CSE, at $r$$\sim$10$^{16}$\,cm (i.e. $\sim$500\,\rstar), and $f$($r$$\lesssim$3$\times$10$^{15}$\,cm)$\lesssim$10$^{-10}$,
where they used H$_2$S and SiS as the S-bearing parent molecules of the model.
This chemical model is also inconsistent with our results given that they do not predict the formation of SO$_2$ in the inner 
parts of the CSE and the abundance in the intermediate and outer parts is aprroximately two orders of magnitude lower than our measures.
In the recent chemical model presented by \cite{lix16}, the authors explore the effect of including SO$_2$ as a parent
molecule. In the absence of a reliable observational estimate of the SO$_2$ abundance in the inner envelope, these authors adopt
$f$(SO$_2$)=2$\times$10$^{-6}$, which is the value estimated from low-\eu\ SO$_2$ transitions arising in the outer envelope 
regions \citep{dec10a}. 
In the innermost parts of the envelope ($r$$<$8\rstar) the SO$_2$ abundance can be enhanced up to $f$(SO$_2$)=4$\times$10$^{-9}$ 
including the effect of shocks and dust grains \citep{gob16}. These authors suggest that the production of SO triggers the 
formation of SO$_2$ at $\sim$4\rstar\ in the gas phase through the reaction with OH.
Although, the SO$_2$ abundance obtained with the inclussion of shocks is approximately two or three
orders of magnitude lower than our estimates.
Photochemistry may also enhance the formation of SO$_2$ in the inner layers of the CSE, which
would require an additional source of UV radiation to dissociate H$_2$O providing 
OH to react with the SO formed leading to the enhancement of SO$_2$. 
This could be plausible if the envelope of \object{IK Tau} is clumpy, as proposed for other objects \citep{agu10}.

\subsection{N-bearing molecules}\label{sec:nbear} 
Besides HCN, HNC, and PN, which are discussed in other sections, we detected CN, NS, and NO.
The CN lines with unblended hyperfine components have widths consistent with \vexp=\vinf\ within errors. 
The spatial distribution of CN in \object{IK Tau} is unknown.
CN has been observed in the outer shells of the C-rich CSE IRC+10216 \citep{luc95}, and chemical models
predict that it is formed as a result of the photodissociation of HCN and HNC in these outer shells \citep[e.g.][]{nej88}.
Given that the HCN size is $\theta_{\mathrm{s}}$(HCN)=3\arcsecp85, CN would be expected to be in a shell external to the HCN.
According to the chemical model by \cite{lix16}, the CN peak abundance occurs at 
r$\sim$1.5$\times$10$^{16}$\,cm. We converted the area between the HCN outer shell
and the CN abundance peak to an equivalent emitting size, obtaining $\theta_{\mathrm{s}}$(CN)$\sim$6\arcsec.
With this size, we calculated the population diagram of CN (Fig.\,\ref{fig:rdallcn}).
We estimated \trot=9$\pm$2 and 
$N_{\mathrm{tot}}$$\sim$1$\times$10$^{14}$\,cm$^{-2}$.
At $r$$\sim$1.5$\times$10$^{16}$\,cm, \tkin\ is $\sim$50\,K and $n$(H$_2$)$\sim$2$\times$10$^{4}$\,cm$^{-3}$, thus, 
CN lines are probably sub-thermally excited ($n_{\mathrm{crit}}$$\gtrsim$10$^6$\,cm$^{-3}$).
With these physical conditions MADEX predicted optically thin lines.
We estimated $f$(CN)$\sim$1$\times$10$^{-7}$ which is consistent with previous estimations \citep{kim10}.
Chemical models predict abundances up to 3$\times$10$^{-7}$ \citep{wil97,lix16}, which are also in good agreement with our 
observations.

As far as we know, our discovery of NS emission toward \object{IK Tau} is the first detection of this molecule in this source.
Given that we did not resolve its hyperfine structure we cannot extract information on the line profiles observed.
There are not observational constraints on the emission size of this molecule.
According to the chemical model by \cite{lix16}, which predicts that NS would be formed through the neutral-neutral 
reaction of NH and S in an external shell of the envelope, similar to the CN shell, we adopted 
a $\theta_{\mathrm{s}}$(NS)$\sim$6\arcsec.
With this size, from its population diagram (Fig.\,\ref{fig:rdallns}), 
we derive a \trot=18$\pm$2\,K and a column density of (2.6$\pm$0.5)$\times$10$^{13}$\,cm.
As for CN, the lines of NS may be sub-thermally excited.
Since, we have not a set of collisional coefficients for NS we could neither estimate the opacities of the lines 
nor their critical densities.
A rough estimation under LTE conditions with MADEX yields optically thin lines ($\tau$$<$0.1).
We derived an abundance of $f$(NS)$\sim$2$\times$10$^{-8}$.
The chemistry of NS in O-rich CSEs was discussed in \cite{wil97}, although, these authors did not give a value for the predicted NS 
abundance. Recently, chemical models by \cite{lix16} predicted $f$(NS)$\sim$8$\times$10$^{-9}$, which is 
(within uncertainties) consistent with our results. 

For NO, we only detected two lines with low $S/N$. One of them is a blend 
of several hyperfine components.
For the NO hyperfine component spectrally resolved we derived \vexp=\vinf.
We estimated a rough value of the NO abundance adopting $\theta_{\mathrm{s}}$$\sim$6\arcsec
(like for CN and NS), 
and an excitation temperature of 30\,K, which is representative of the \tkin\ in the outer shells of the CSE 
($r$$\sim$[2--5]$\times$10$^{16}$\,cm).
With these considerations, we obtained $f$(NO)$\sim$2$\times$10$^{-6}$, in agreement with the predictions of the chemical 
model by \cite{lix16}. 

\subsection{P-bearing molecules}\label{sec:pbear} 
PN and PO were detected for the first time toward \object{IK Tau} by \cite{deb13}, and we detected one additional line of PN, and 
seven additional lines of PO.
Concerning PO, which has hyperfine structure, 
we observed several spectrally resolved (as well as unresolved) lines with linewidths consistent with \vexp$\sim$9\,\kms.
This suggests that PO emission lines arise from $r$$<$8\rstar.
\cite{deb13} mapped the brightness distribution of both PN and PO toward \object{IK Tau}, and found 
$\theta_{\mathrm{s}}$$\lesssim$0\arcsecp7. With this size we calculated the rotational diagram of both molecules
(Fig.\,\ref{fig:rdallphos}), and we derived low rotational temperatures (\trot$\sim$20\,K),  
$N_{\mathrm{tot}}$(PN)$\sim$8$\times$10$^{14}$\,cm$^{-2}$ and $N_{\mathrm{tot}}$(PO)$\sim$2$\times$10$^{15}$\,cm$^{-2}$.
The low \trot\ deduced for PN and PO is probably indicative of sub-thermal excitation since at the inner wind layers 
($<$0.7\arcsec)
where the emission is produced, the gas kinetic temperature is expected to be well above 200\,K.
We used our radiative transfer code (see Sect.\,\ref{sec:madex}) to confirm 
our results. MADEX predicted optically thin ($\tau$$\lesssim$0.4) and sub-thermally excited lines for PN.
In the case of PO, MADEX predicted optically thin ($\tau$$<$0.1) PO lines, under LTE approximation given that there is not 
a set of PO collisional coefficients available.
We derived $f$(PN)$\sim$7$\times$10$^{-7}$ and $f$(PO)$\sim$2$\times$10$^{-6}$, which are consistent 
with previous estimates considering uncertainties \citep{deb13}.

Concerning chemical models, these P-bearing molecules are adopted as parent molecules and their abundances 
are assumed from observations of the inner region of the CSE \citep{deb13,lix16}.
TE calculations predict abundances for PN compatible with our results, and one order of magnitude lower than our measurements for PO
\citep{tsu73,agu07,mil08}.

\subsection{HCO$^+$}\label{sec:ions} 
We detected the $J$=1--0 line of the HCO$^+$ ion. The $J$=3--2 line was not detected probably due to an unsufficient
sensitivity. The $J$=2--1 and the $J$=4--3 lines lie in wavelength ranges that were not observed. 
The flat-topped profile of the HCO$^{+}$ $J$=1--0 line indicates \vexp$\sim$\vinf, optically thin, and spatially unresolved 
emission.
We estimated a very rough value of the $f$(HCO$^{+}$) adopting an emission region and an excitation temperature equal to those 
adopted for NO (see Sect.\,\ref{sec:nbear}) since both molecules are expected to be formed in the outer shells of the CSE. 
With these values, we obtained $f$(HCO$^+$)$\sim$10$^{-8}$.

According to chemical models, this molecule is formed efficiently in the outer layers of O-rich CSEs as a result of reactions 
that involve CO, H$_2$O, and their photodissociation products, with abundances consistent with our results 
\citep[][and references therein]{wil97,san15}.

\subsection{The organic precursor missing link to carbon chemistry: H$_2$CO}\label{sec:organ}
This is the first detection of the organic precursor molecule H$_2$CO toward \object{IK Tau}.
We detected ortho and para lines with flat-topped profiles, indicating optically thin and 
spatially unresolved emission, even for the lines at high 
frequencies with $\theta_{\mathrm{b}}$$\sim$8\arcsec (see Eq.\,(\ref{eq:hpbw})).
We adopted a size of $\theta_{\mathrm{s}}$(H$_2$CO)=2\arcsec, which is an educated guess, taking into account that
most of the molecules detected are expected to emit in that region of the CSE.

In the population diagram of H$_2$CO (see Fig.\,\ref{fig:rdallh2co}), 
the $K_{\mathrm{a}}$=0, $K_{\mathrm{a}}$=1, and $K_{\mathrm{a}}$=2 ladders were fitted separately.
For $K_{\mathrm{a}}$=2 only two data were collected.
The average rotational temperature indicated by the fits of the different $K_{\mathrm{a}}$ ladders is $\sim$10\,K.
For the $K_{\mathrm{a}}$=3 ladder we only detected two lines
with the same upper energy level, therefore, we adopted an average excitation temperature of $\sim$10\,K 
to derive a rough value of the $K_{\mathrm{a}}$=3 column density. 
Formaldehyde is an asymmetric rotor with its dipolar moment oriented along the $a$ axis.
Therefore, transitions between different $K_{\mathrm{a}}$ levels are weakly connected through radiative processes.
This explains why each of the $K_{\mathrm{a}}$ ladders appears as separated lines in the rotational diagram.
The column densities derived result in an ortho-to-para ratio of $\sim$3:1, which has been computed dividing the 
sum of the $K_{\mathrm{a}}$=1 and 3 (ortho transitions) by the sum of the $K_{\mathrm{a}}$=0 and 2 (para transitions) column densities 
(see Fig.\,\ref{fig:rdallh2co}).
The total column density is $N_{\mathrm{tot}}$(H$_2$CO)$\sim$3$\times$10$^{14}$\,cm$^{-2}$.
Adding both o-H$_2$CO and p-H$_2$CO, we obtained a fractional abundance of $f$(H$_2$CO)$\sim$2$\times$10$^{-7}$.
In this case, MADEX predicted optically thin lines ($\tau$$<$0.3) and sub-thermal excitation 
which is consistent with \tkin($r$=1\arcsec)$>$\trot.

The origin of formaldehyde, as well as other C-bearing species, in O-rich envelopes has puzzled the scientific community 
since H$_2$CO was first detected in the O-rich CSE \object{OH231.8+4.2}, with a fractional abundance of 4$\times$10$^{-8}$ 
\citep{lin92,cha95}. 
\cite{mil93} proposed a formation route which requires methane to produce formaldehyde in the 
external envelope with abundances up to 10$^{-7}$ depending on the mass loss rate of the star. 
Methane is a highly symmetric molecule with no permanent dipole, thus, indirect evidence for the presence of methane can be 
provided by the search of expected products of CH$_4$ chemistry, such as C$_2$H and CH$_3$OH.
A previous search for these molecules toward \object{IK Tau} and other O-rich CSEs has resulted in non-detections
of C$_2$H ($f$(C$_2$H)$<$9.7$\times$10$^{-9}$) and methanol ($f$(CH$_3$OH)$<$3.2$\times$10$^{-8}$) \citep{cha97,mar05}. 
We also did not detect emission of these molecules consistent with the upper limits provided by \cite{mar05}.
If the correct scenario were that H$_2$CO is formed in the outer envelope, the size adopted for the 
calculation of the population diagram would result in an abundance overestimate. 
In that case, assuming $\theta_{\mathrm{s}}$=6\arcsec, like 
for CN, NS, or NO, we derive an abundance of $f$(H$_2$CO)$\sim$2$\times$10$^{-8}$.
Furthermore, a clumpy envelope may lead to an enhanced photochemistry in the inner layers of the CSE, 
which result in the formation of carbon molecules in the inner and intermediate layers of O-rich CSEs, and in particular 
formaldehyde with $f$(H$_2$CO)$\sim$10$^{-9}$ \citep{agu10}. 

\subsection{Isotopic ratios}\label{sec:isot}
Isotopic ratios of different species can be measured from the column densities derived in the rotational diagrams
(see Table\,\ref{tab:isorat}). However, these ratios have to be considered as lower limits when the molecule used to calculate the 
ratio has optically thick lines. In case that the opacities are moderately high, the isotopic ratio can be corrected 
using the approach by \cite{gol99}.

For $^{12}$C/$^{13}$C ratio, we measured values of eight to ten depending on the molecule used (i.e. CO, CS or HCN).
The opacity correction yielded a $^{12}$C/$^{13}$C ratio of $\sim$10 from CO.
This value is in good agreement with that obtained by \cite{ram14} from their radiative transfer model of the 
$^{12}$CO and $^{13}$CO emission in \object{IK Tau}. This ratio is also compatible with other estimates in M-type stars like TX\,Cam or 
W\,Hya, consistent with a standard evolution for an M-type star \citep[][and references therein]{ram14}.

For the $^{16}$O/$^{17}$O and $^{16}$O/$^{18}$O ratios, we estimated lower limits of 90 and 60, respectively. 
MADEX predicted opacities as high as $\tau$$\sim$10 for a few SiO lines that would result in opacity corrected 
values of one order of magnitude higher, in agreement with previous estimates \citep{dec10b}.

The opacity corrected isotopic ratios of $^{28}$Si/$^{29}$Si and $^{28}$Si/$^{30}$Si are $\sim$18 and $\sim$34, respectively. 
Both isotopic ratios are (within uncertainties) in reasonable agreement with previous estimations toward \object{IK Tau} \citep{dec10a}, 
and also with the solar ratios ([$^{28}$Si/$^{29}$Si]$\sim$20 and [$^{28}$Si/$^{30}$Si]$\sim$30, \cite{asp09}). 
Therefore, it seems that in the case of \object{IK Tau}, Si isotopic ratios do not indicate significant alterations in the post-main 
sequence evolution.

Finally, we measured the isotopic ratio of $^{32}$S/$^{34}$S using SO, SiS, SO$_2$, and CS obtaining values between 10 and 13. 
We corrected the effect of optically thick emission and we estimated a $^{32}$S/$^{34}$S ratio of $\sim$15.
As far as we know, there are no previous observational constraints to this isotopic ratio toward \object{IK Tau}. 
The solar $^{32}$S/$^{34}$S ratio is $\sim$22 \citep{asp09}. Recently, \cite{dan16} reported [$^{32}$S/$^{34}$S]$\sim$32 
toward the O-rich CSE of R\,Dor. Both, Sun and R\,Dor isotopic ratios are, within uncertainties, compatible with 
our estimations.

\subsection{Qualitative comparison with other O-rich objects}\label{sec:other}
The molecular content of only a few O-rich CSEs has been studied so far. 
In particular, the best studied objects are the AGB CSE \object{IK Tau}, the CSE of the hypergiant \object{VY CMa} and 
the peculiar object \object{OH231.8+4.2} \citep[][and references therein]{alc13,deb13,mat14,san15,ziu07}.

\object{IK Tau} has, in terms of chemical composition, more similarities with \object{VY CMa}.
AlOH and H$_3$O$^{+}$ are the only molecules present in the CSE of \object{VY CMa} that are not found 
in the CSE of \object{IK Tau}. Formaldehyde is found in the CSE of \object{IK Tau} but it is not found in the CSE of \object{VY CMa}.
Regardless of the possible chemical processes at work in the CSE of \object{IK Tau},
the presence of CO, CN, CS, HCN, HNC, HCO$^{+}$, and H$_2$CO 
in \object{IK Tau} indicates that the emission of C-bearing molecules in \object{VY CMa} is not so unique \citep{ziu09}. 

\object{OH231.8+4.2} displays emission of several molecules that are not found toward \object{IK Tau}:
HNCO, HNCS, OCS, H$^{13}$CO$^{+}$, SO$^{+}$, N$_2$H$^+$, and H$_3$O$^{+}$.
The remarkable chemistry of \object{OH231.8+4.2} probably reflects the molecular regeneration process within its envelope after 
the passage of fast ($\sim$100\,\kms) shocks that accelerated and dissociated molecules in the AGB wind $\sim$800\,yr ago
\citep{san15}.
In \object{IK Tau} there is no evidence of a similar molecular destruction process by fast ($\sim$100\,\kms) velocity shocks.
Instead, slower shocks due to stellar pulsation may have an impact on the chemistry of AGB CSEs \citep{gob16}.
However, the fact that these molecules are not observed toward \object{IK Tau} point out that slow shocks are not able to 
enhance the formation of these particular species, which are unexpectedly abundant in \object{OH231.8+4.2} \citep{vel15}.
Nevertheless, slow shocks could enhance the formation of molecules like HCN or CS (see Sect.\,\ref{sec:crich}).
Another difference with respect to \object{OH231.8+4.2} is that \object{IK Tau} displays emission of NaCl and more intense lines of
vibrationally excited SiO (Velilla Prieto et al. in prep.). The emission of these lines arises from very warm and inner regions of 
the CSE.
The absence of NaCl and the weakness of the vibrationally excited SiO lines toward \object{OH231.8+4.2},
probably indicates that the mass loss rate of \object{OH231.8+4.2} is decreasing at present \citep[as suggested by][]{san02}, which results 
in the progressive growth of a central cavity around the star.

\section{Conclusion}\label{sec:conclusion}
In this work we present the detection toward \object{IK Tau} of $\sim$350 rotational lines corresponding to 
a list of H-, O-, C-, N-, S-, Si- and P-bearing molecules, which evidences an active chemistry for an O-rich AGB CSE. 

We detected for first time in this source emission of 
HCO$^{+}$, NO, H$_2$CO, and NS. We also detected for the first time toward \object{IK Tau} rotational lines of
C$^{18}$O, Si$^{17}$O, Si$^{18}$O, $^{29}$SiS,  $^{30}$SiS, Si$^{34}$S,
H$^{13}$CN, $^{13}$CS, C$^{34}$S, H$_2$$^{34}$S, $^{34}$SO, $^{34}$SO$_2$, and H$_2$O $\nu_{\mathrm{2}}$=2,
as well as several rotational lines of SiO isotopologues in vibrationally excited states.
In addition, we significantly increased the number of lines detected for those molecules that were previously identified  
toward \object{IK Tau}.
This has allowed us to deduce characteristic values of 
the rotational temperatures, column densities, and ultimately fractional 
abundances of the molecules present in its envelope.
From our work we extract the following conclusions:
\begin{itemize}
\item The intensity of the rotational lines of molecules in the ground vibrational state do not show a significant 
variability as a function of time, for the spectral ranges that we could observe in different epochs. 
The small variations found for these lines can be explained owing to calibration or pointing uncertainties within a 
25\%. We confirmed the time variability of the intensity of the lines of molecules in vibrationally excited states (e.g. SiO $v=1$)
by an average factor of 60\%.

\item Most of the molecules display rotational temperatures between 15 and 40\,K. NaCl and SiS 
isotopologues display rotational temperatures of $\sim$65\,K. 
\item We detected a warm component of SO$_2$ traced out by lines with upper energy levels between
160 and 730\,K which display \vexp$<$\vinf. 
This points out 
that SO$_2$ is present close to the stellar surface ($\lesssim$8\,\rstar) with an abundance of 
$f$(SO$_2$)$\sim$10$^{-6}$.
\item Among the species detected, we highlight the detection of H$_2$CO and NS for the first time in this 
source with abundances of $f$(H$_2$CO)$\sim$[10$^{-7}$--10$^{-8}$] and $f$(NS)$\sim$10$^{-8}$.
We also estimated fractional abundances for the first time detected (toward \object{IK Tau}) molecules HCO$^+$ and NO
obtaining $f$(HCO$^+$)$\sim$10$^{-8}$ and $f$(NO)$\sim$10$^{-6}$.
\item The detection of several C-bearing species like HCN, CS, H$_2$CO or CN with abundances of $\sim$10$^{-7}$
indicates an active carbon chemistry which is not expected given that most of the available carbon should be
locked up into CO. 
\item The greatest discrepancies between our results and previous chemical models are found for PO, NaCl, and SO$_2$.
\end{itemize}

It would be necessary to obtain very high angular resolution observations to characterise the
molecular emission in the inner parts of the CSE and the abundances and distribution of the molecules 
formed in this region.
Further investigation is required to understand the nature of the discrepancies found between 
our derived values and chemical models, in particular, the discrepancies for S-bearing molecules and 
C-bearing molecules.
The inclusion of photo-induced or shock-induced chemistry or maybe other processes is necessary to enhance 
the formation of these molecules up to values comparable to the abundances observed.
Additionally, there are $\sim$40 lines that still remain unidentified.
We expect that future observations, supported by improvements in the molecular catalogues 
and chemical models, lead to fully understand the envelope of \object{IK Tau} and, more generally, in O-rich AGB envelopes.

\begin{acknowledgements}
We acknowledge the IRAM staff for the support and help offered during all the observational runs.
We acknowledge the Spanish MICINN/MINECO for funding support through grants
AYA2009-07304, AYA2012-32032, the ASTROMOL Consolider project
CSD2009-00038 and also the European Research Council funding support (ERC grant 610256:
NANOCOSMOS).  
L.\,V.\,P.\, also acknowledges the support of the Universidad Complutense de
Madrid PhD programme.  
This research has made use of the The JPL Molecular Spectroscopy catalog,
The Cologne Database for Molecular Spectroscopy, the SIMBAD database
operated at CDS (Strasbourg, France), the NASA's Astrophysics Data
System, the IRAM GILDAS software, and Aladin.
\end{acknowledgements}


\appendix
\onecolumn
\section{Table of measured lines}\label{sec:app_measures}
\vspace{-5mm}
\begin{longtable}{lllllll}
\caption{\label{tab:measures} Measured and spectroscopic parameters of the detected lines.}\\
\hline\hline
Molecule & Transition & $\nu _{\mathrm{rest}}$ & $E_{\mathrm{u}}$ & $A_{\mathrm{ul}}$ & $\int T_{\mathrm{ant}}^{*}\,d\mathrm{v}$ & v$_{\mathrm{exp}}$ \\    
         & QNs        & (MHz)                  & (K)              &  (s$^{-1}$)       & (K km s$^{-1}$)                          & (km s$^{-1}$) \\ 
\hline
\endfirsthead

\caption{Continued.}\vspace{-1mm}\\
\hline\hline
Molecule & Transition & $\nu _{\mathrm{rest}}$ & $E_{\mathrm{u}}$ & $A_{\mathrm{ul}}$ & $\int T_{\mathrm{ant}}^{*}\,d\mathrm{v}$ & v$_{\mathrm{exp}}$ \\    
         & QNs        & (MHz)                  & (K)              &  (s$^{-1}$)       & (K km s$^{-1}$)                          & (km s$^{-1}$) \\ 
\hline
\endhead

\endfoot

\hline
\hline
\multicolumn{7}{l}{{\tablefoot{
(Col. 2) Quantum numbers;
(Col. 6) Integrated intensity (in antenna temperature) of the line with its formal uncertainty; 
(Col. 7) Expansion velocity derived from the linewidth with its uncertainty which includes the width of half a channel (1\,MHz) except for the 
lines with a low $S/N$ where a full channel (2\,MHz) is included.
The uncertainties given (within parentheses) do not include additional uncertainties derived from the absolute flux calibration. 
(HB) Hard blend, thus, we cannot estimate the contribution of the line to the spectral feature observed.
(BF) Bad fit.}}}
\endlastfoot
\hline
CO&1--0&115271.202&5.5&7.205$\times$10$^{-8}$&49.14\,(0.04)&18\,(3)\\
CO&2--1&230538.000&16.6&6.916$\times$10$^{-7}$&121.61\,(0.06)&16.7\,(1.3)\\
CO&3--2&345795.990&33.2&2.501$\times$10$^{-6}$&103.38\,(0.07)&16.7\,(1.0)\\
\hline
$^{13}$CO&1--0&110201.354&5.3&6.336$\times$10$^{-8}$&5.08\,(0.02)&19\,(3)\\
$^{13}$CO&2--1&220398.684&15.9&6.082$\times$10$^{-7}$&13.11\,(0.05)&18.6\,(1.4)\\
$^{13}$CO&3--2&330587.964&31.7&2.199$\times$10$^{-6}$&15.37\,(0.12)&18.1\,(1.0)\\
\hline
C$^{18}$O&1--0&109782.176&5.3&6.263$\times$10$^{-8}$&0.07\,(0.02)&14\,(6)\\
C$^{18}$O&2--1&219560.358&15.8&6.013$\times$10$^{-7}$&0.39\,(0.03)&19.8\,(1.4)\\
\end{longtable}
\clearpage


\begin{longtable}{llllllll}
\caption{\label{tab:variability} Line parameters for transitions that may present variability.}\\
\hline\hline
Molecule & Transition & $\nu _{\mathrm{rest}}$ & $E_{\mathrm{u}}$ & $A_{\mathrm{ul}}$ & $\int T_{\mathrm{ant}}^{*}\,d\mathrm{v}$ & v$_{\mathrm{exp}}$ & Julian date  \\    
         & QNs        & (MHz)                  & (K)              &  (s$^{-1}$)       & (K km s$^{-1}$)                          & (km s$^{-1}$)      &              \\ 
\hline
\endfirsthead

\caption{Continued.}\\
\hline\hline
Molecule & Transition & $\nu _{\mathrm{rest}}$ & $E_{\mathrm{u}}$ & $A_{\mathrm{ul}}$ & $\int T_{\mathrm{ant}}^{*}\,d\mathrm{v}$ & v$_{\mathrm{exp}}$ & Julian date  \\    
         & QNs        & (MHz)                  & (K)              &  (s$^{-1}$)       & (K km s$^{-1}$)                          & (km s$^{-1}$)      &              \\ 
\hline
\endhead

\endfoot

\noalign{\vskip 1mm} 
\hline
\hline
\multicolumn{8}{l}{{\tablefoot{
(Col. 2) Quantum numbers of the corresponding line;
(Col. 6) Integrated intensity (in antenna temperature) of the line with its formal uncertainty; 
(Col. 7) Expansion velocity derived from the linewidth with its uncertainty which includes the width of half a channel (1\,MHz) except for the 
lines with a low $S/N$ where a full channel (2\,MHz) is included.
(Col. 8) Julian date.
The uncertainties given (within parentheses) do not include additional uncertainties derived from the absolute flux calibration 
and baseline subtraction. The ellipsis symbol indicate that the field is equal to the same field of the previous row.}}}
\endlastfoot
\hline
SiO $v$=1&2--1&86243.429&1775.4&2.903$\times$10$^{-5}$&154.73\,(0.04)&7\,(3)&2455018.5\\
SiO $v$=1&3--2&129363.350&1781.6&1.050$\times$10$^{-4}$&50.01\,(0.02)&5\,(2)&2455018.5\\
SiO $v$=1&4--3&172481.120&1789.9&2.580$\times$10$^{-4}$&12.98\,(0.02)&5\,(2)&2455578.5\\
SiO $v$=1&5--4&215596.021&1800.3&5.154$\times$10$^{-4}$&7.52\,(0.03)&4\,(1)&2456579.5\\ 
SiO $v$=1&6--5&258707.335&1812.7&9.042$\times$10$^{-4}$&1.04\,(0.07)&8\,(1)&2455309.5\\
\ldots&\ldots&\ldots&\ldots&\ldots&1.99\,(0.05)&4\,(1)&2456579.5\\
SiO $v$=1&7--6&301814.346&1827.2&1.452$\times$10$^{-3}$&4.40\,(0.08)&6\,(1)&2455550.5\\
\ldots&\ldots&\ldots&\ldots&\ldots&5.35\,(0.10)&6\,(1)&2455574.5\\
\end{longtable}
\clearpage

\section{Unidentified lines}\label{sec:app_uis}
\begin{longtable}{llll} %
\caption{\label{tab:uis} Unidentified features and, in some cases, tentative line identification.}\\
\hline\hline
$\nu _{\mathrm{rest}}$ & Tentative assignment & $\int T_{\mathrm{ant}}^{*}\,d\mathrm{v}$ & v$_{\mathrm{exp}}$ \\    
(MHz)                  & -                    & (K km s$^{-1}$)                          & (km s$^{-1}$) \\ 
\hline
\endfirsthead

\caption{Continued}\\
\hline\hline
$\nu _{\mathrm{rest}}$ & Tentative assignment & $\int T_{\mathrm{ant}}^{*}\,d\mathrm{v}$ & v$_{\mathrm{exp}}$ \\    
(MHz)                  & -                    & (K km s$^{-1}$)                          & (km s$^{-1}$) \\ 
\hline
\endhead

\endfoot

\hline
\multicolumn{4}{l}{{\tablefoot{
(Col. 3) Integrated intensity (in antenna temperature) of the line with its formal uncertainty; 
(Col. 4) Expansion velocity derived from the linewidth with its uncertainty which includes the width of half a channel (1\,MHz) except for the 
lines with a low $S/N$ where a full channel (2\,MHz) is included.
The uncertainties given (within parentheses) do not include additional uncertainties derived from the absolute flux calibration 
and baseline subtraction.}}}
\endlastfoot
 81166                  &                  --- & 0.08\,(0.03) & 16\,(7)\\ 
 83825                  & SiO $v$=5 $J$=2--1   & 0.30\,(0.03) & 35\,(5)\\
 84478                  &                  --- & 0.06\,(0.02) & 7\,(7)\\
 86190                  & OH $v$=1 $J,\Omega,p,F$=8/2,3/2,1,9--8/2,3/2,-1,9 & 0.71\,(0.02) & 7\,(3)\\
 87385                  &                  --- & 0.23\,(0.02) & 17\,(3)\\
108892                  &                  --- & 0.09\,(0.01) & 11\,(4)\\
128883                  & TiO$_2$ $J_{\mathrm{K,k}}$=19$_{\mathrm{10,10}}$--20$_{\mathrm{9,11}}$ & 0.10\,(0.02) & 13\,(4)\\
131195                  &  SiN $N,J,F$=3,7/2,7/2-2,5/2,7/2 & 0.05\,(0.01) & 7\,(4)\\
\end{longtable}
\clearpage

\begin{table}[hbtp!] %
\caption{Artifacts identified in the spectra between 167475--174800\,MHz.
}
\label{tab:loproblems}
\centering    
\begin{tabular}{l l l}
\hline\hline
$\nu _{\mathrm{rest}}$ & $\int T_{\mathrm{ant}}^{*}\,d\mathrm{v}$ & v$_{\mathrm{exp}}$ \\ 
(MHz)                  & (K km s$^{-1}$)                          & (km s$^{-1}$) \\ 
\hline
168422  & 0.007\,(0.002) &9\,(1)\\
169104  & 0.008\,(0.002) &23\,(2)\\
169729  & 0.012\,(0.002) &6.1\,(0.9)\\
170407  & 0.008\,(0.002) &4.0\,(0.9)\\
171000  & 0.005\,(0.002) &22.8\,(0.6)\\
171145  & 0.037\,(0.002) &10.9\,(0.7)\\
171174  & 0.010\,(0.002) &20\,(1)\\
171393  & 0.032\,(0.002) &26.0\,(0.4)\\
171591  & 0.009\,(0.002) &13\,(2)\\
171638  & 0.044\,(0.003) &17.8\,(0.8)\\
171758  & 0.016\,(0.003) &23\,(2)\\
171874  & 0.013\,(0.003) &4.6\,(0.4)\\
172605  & 0.063\,(0.004) &8.13\,(0.08)\\
172725  & 0.049\,(0.003) &3.9\,(0.2)\\
172799  & 0.010\,(0.003) &21\,(1)\\
173242  & 0.015\,(0.003) &3\,(31)\\
173444  & 0.048\,(0.004) &24.9\,(0.7)\\
173562  & 0.122\,(0.004) &18.7\,(0.3)\\
173812  & 0.135\,(0.006) &18.8\,(0.3)\\
173933  & 0.051\,(0.005) &22\,(1)\\
174267  & 0.014\,(0.004) &7\,(2)\\
\hline
\hline
\end{tabular}
\end{table}

\begin{figure*}[!hbtp]
\centering
\includegraphics{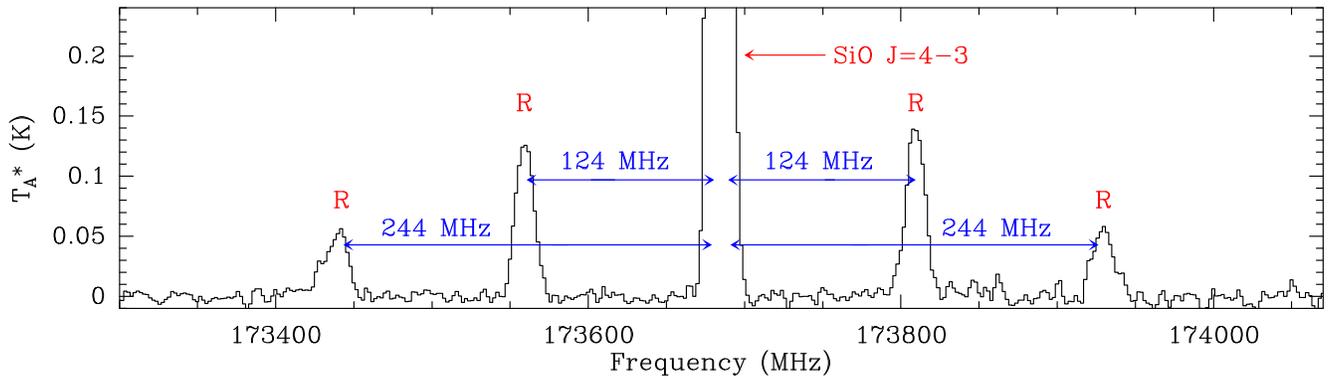}
\caption{Example of the artifacts (R) produced by the receivers (see Table\,\ref{tab:loproblems}).
Several symmetrical spurious replicas of the real SiO $J$=4--3 line appear at both sides of it (equidistant). 
The intensities of the replicas decrease with the frequency distance to the real feature.
}                                                                                               
\label{fig:ikprobs}
\end{figure*}

\onecolumn
\section{IRAM-30\,m survey of \object{IK Tau}}\label{sec:app_survey}

\begin{figure*}[htp] 
\centering
\includegraphics[width=0.70\hsize]{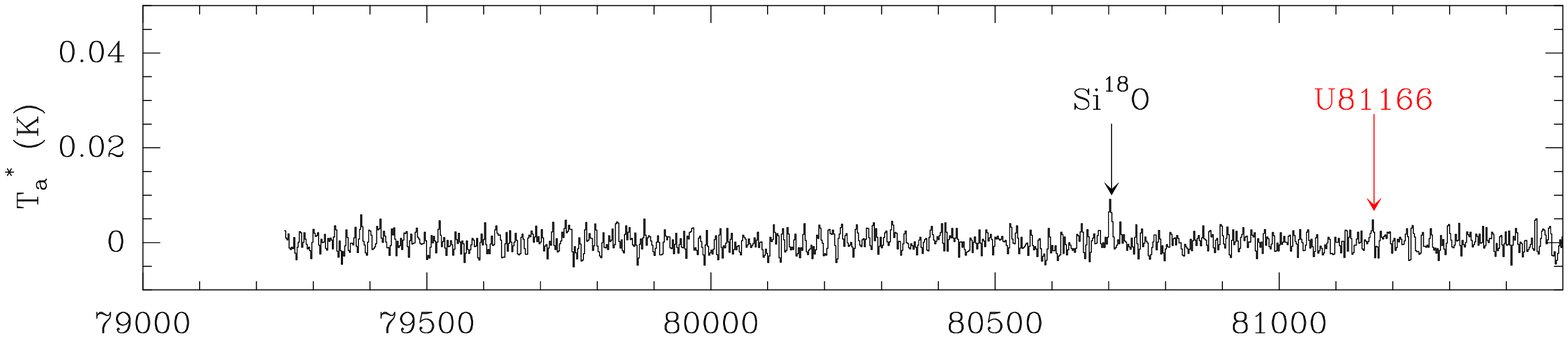}
\includegraphics[width=0.70\hsize]{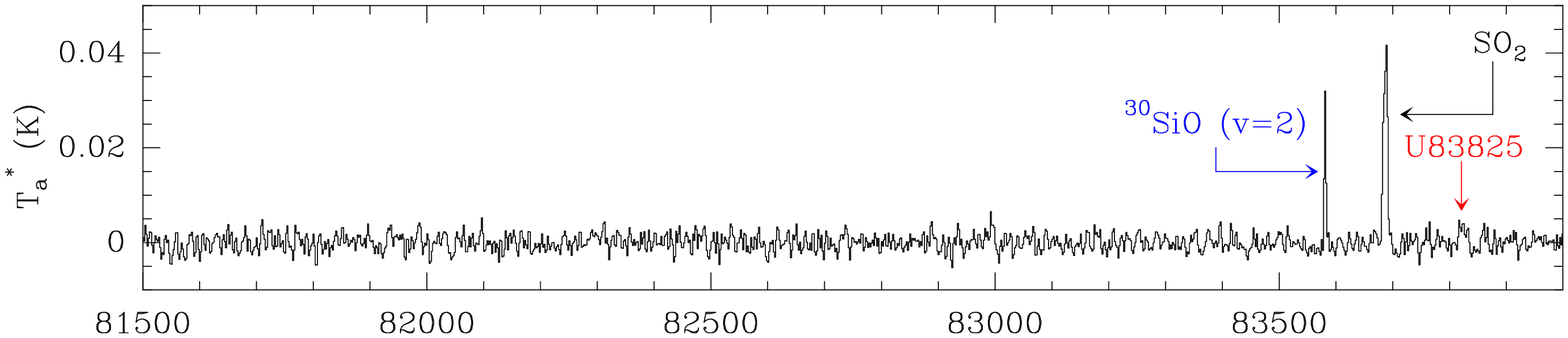}
\includegraphics[width=0.70\hsize]{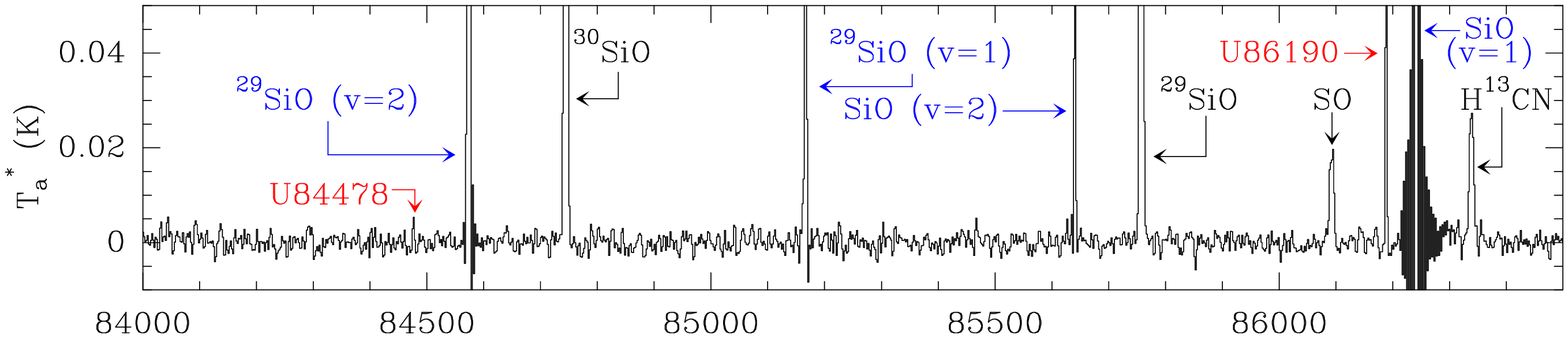}
\includegraphics[width=0.70\hsize]{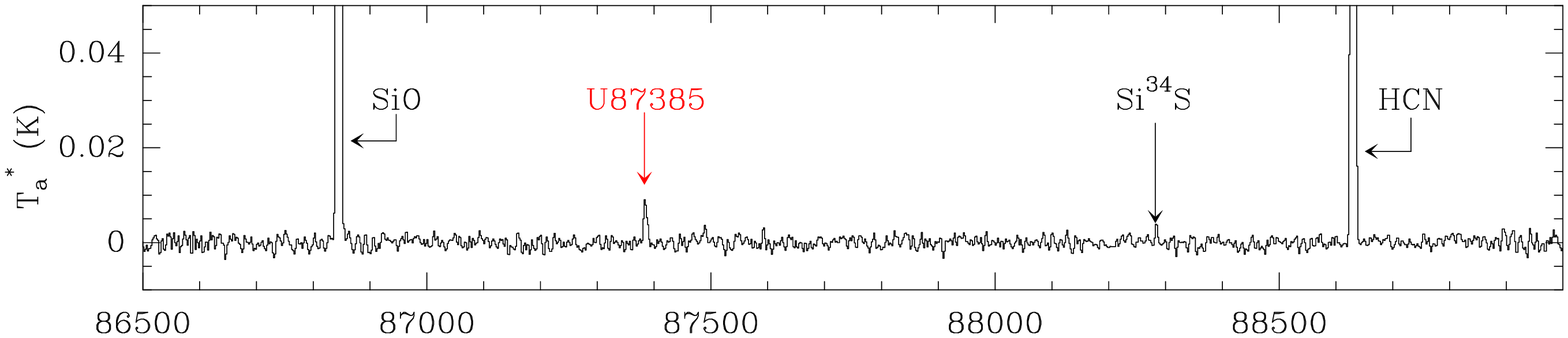}
\includegraphics[width=0.70\hsize]{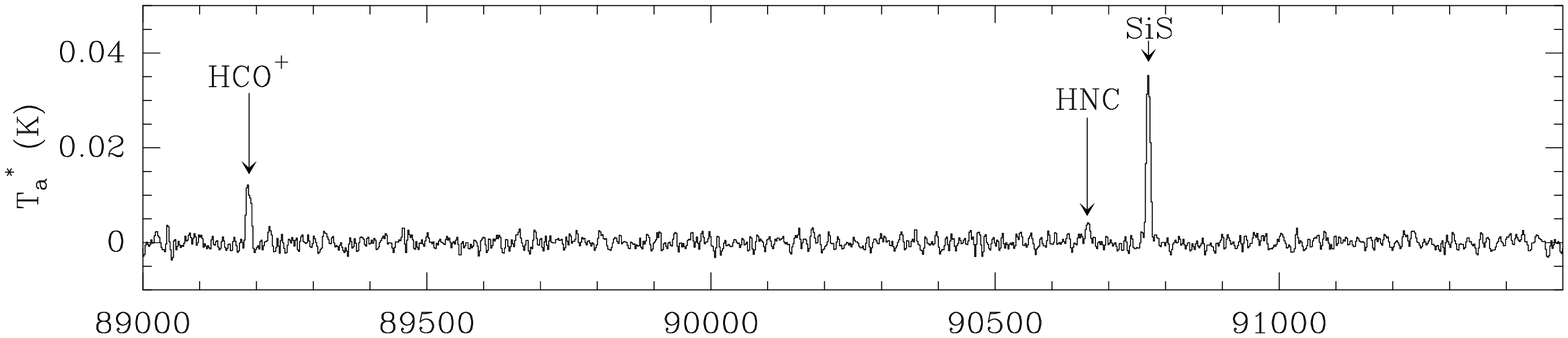}
\includegraphics[width=0.70\hsize]{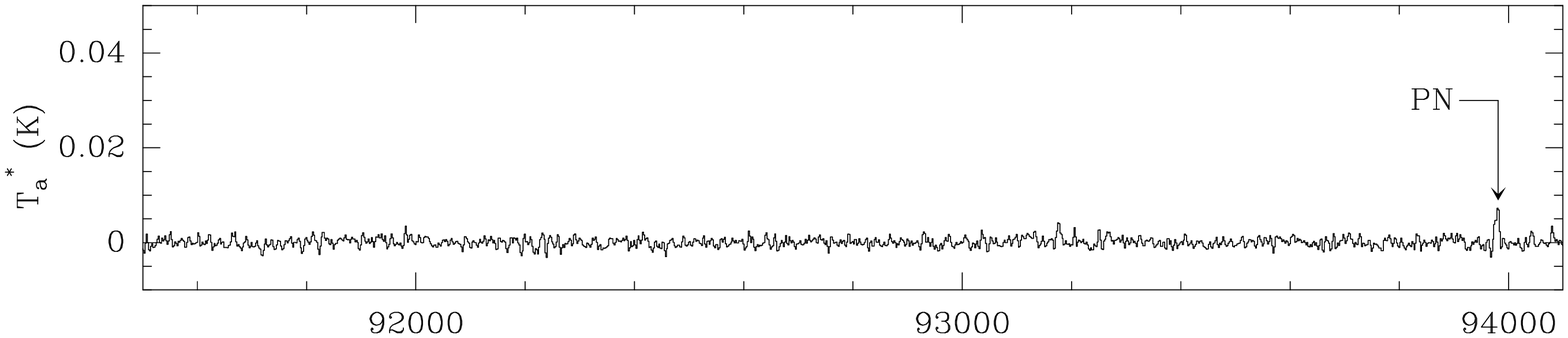}
\includegraphics[width=0.70\hsize]{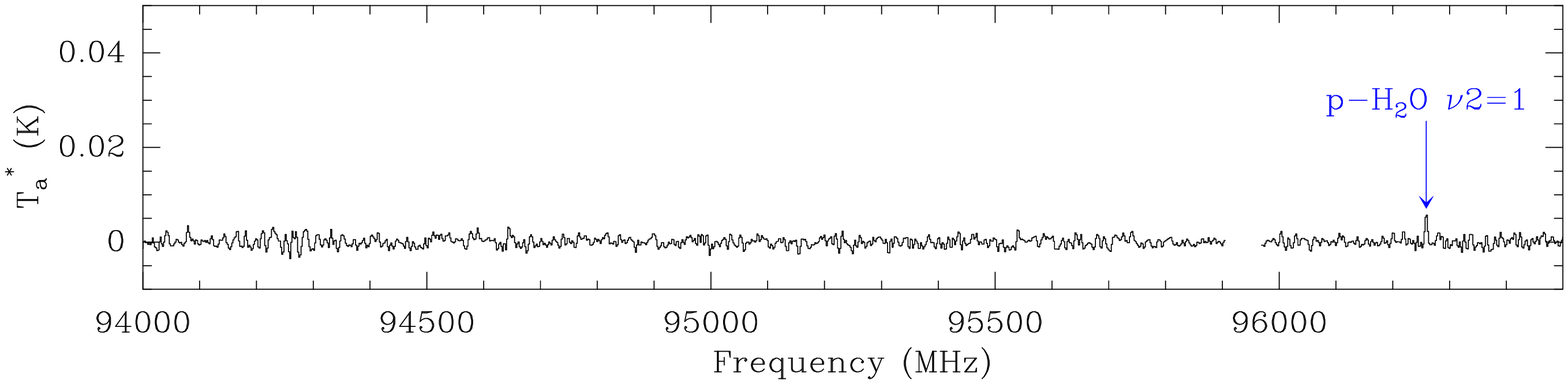}
\caption{Line identification of the IRAM-30\,m line survey of IK\,Tau. We marked in red the unidentified lines (UIs) 
with the central frequency of the line. The lines which display time variability of their intensity are marked in blue.
Image band and spurious feaures/artifacts have been blanked off (see Table\,\ref{tab:loproblems} and 
Fig.\,\ref{fig:ikprobs}.)}
\label{fig:survey1}
\end{figure*}

\setcounter{figure}{0}
\begin{figure*}[p] 
\centering
\includegraphics[width=0.70\hsize]{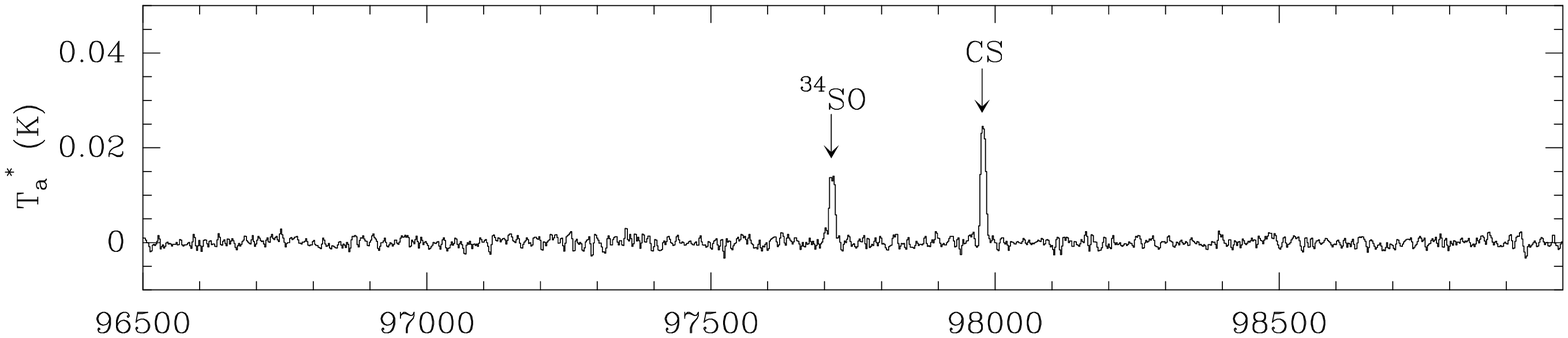}
\includegraphics[width=0.70\hsize]{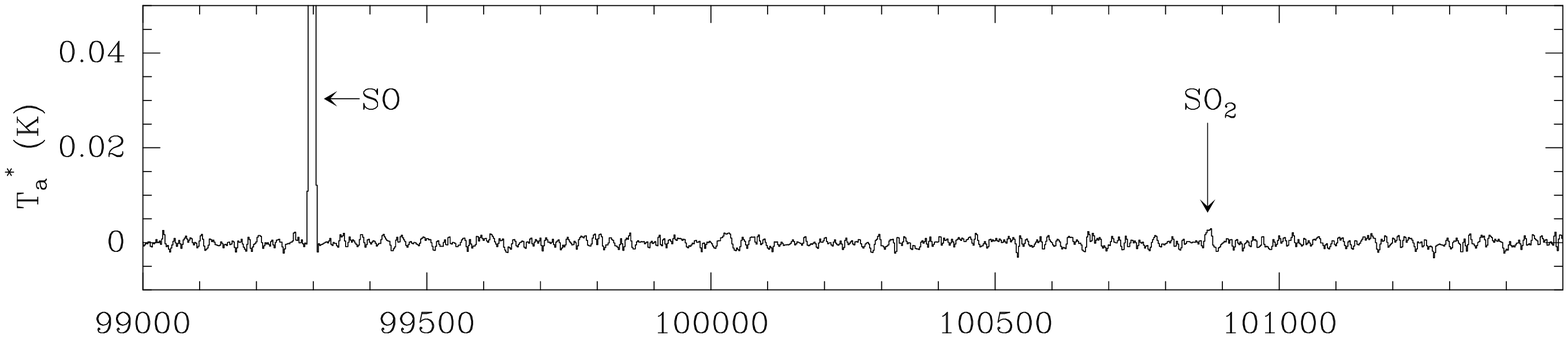}
\includegraphics[width=0.70\hsize]{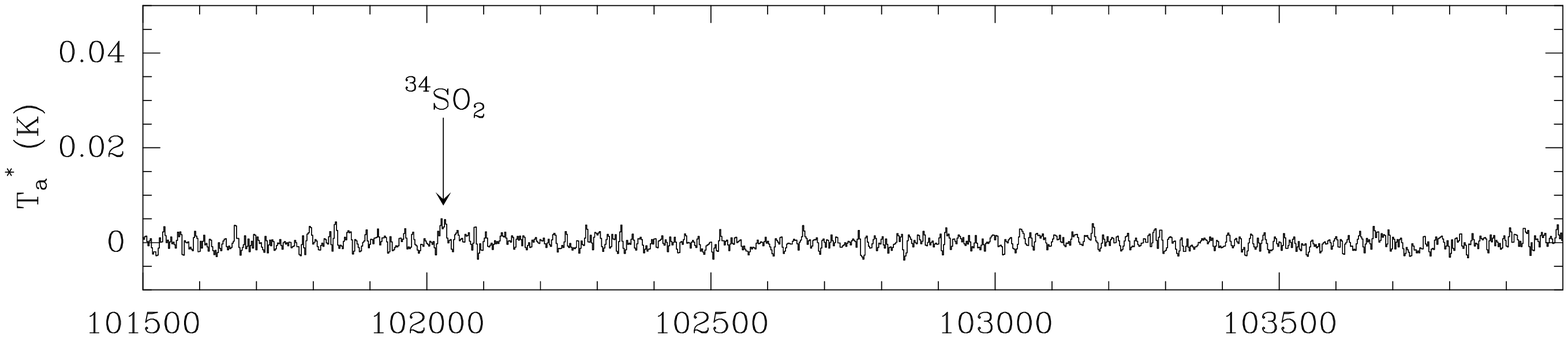}
\includegraphics[width=0.70\hsize]{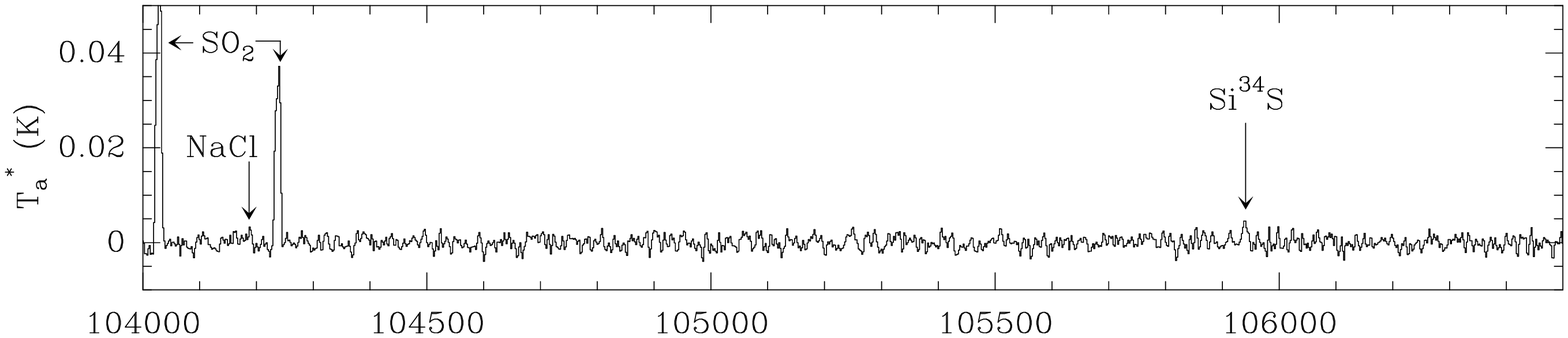}
\includegraphics[width=0.70\hsize]{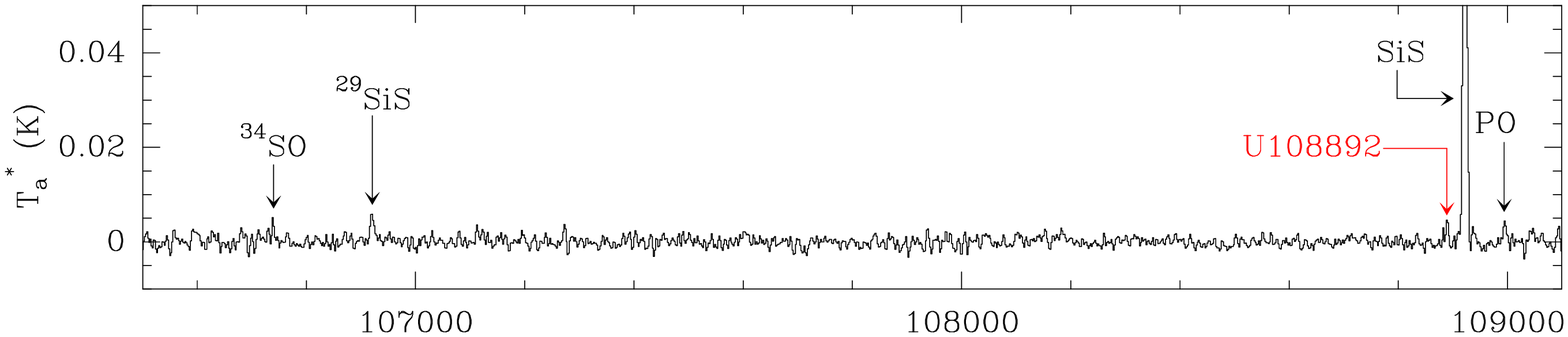}
\includegraphics[width=0.70\hsize]{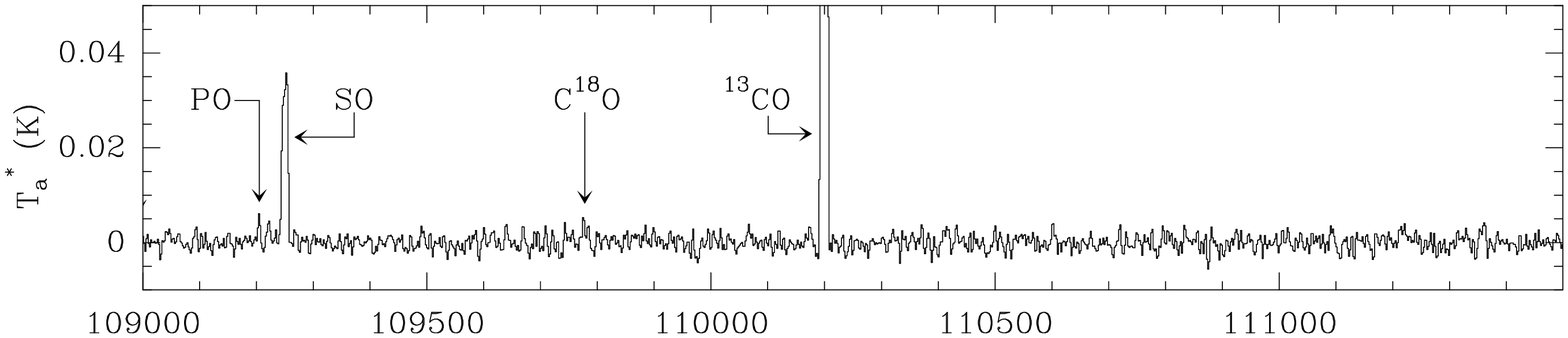}
\includegraphics[width=0.70\hsize]{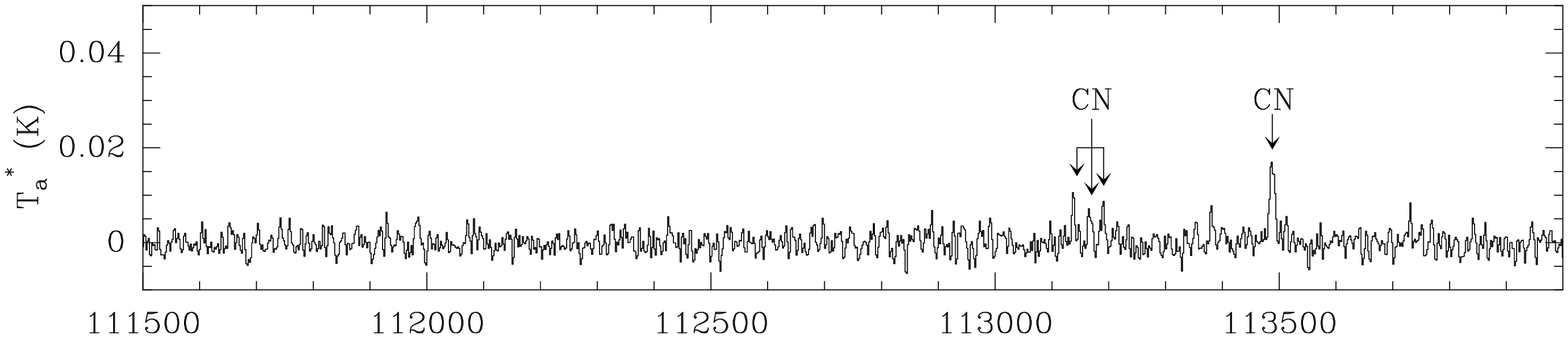}
\includegraphics[width=0.70\hsize]{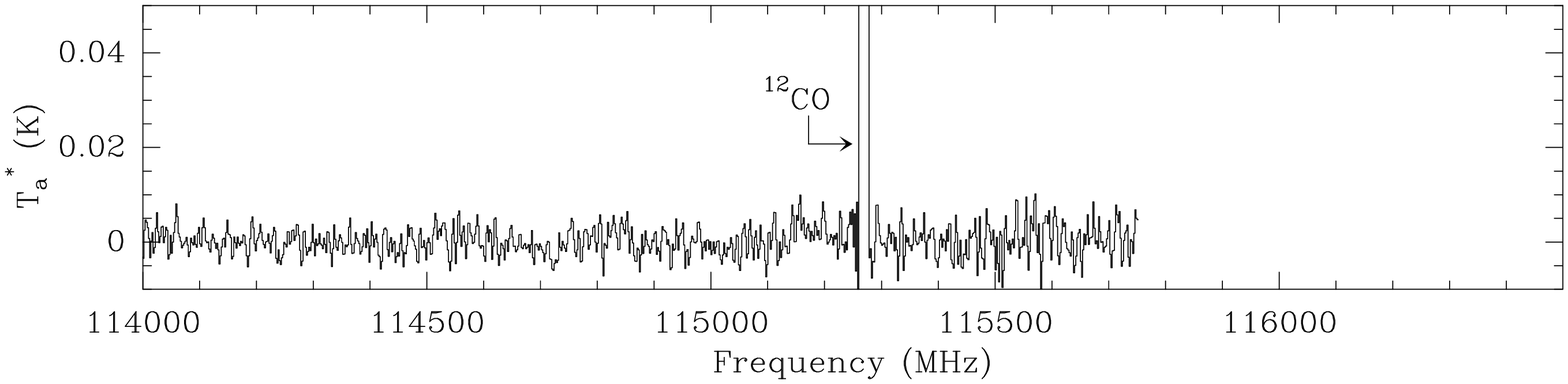}
\caption{(Continued)}
\end{figure*}

\setcounter{figure}{0}
\begin{figure*}[p] 
\centering
\includegraphics[width=0.70\hsize]{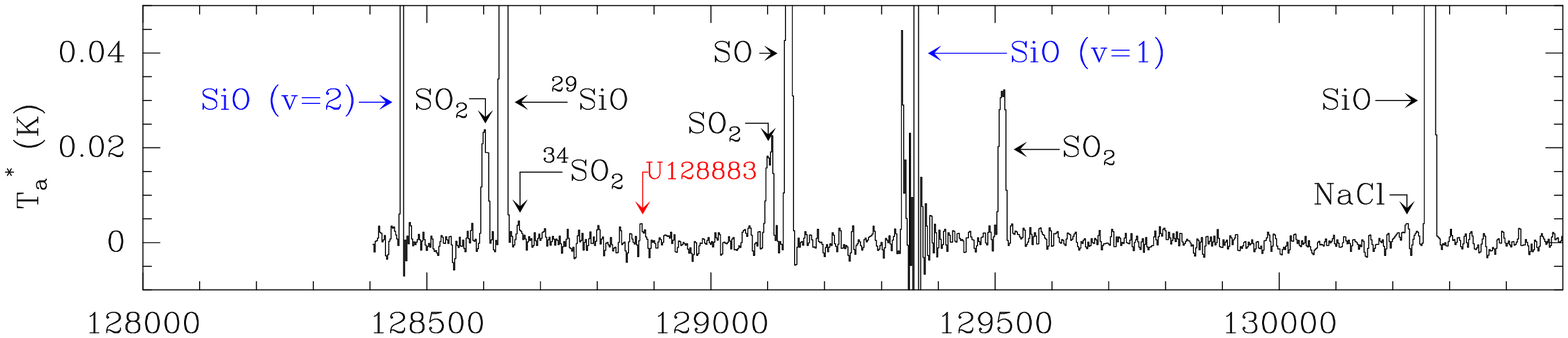}
\includegraphics[width=0.70\hsize]{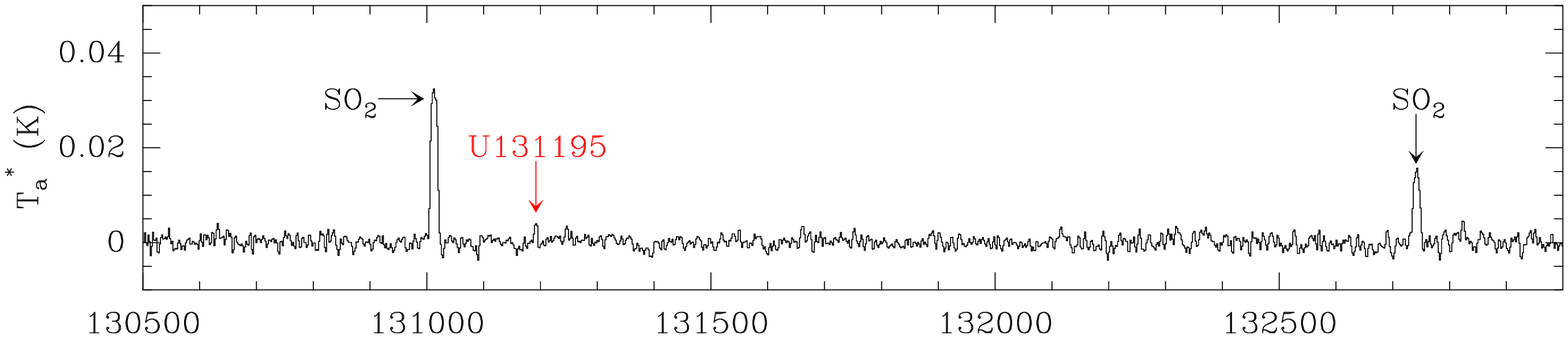}
\includegraphics[width=0.70\hsize]{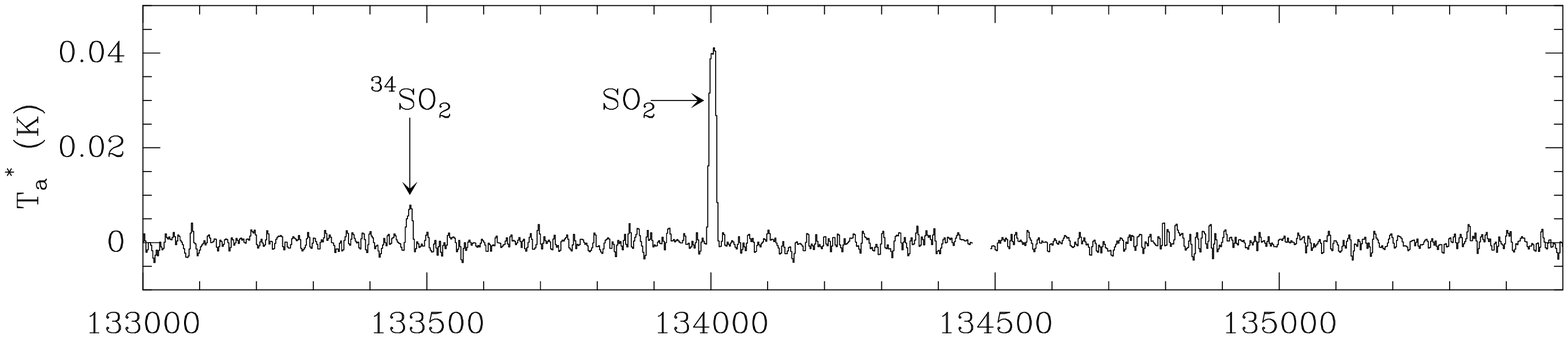}
\includegraphics[width=0.70\hsize]{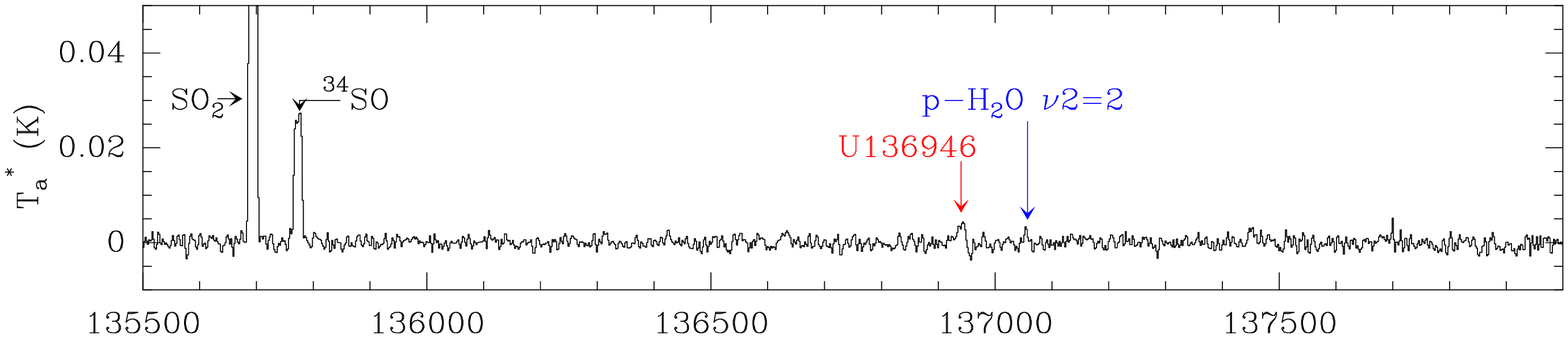}
\includegraphics[width=0.70\hsize]{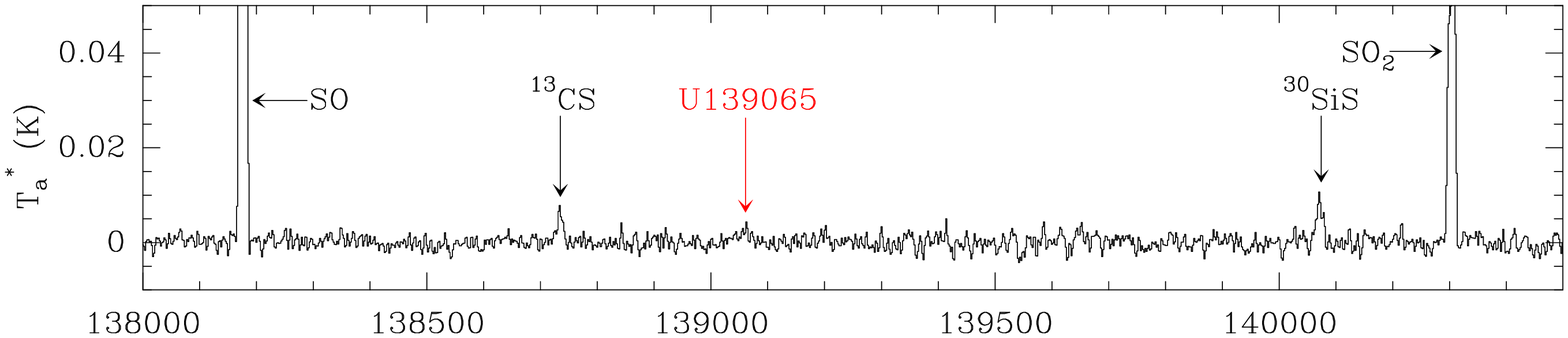}
\includegraphics[width=0.70\hsize]{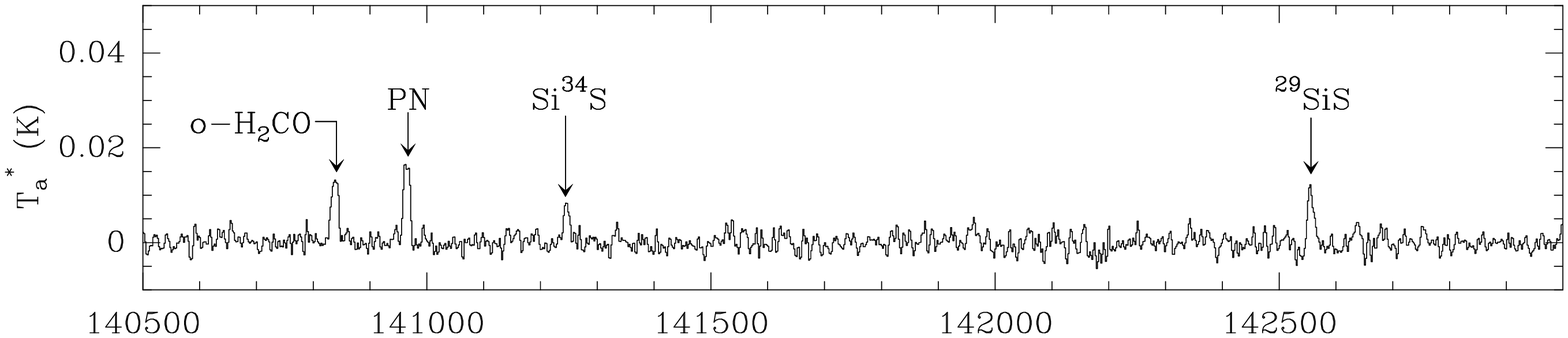}
\includegraphics[width=0.70\hsize]{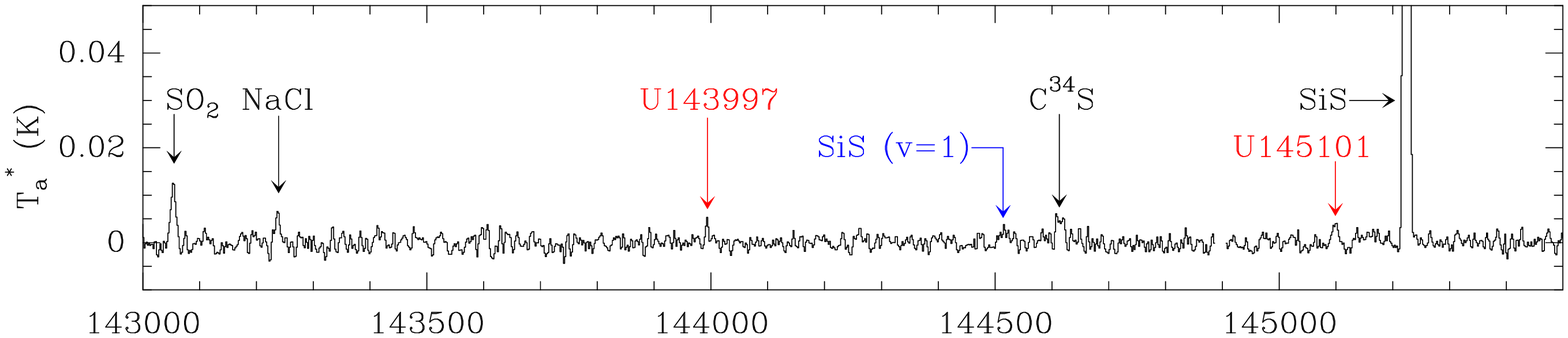}
\includegraphics[width=0.70\hsize]{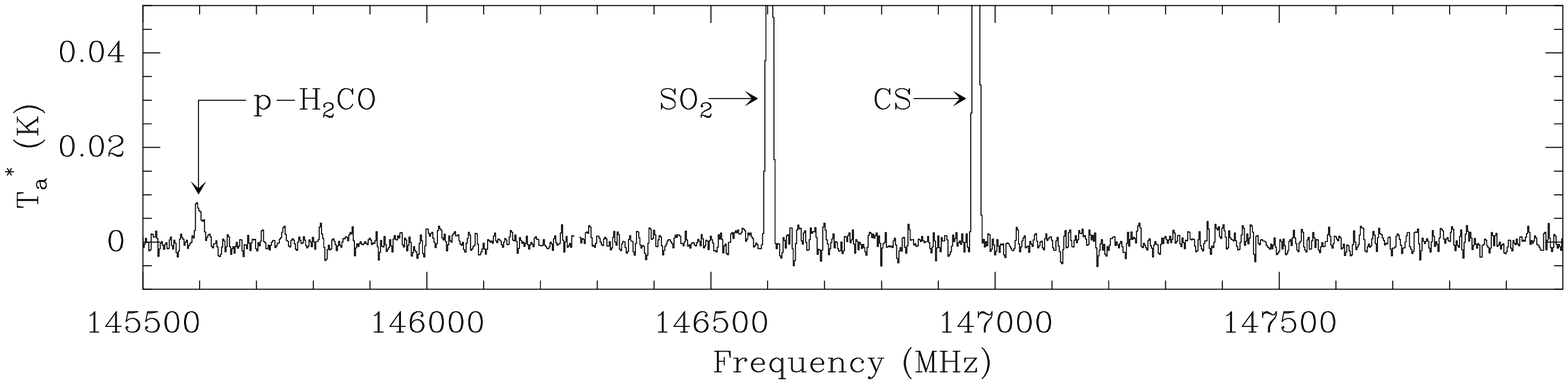}

\caption{(Continued)}
\end{figure*}

\setcounter{figure}{0}
\begin{figure*}[p] 
\centering
\includegraphics[width=0.70\hsize]{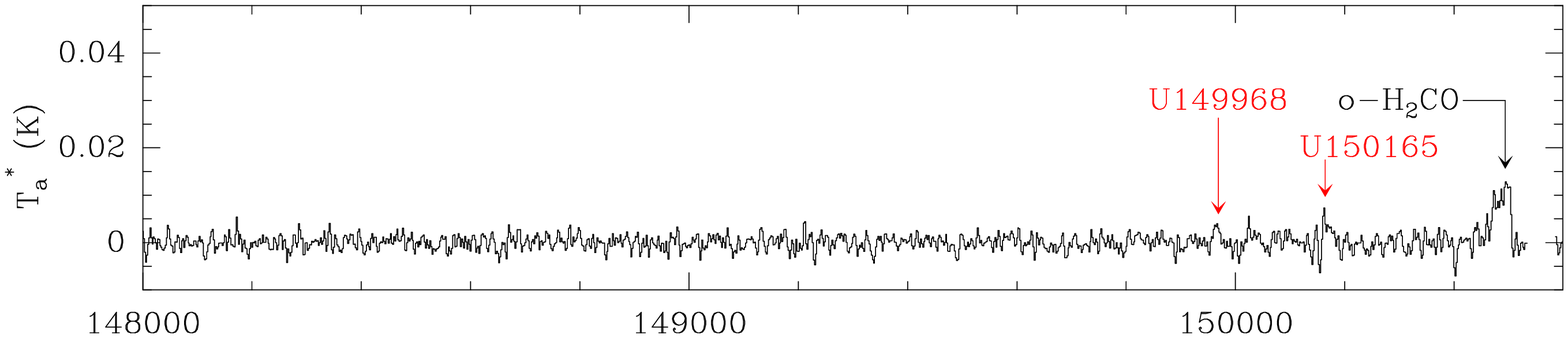}
\includegraphics[width=0.70\hsize]{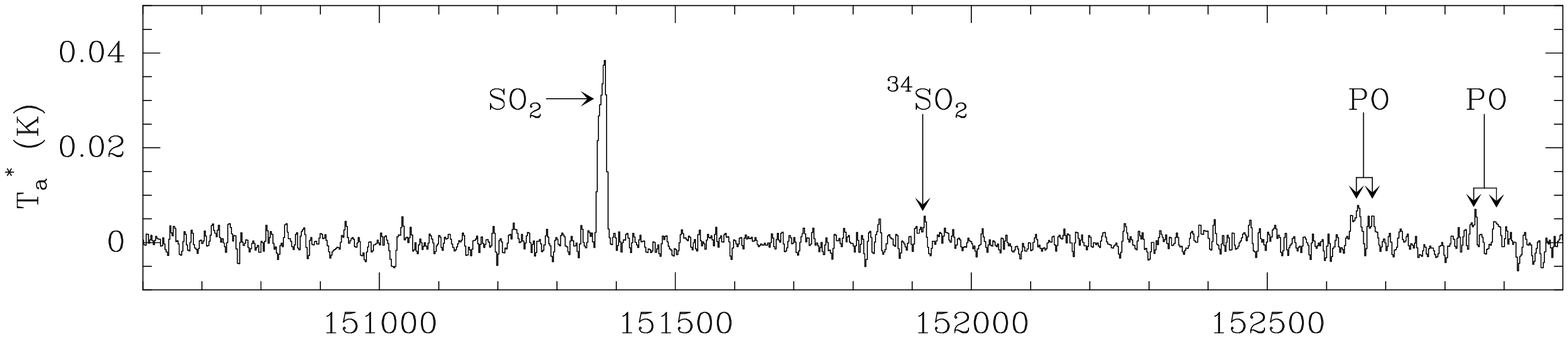}
\includegraphics[width=0.70\hsize]{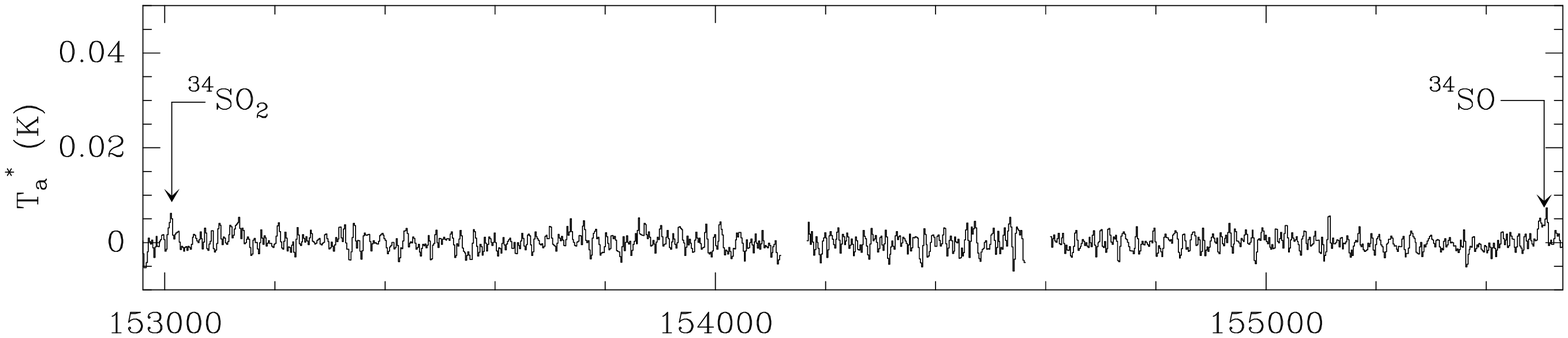}
\includegraphics[width=0.70\hsize]{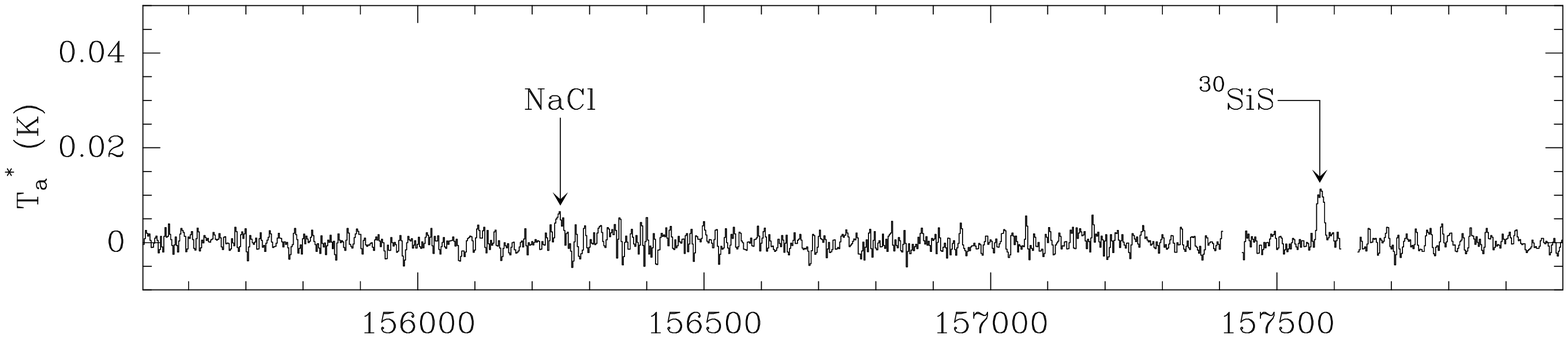}
\includegraphics[width=0.70\hsize]{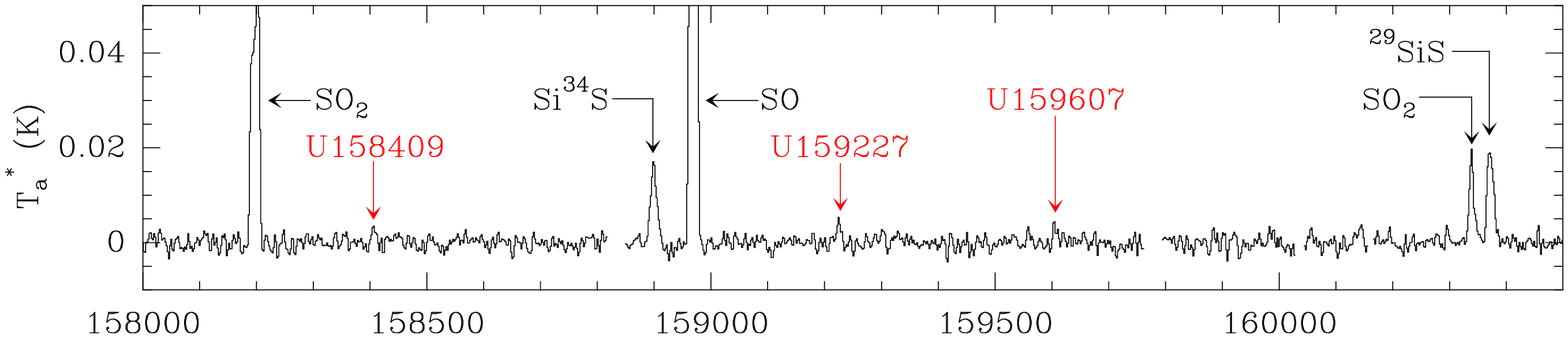}
\includegraphics[width=0.70\hsize]{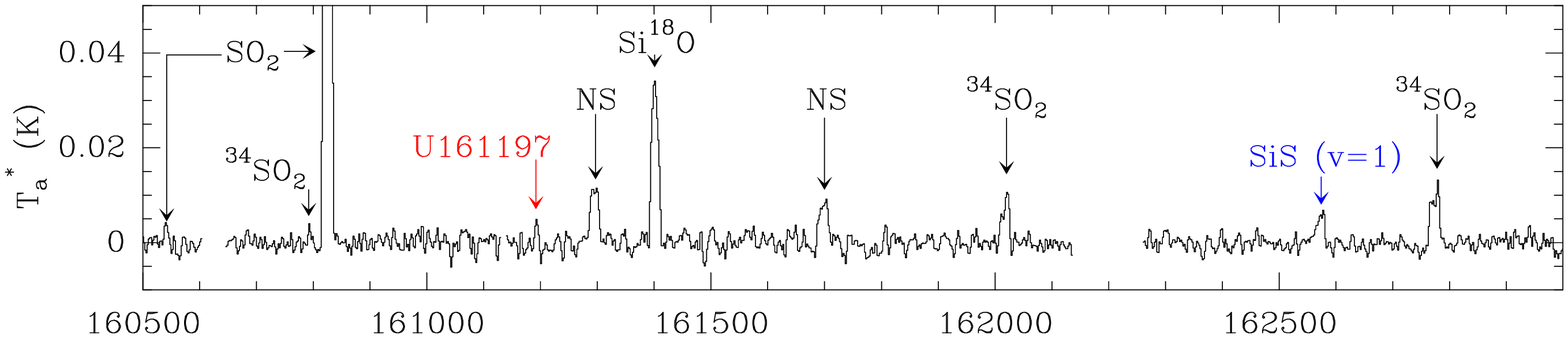}
\includegraphics[width=0.70\hsize]{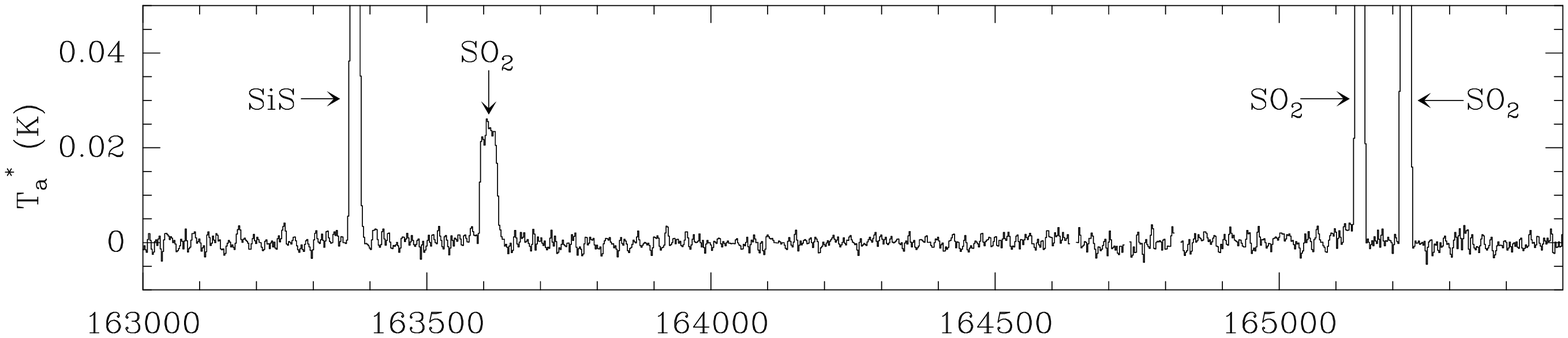}
\includegraphics[width=0.70\hsize]{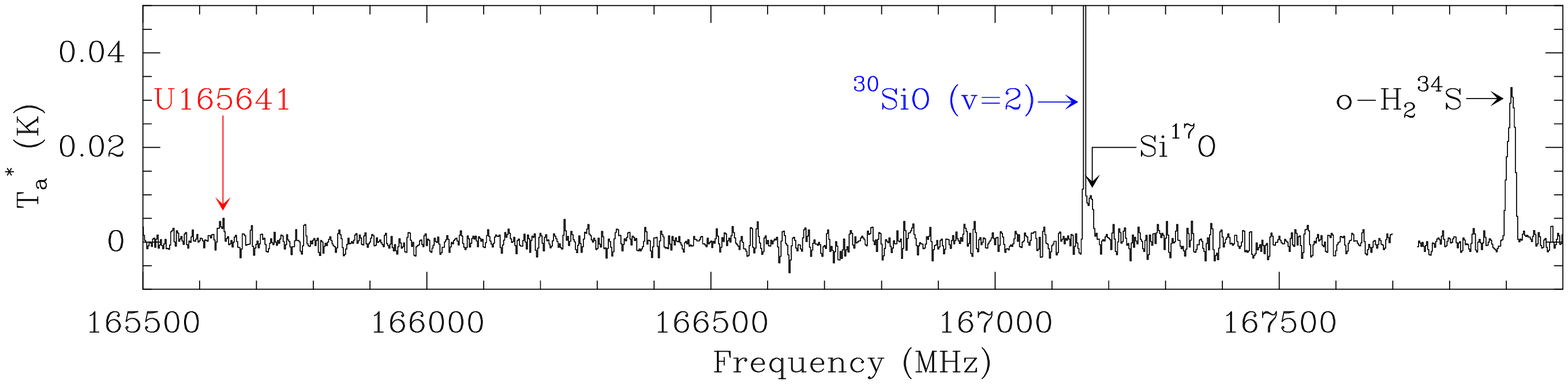}

\caption{(Continued)}
\end{figure*}

\setcounter{figure}{0}
\begin{figure*}[p] 
\centering
\includegraphics[width=0.70\hsize]{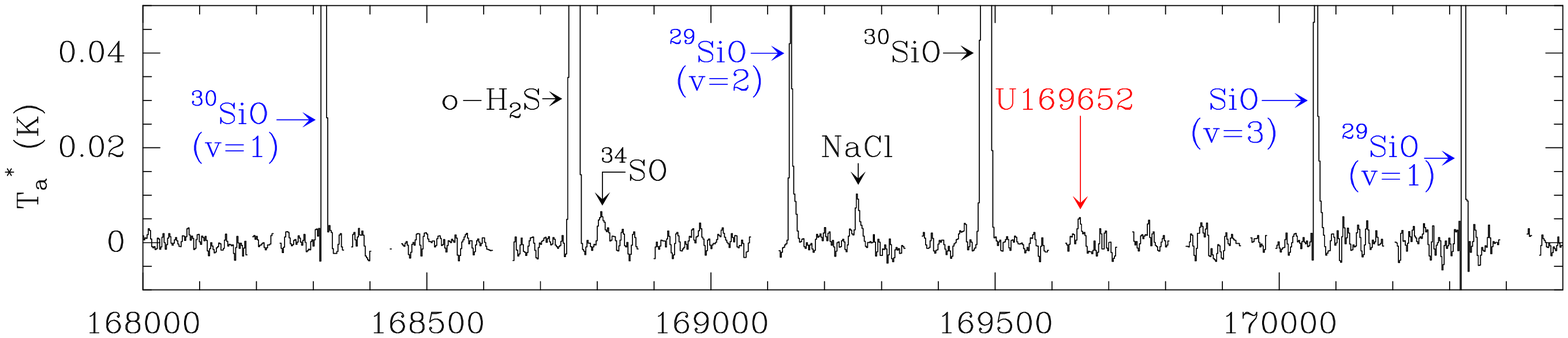}
\includegraphics[width=0.70\hsize]{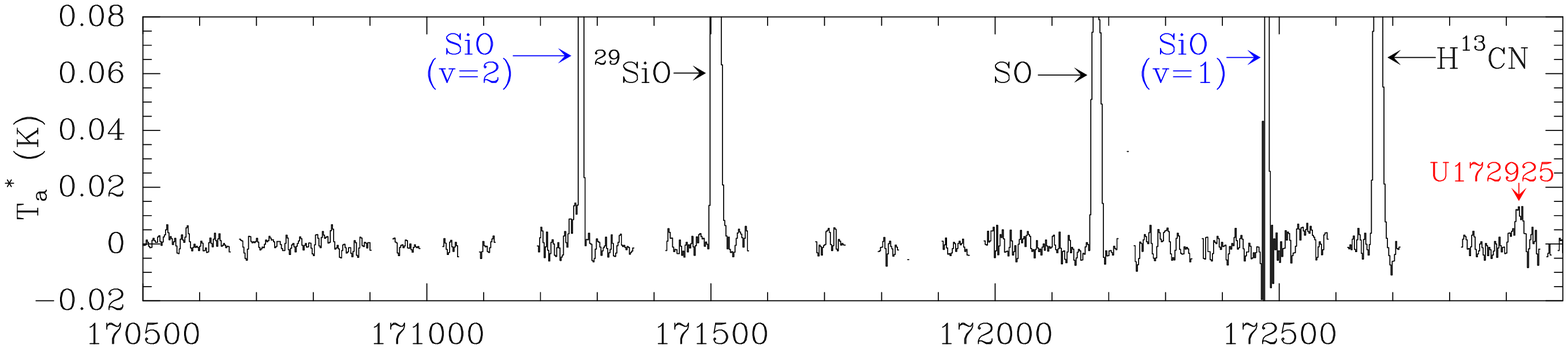}
\includegraphics[width=0.70\hsize]{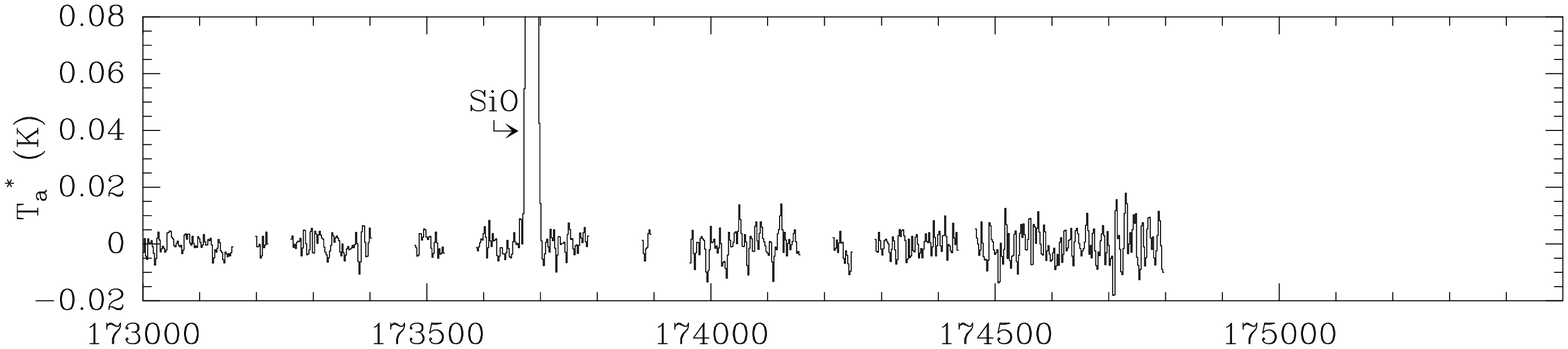}
\includegraphics[width=0.70\hsize]{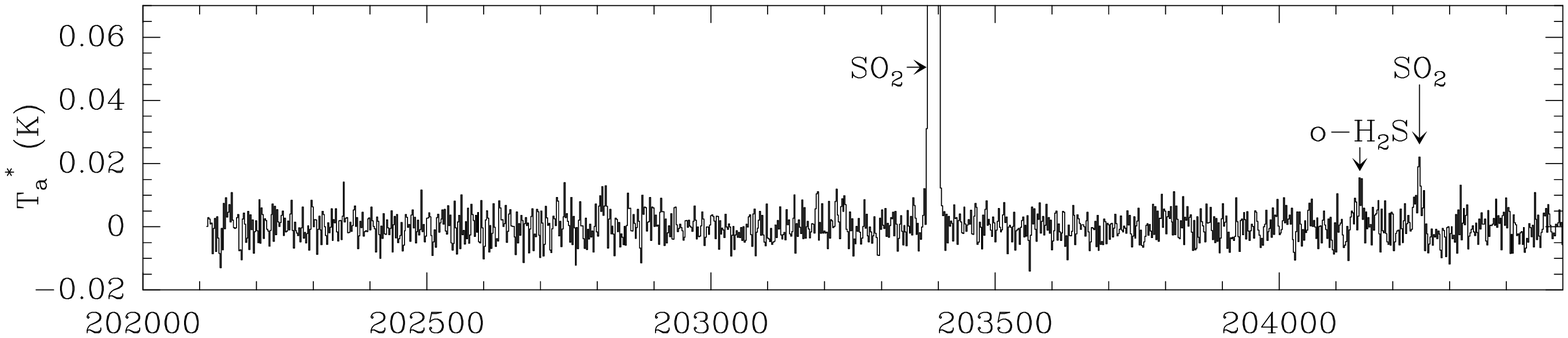}
\includegraphics[width=0.70\hsize]{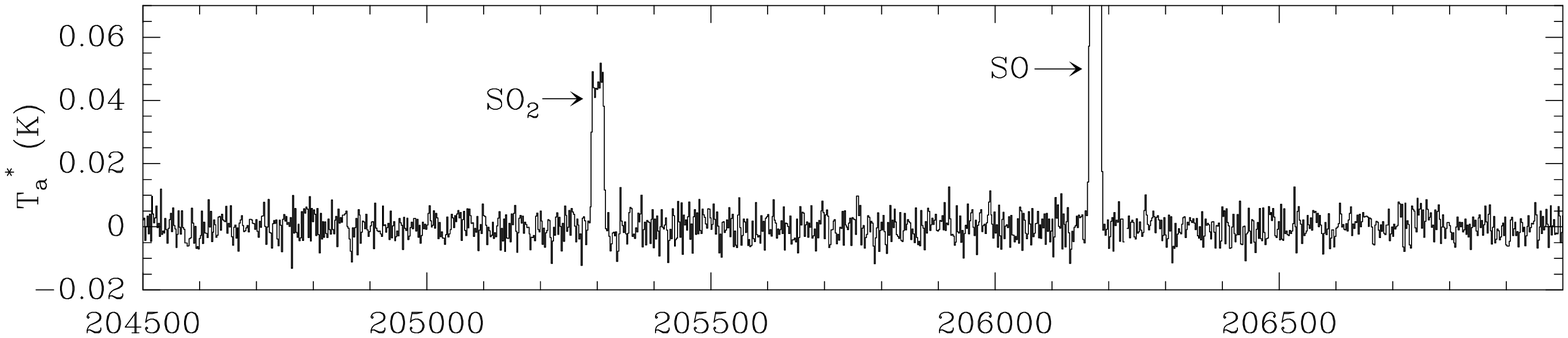}
\includegraphics[width=0.70\hsize]{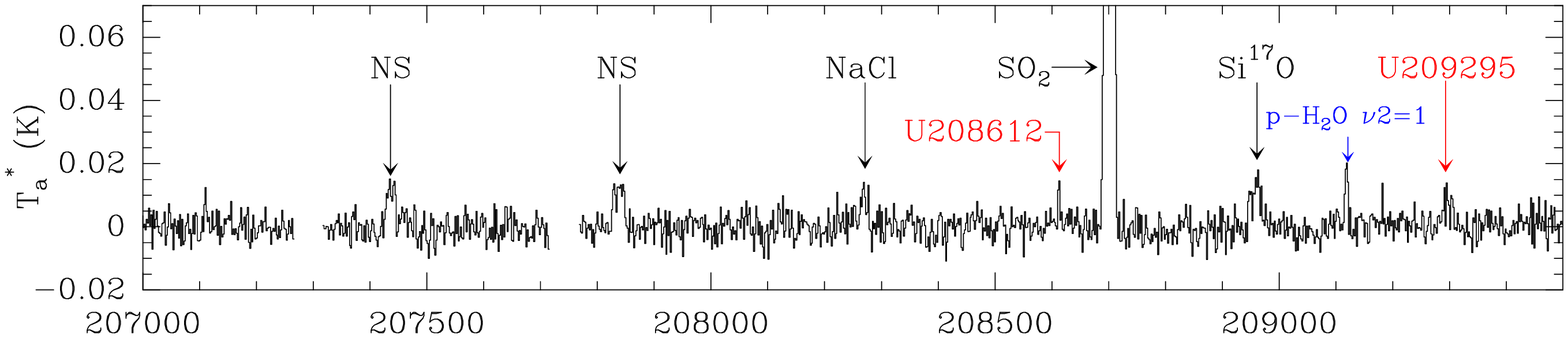}
\includegraphics[width=0.70\hsize]{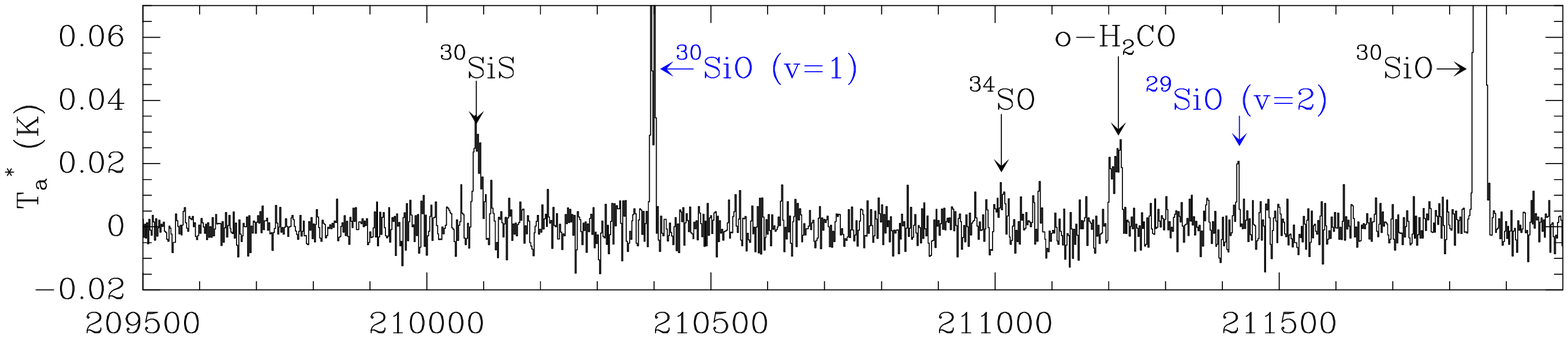}
\includegraphics[width=0.70\hsize]{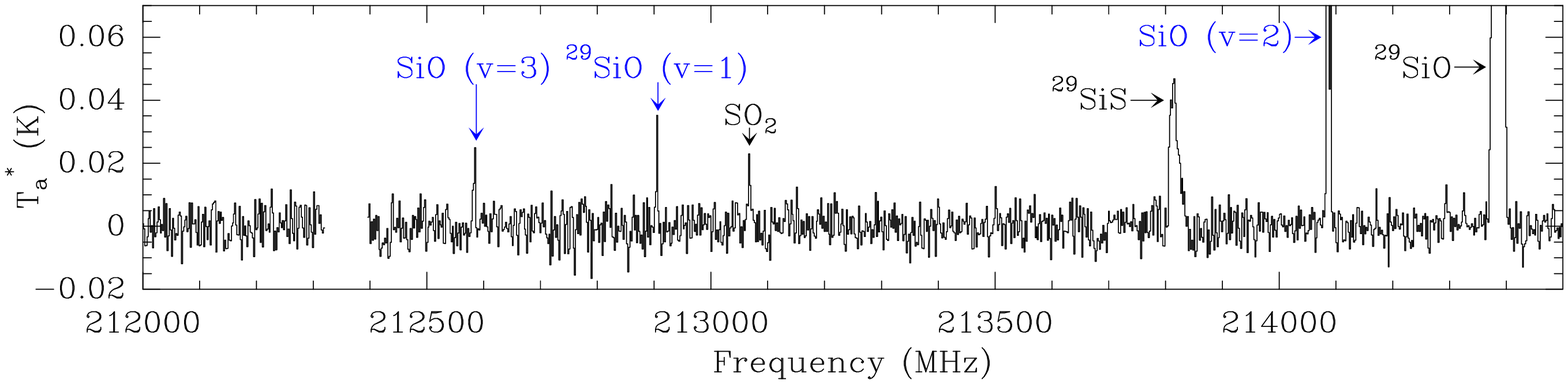}
\caption{(Continued)}
\end{figure*}

\setcounter{figure}{0}
\begin{figure*}[p] 
\centering
\includegraphics[width=0.70\hsize]{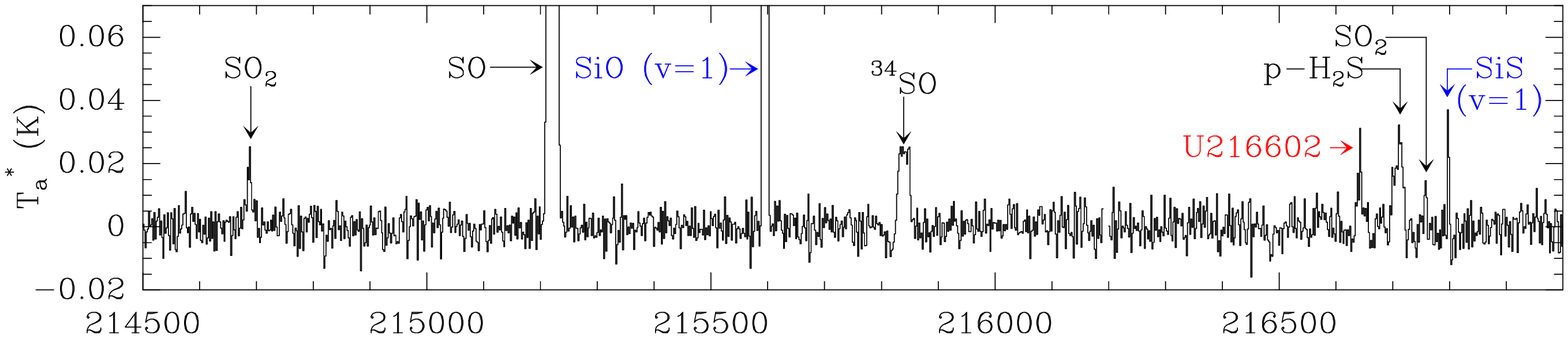}
\includegraphics[width=0.70\hsize]{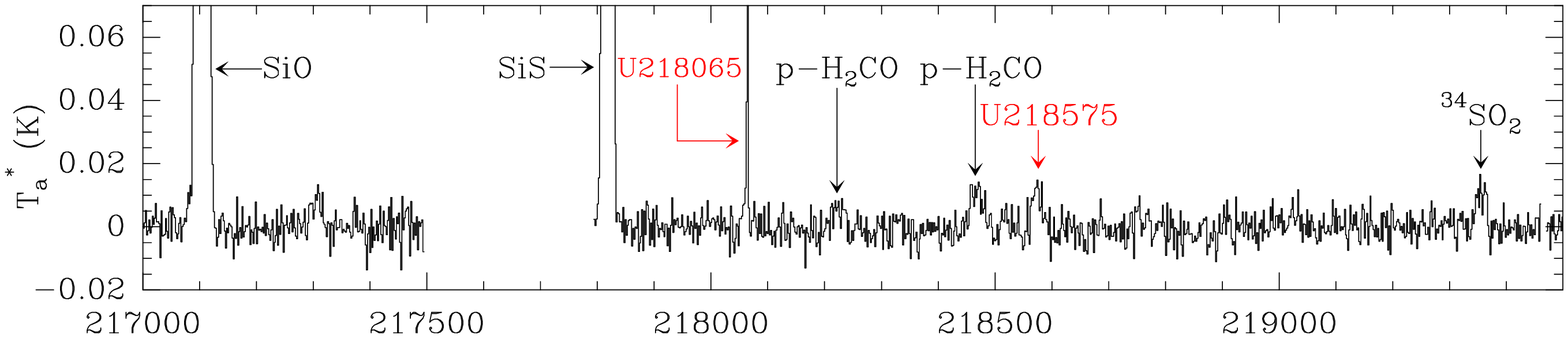}
\includegraphics[width=0.70\hsize]{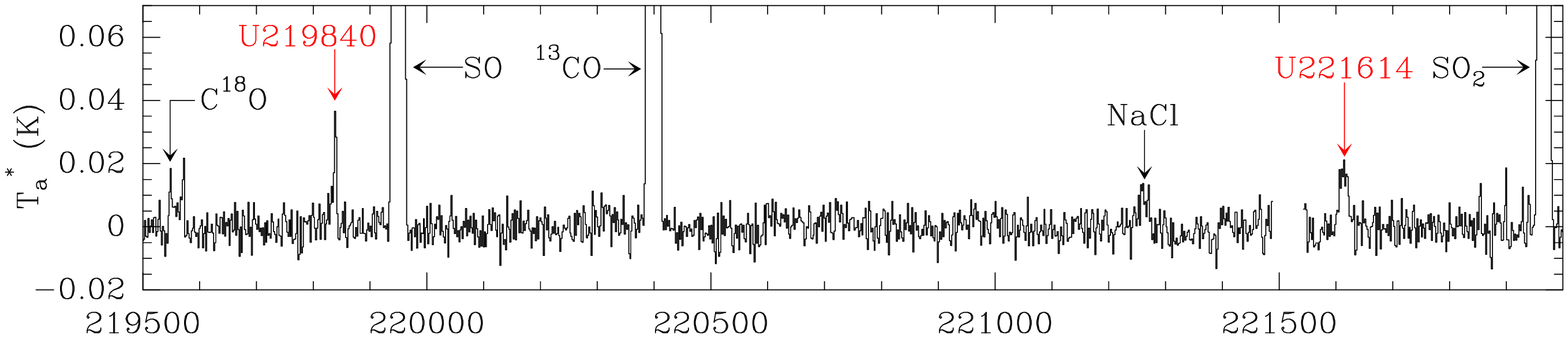}
\includegraphics[width=0.70\hsize]{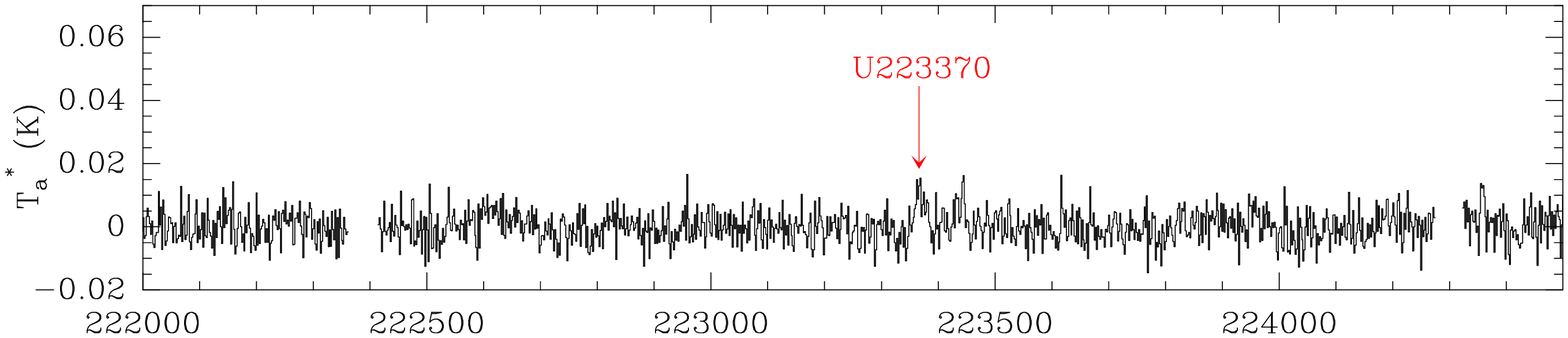}
\includegraphics[width=0.70\hsize]{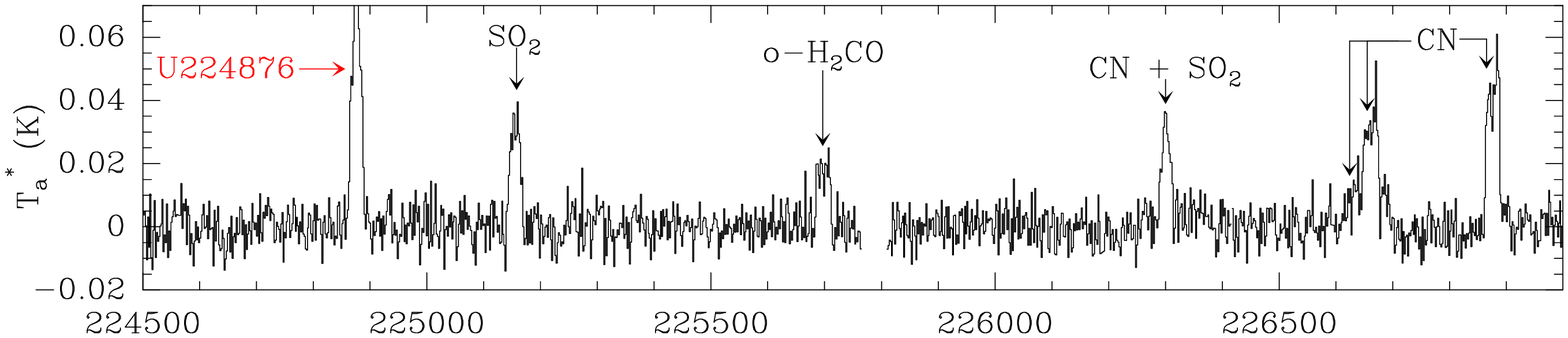}
\includegraphics[width=0.70\hsize]{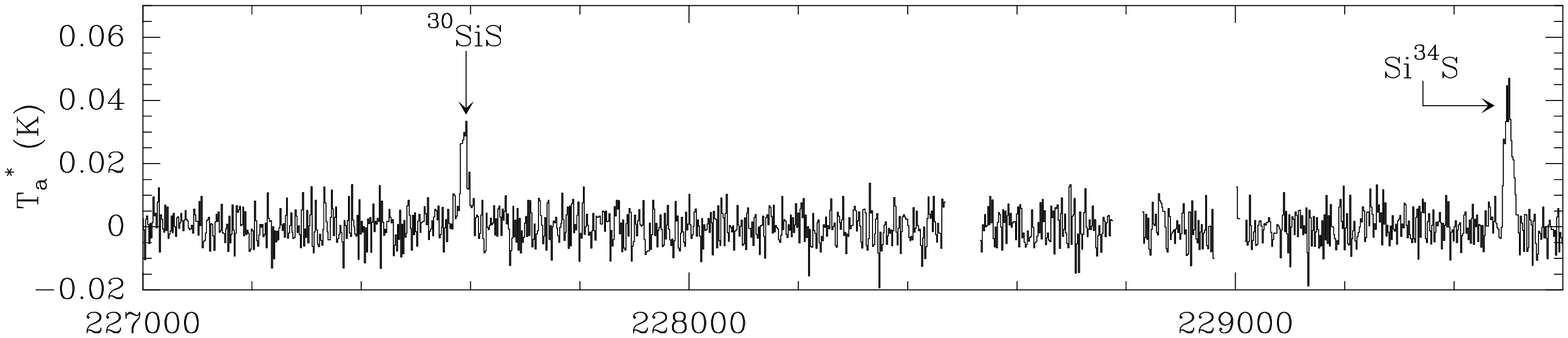}
\includegraphics[width=0.70\hsize]{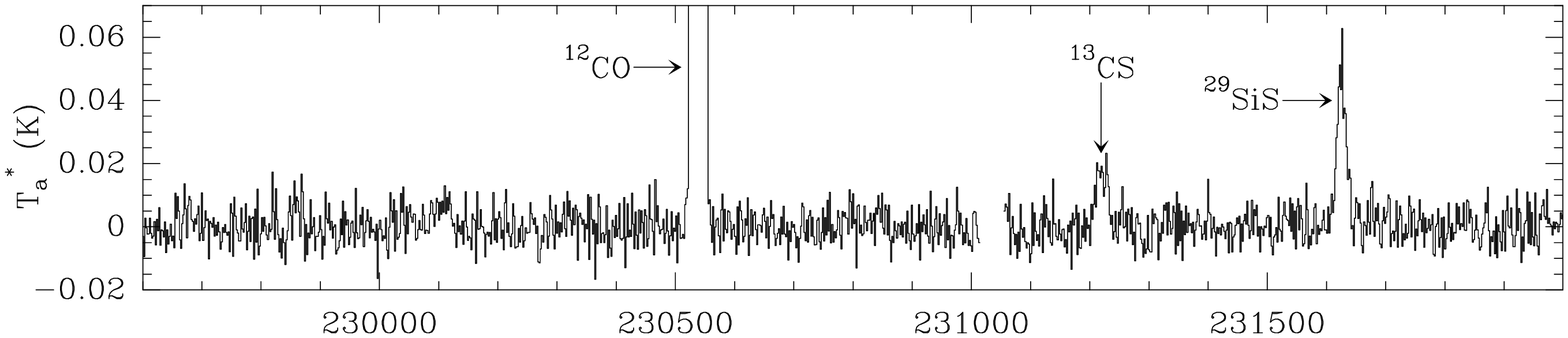}
\includegraphics[width=0.70\hsize]{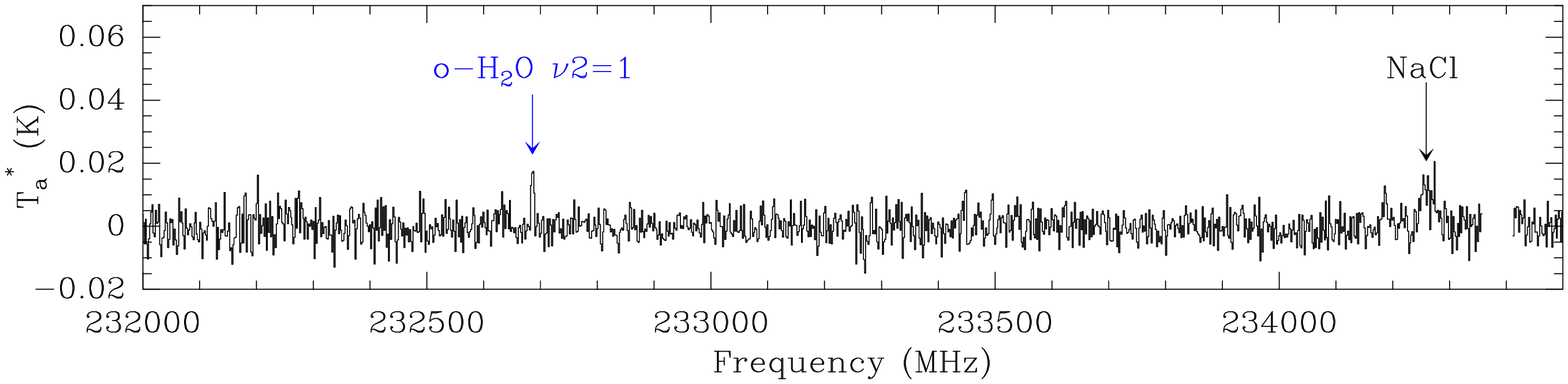}
\caption{(Continued)}
\end{figure*}

\setcounter{figure}{0}
\begin{figure*}[p] 
\centering
\includegraphics[width=0.70\hsize]{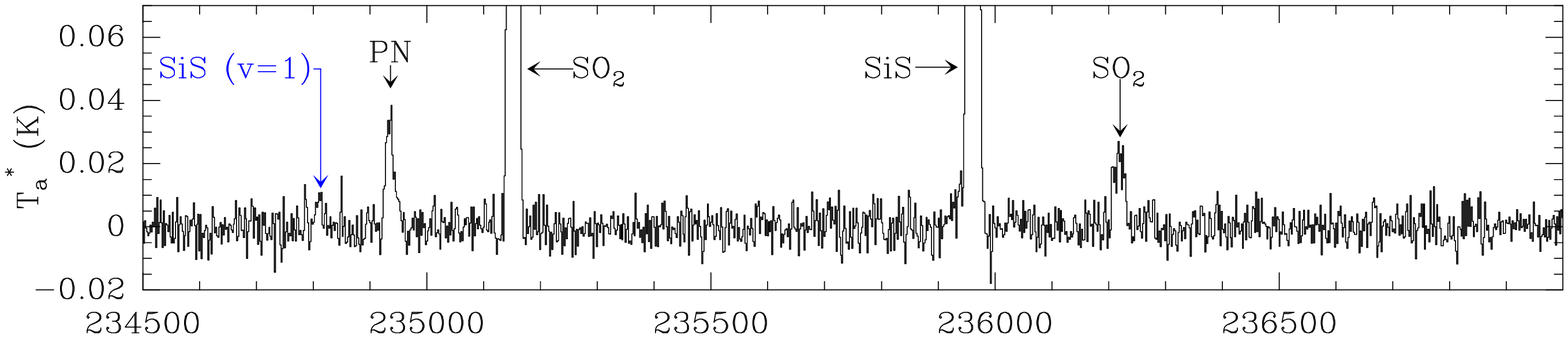}
\includegraphics[width=0.70\hsize]{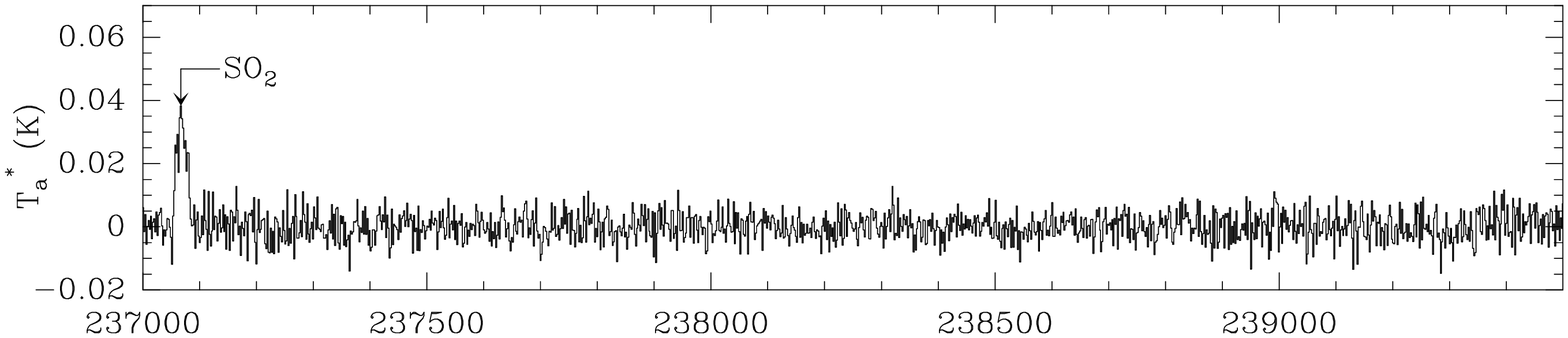}
\includegraphics[width=0.70\hsize]{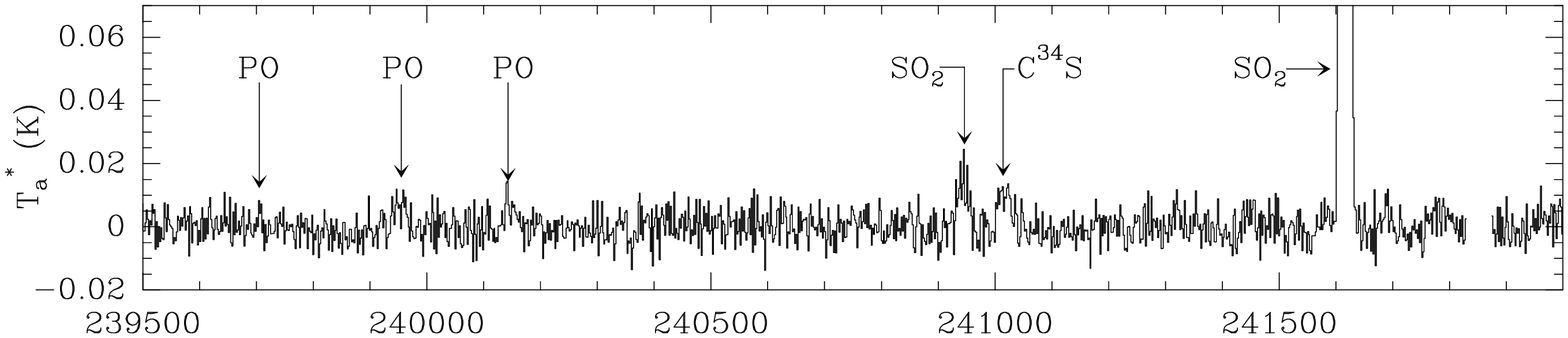}
\includegraphics[width=0.70\hsize]{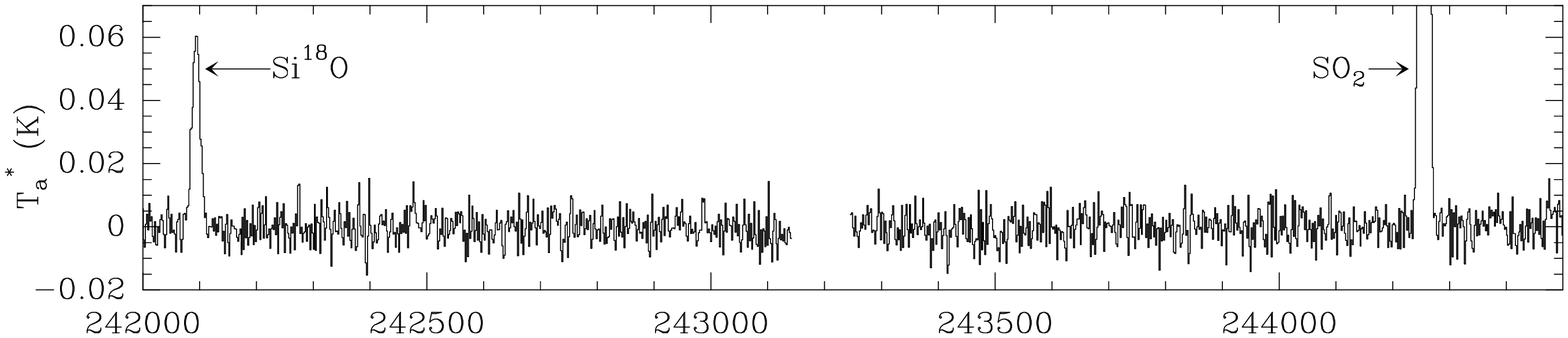}
\includegraphics[width=0.70\hsize]{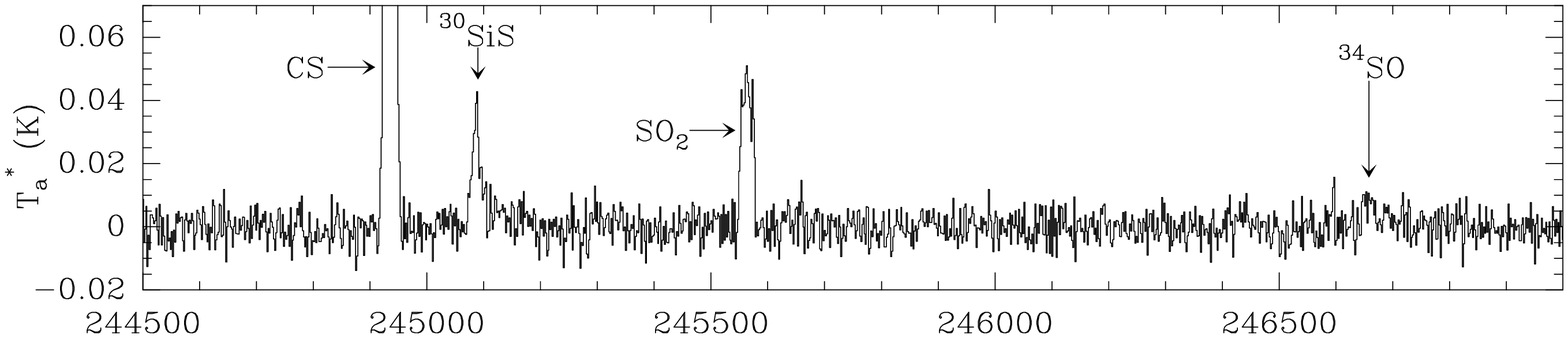}
\includegraphics[width=0.70\hsize]{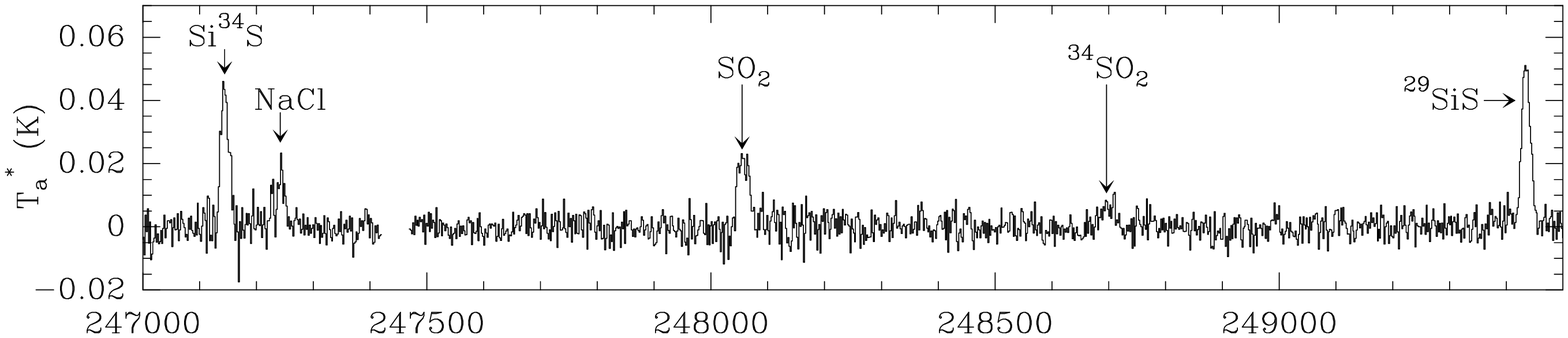}
\includegraphics[width=0.70\hsize]{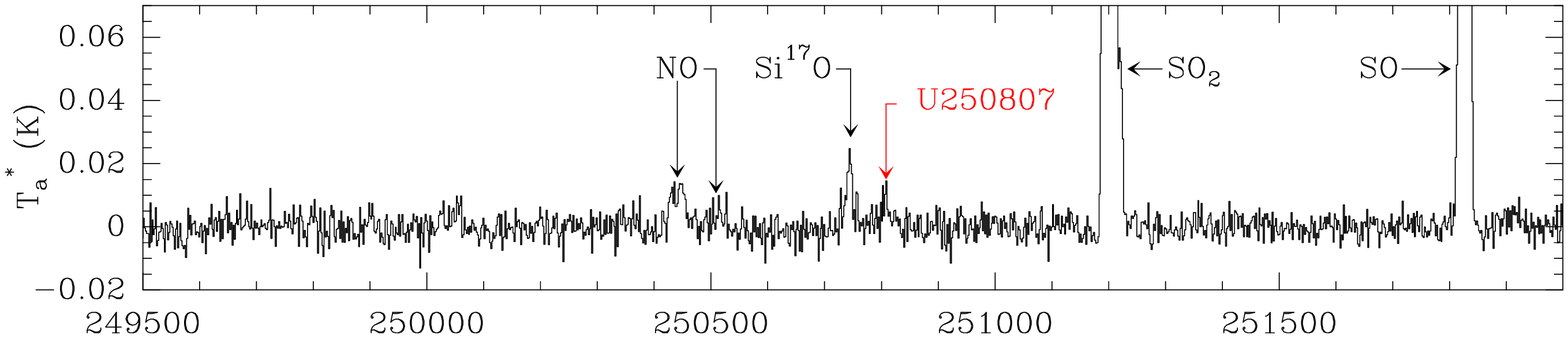}
\includegraphics[width=0.70\hsize]{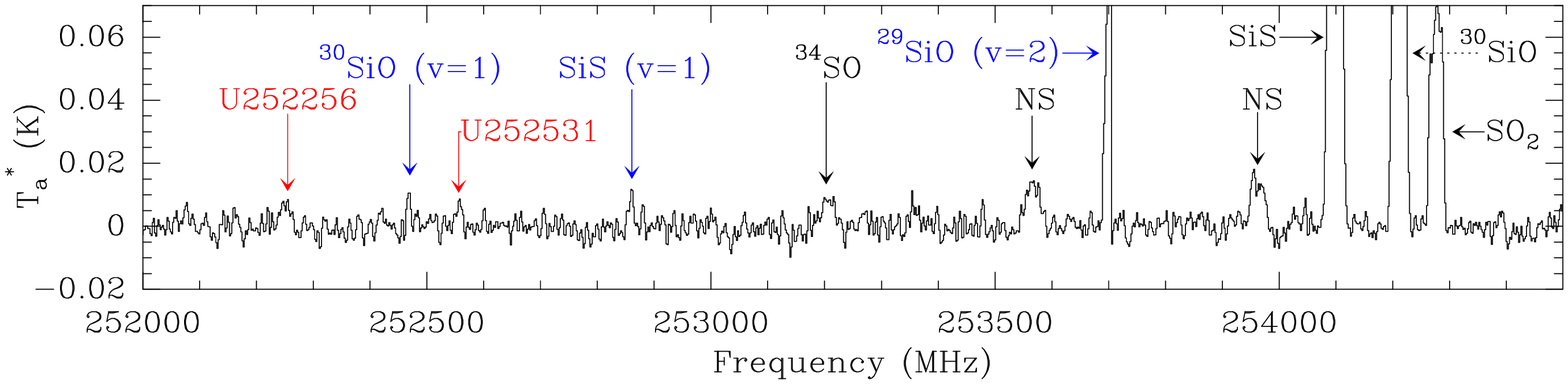}
\caption{(Continued)}
\end{figure*}

\setcounter{figure}{0}
\begin{figure*}[p] 
\centering
\includegraphics[width=0.70\hsize]{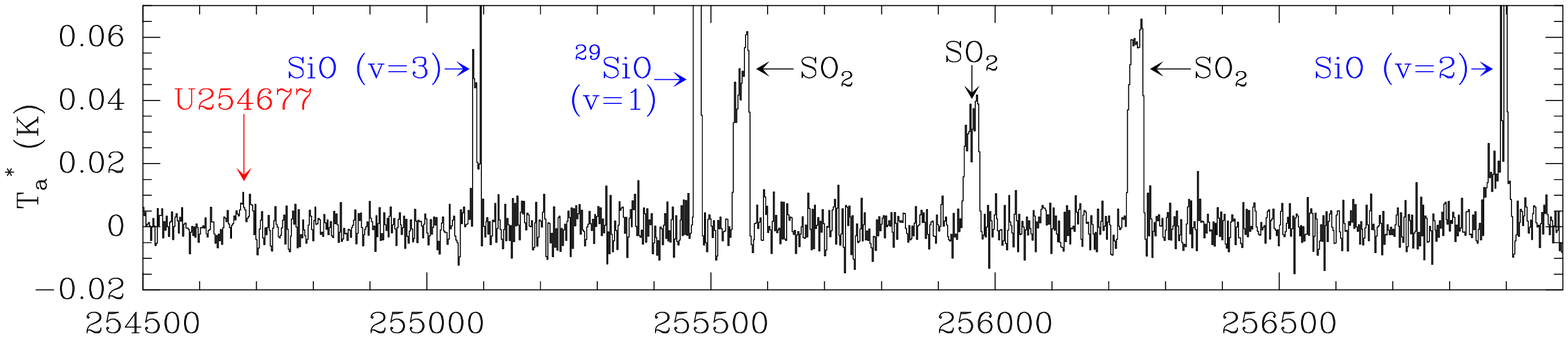}
\includegraphics[width=0.70\hsize]{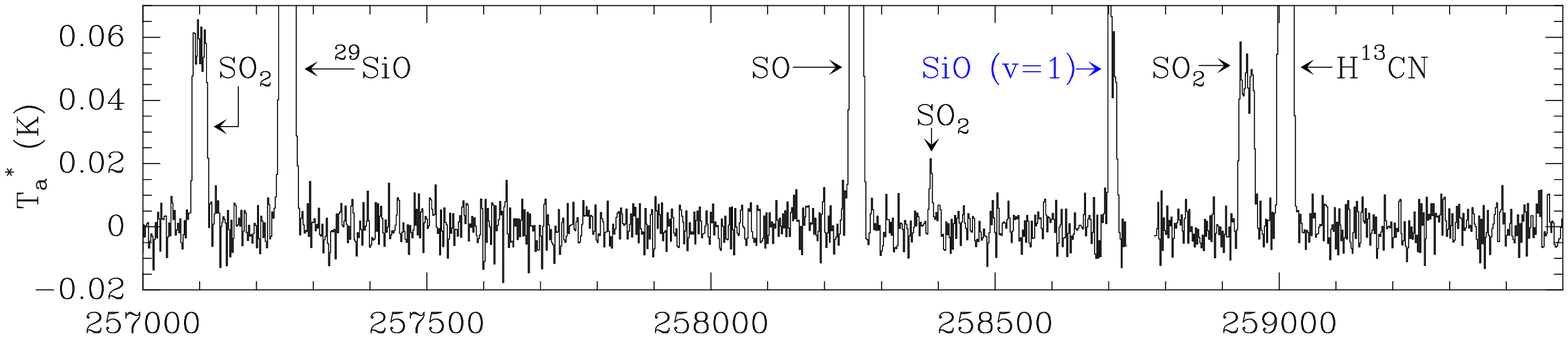}
\includegraphics[width=0.70\hsize]{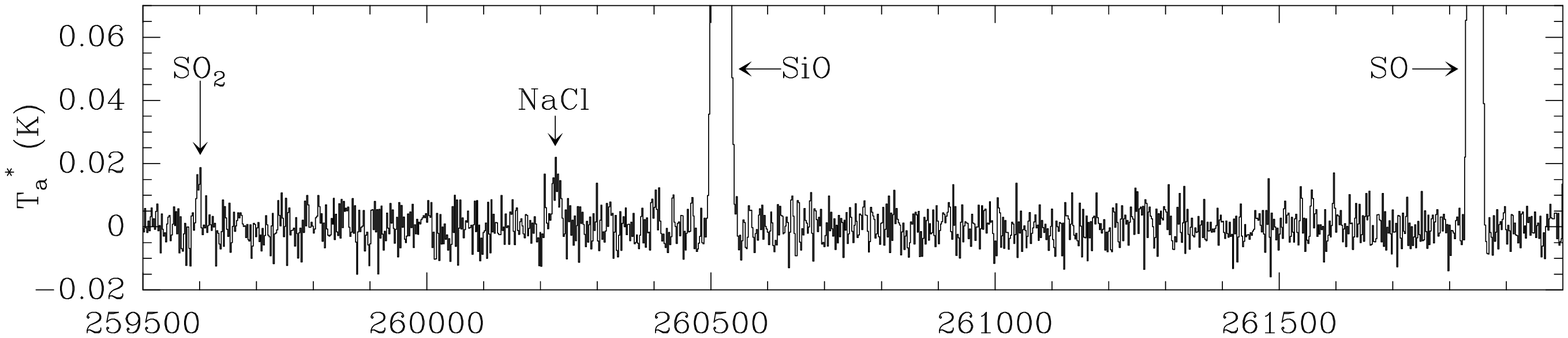}
\includegraphics[width=0.70\hsize]{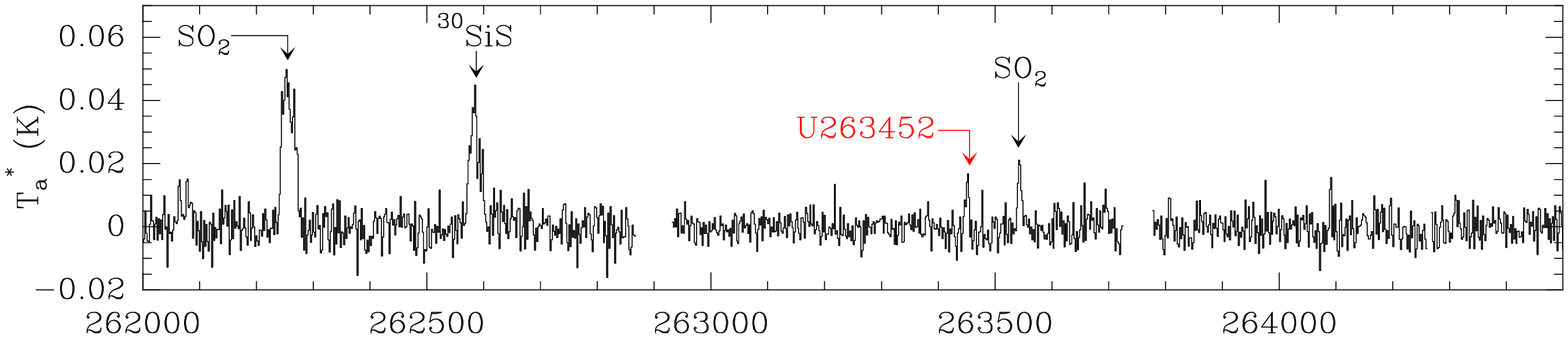}
\includegraphics[width=0.70\hsize]{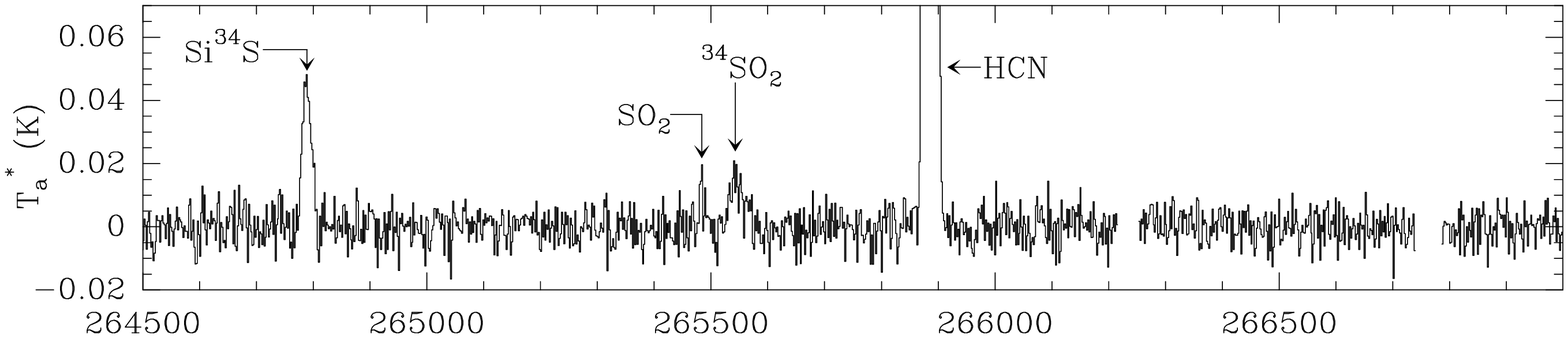}
\includegraphics[width=0.70\hsize]{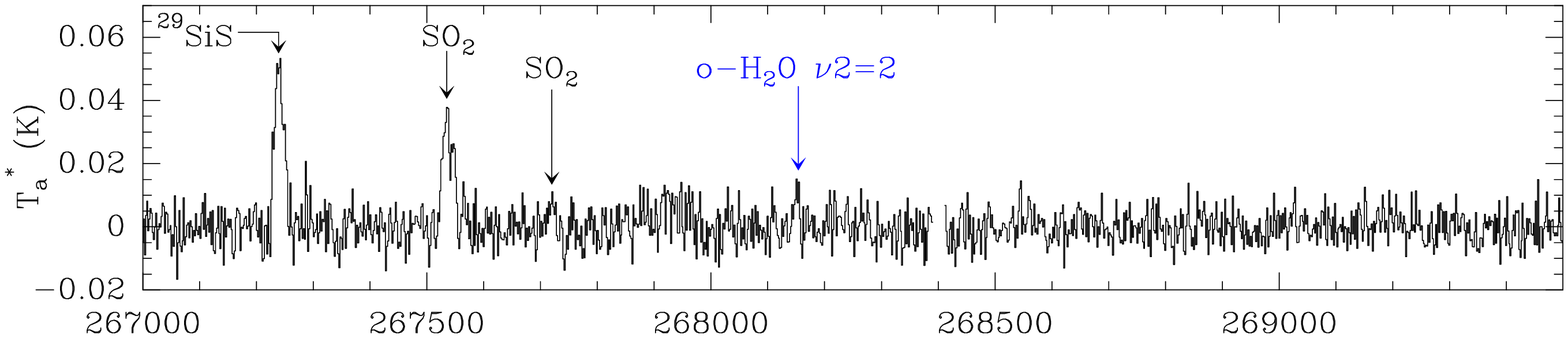}
\includegraphics[width=0.70\hsize]{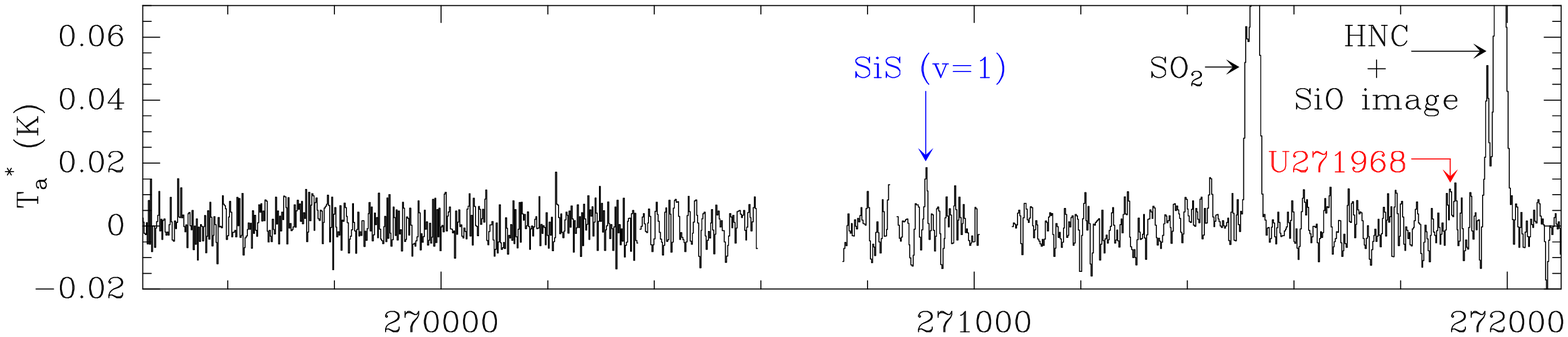}
\includegraphics[width=0.70\hsize]{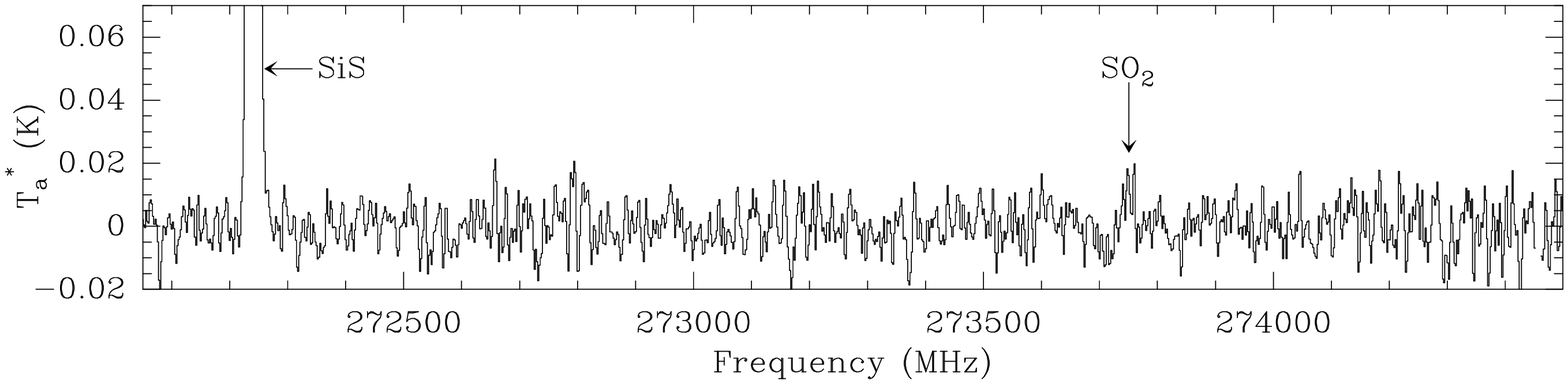}
\caption{(Continued)}
\end{figure*}

\setcounter{figure}{0}
\begin{figure*}[p] 
\centering
\includegraphics[width=0.70\hsize]{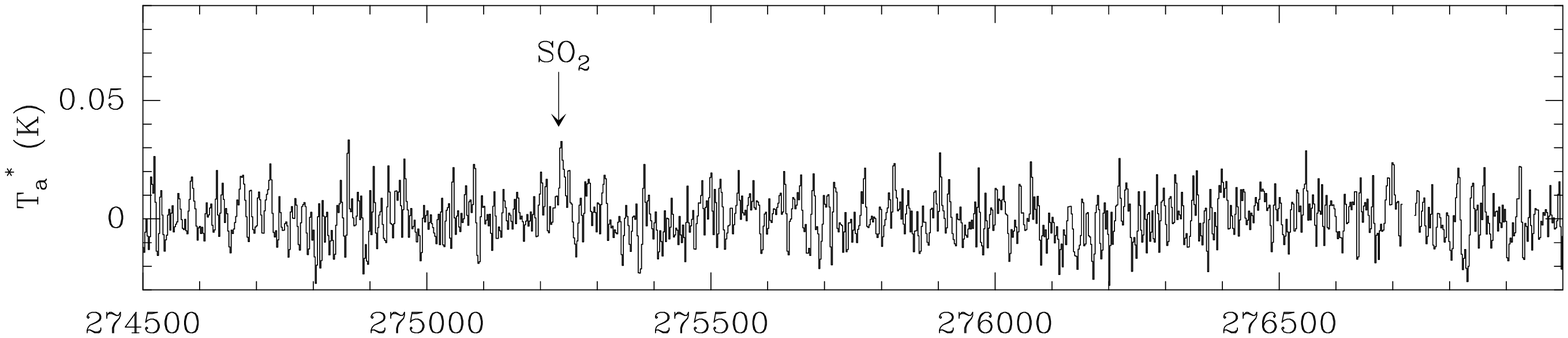}
\includegraphics[width=0.70\hsize]{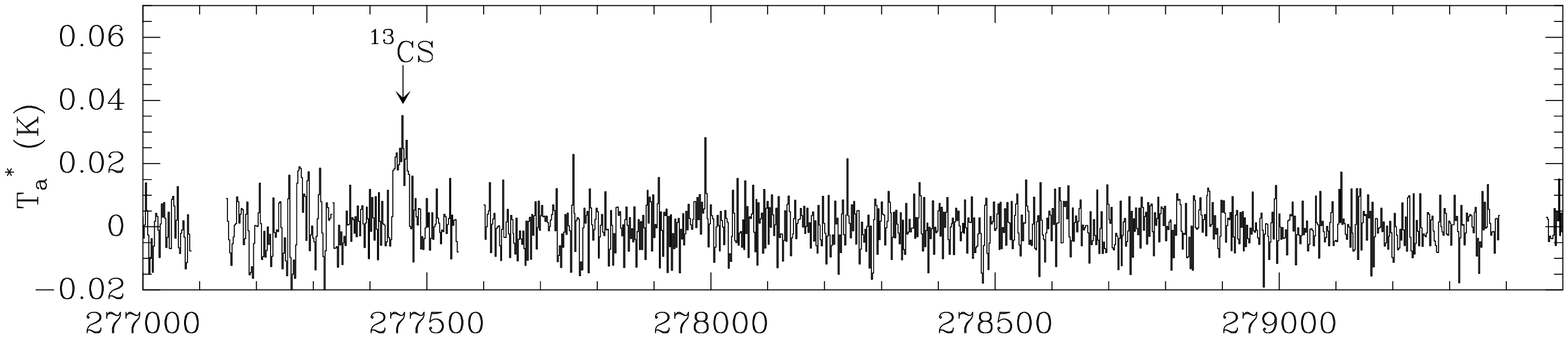}
\includegraphics[width=0.70\hsize]{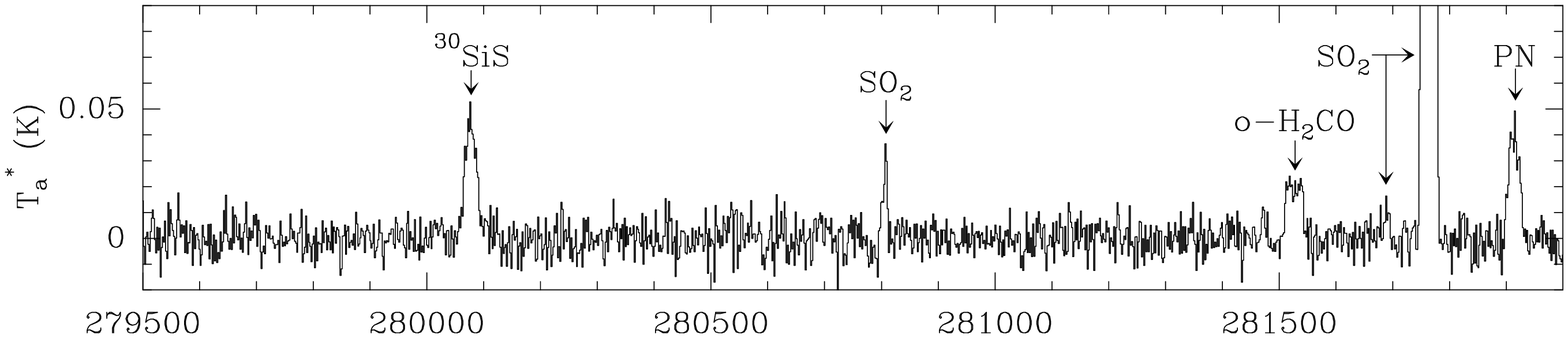}
\includegraphics[width=0.70\hsize]{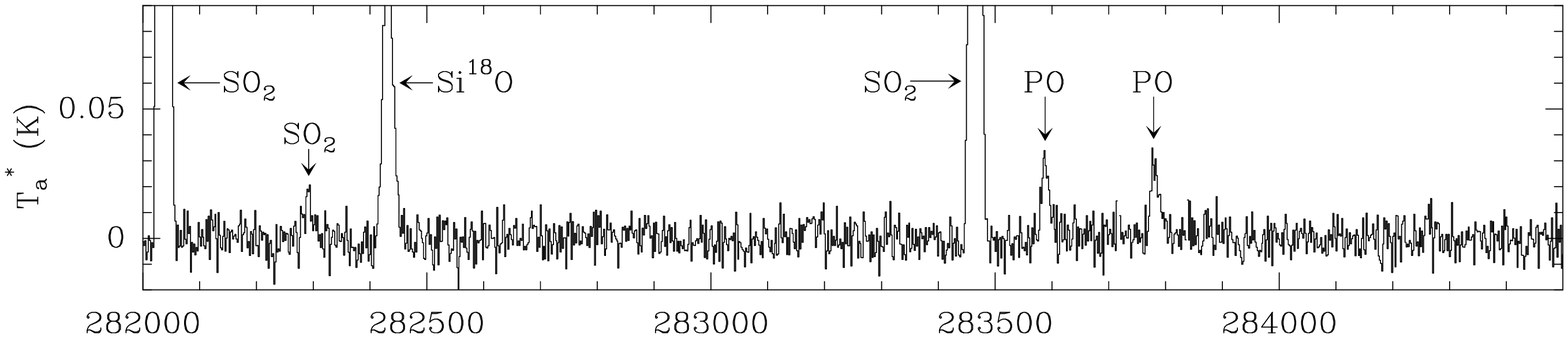}
\includegraphics[width=0.70\hsize]{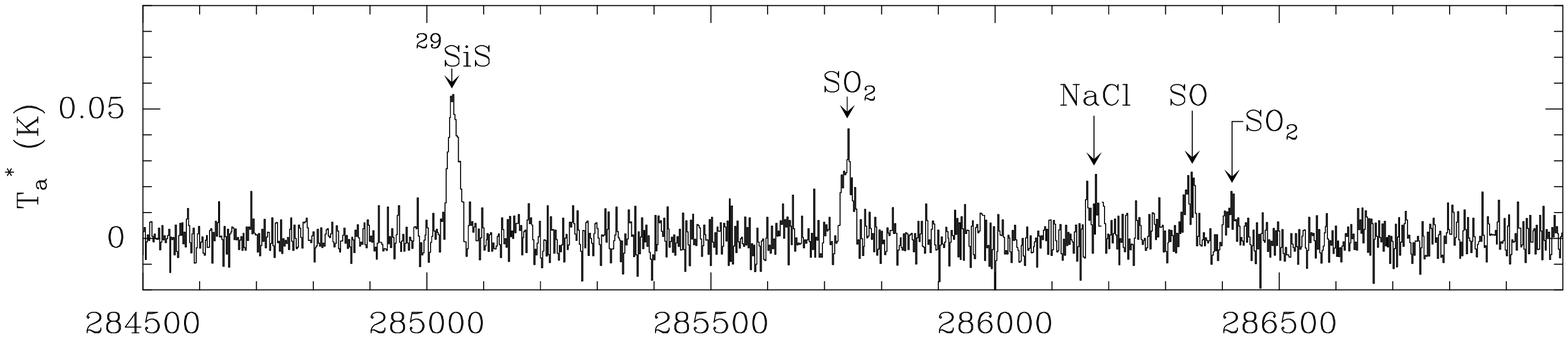}
\includegraphics[width=0.70\hsize]{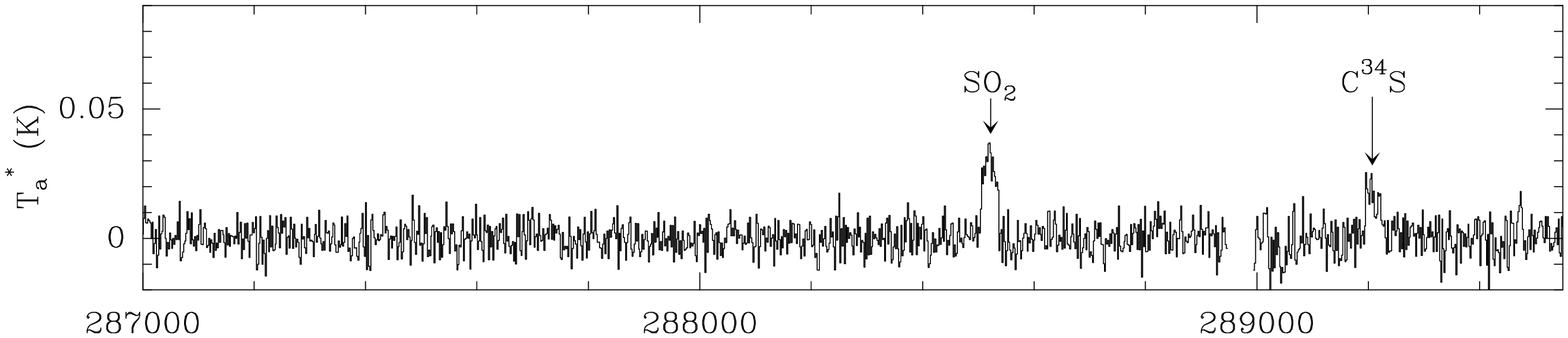}
\includegraphics[width=0.70\hsize]{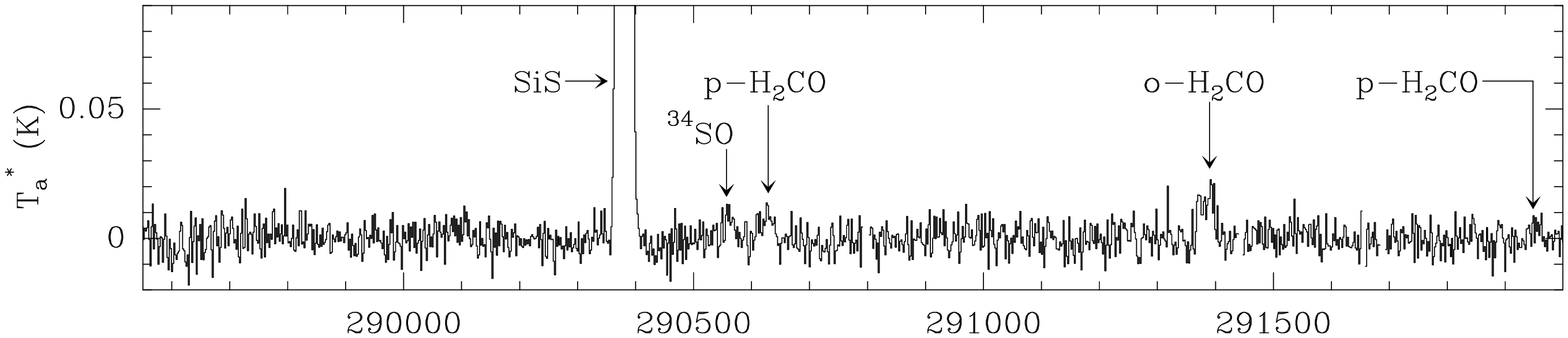}
\includegraphics[width=0.70\hsize]{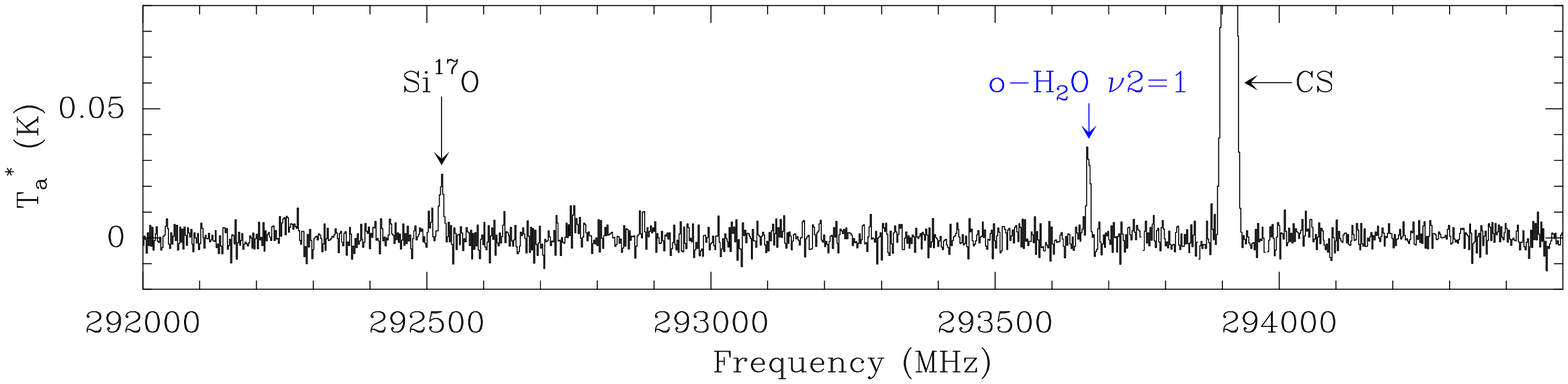}
\caption{(Continued)}
\end{figure*}

\setcounter{figure}{0}
\begin{figure*}[p] 
\centering
\includegraphics[width=0.70\hsize]{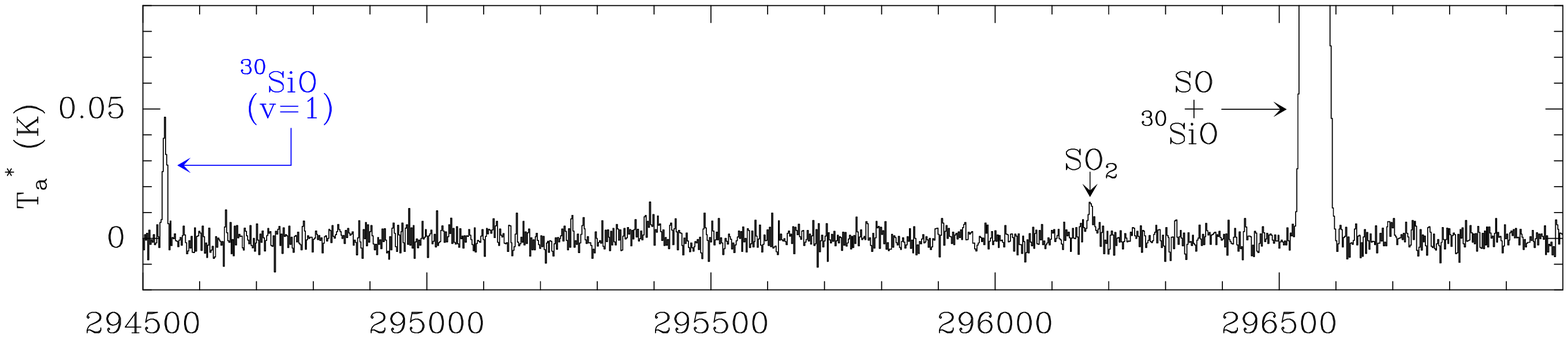}
\includegraphics[width=0.70\hsize]{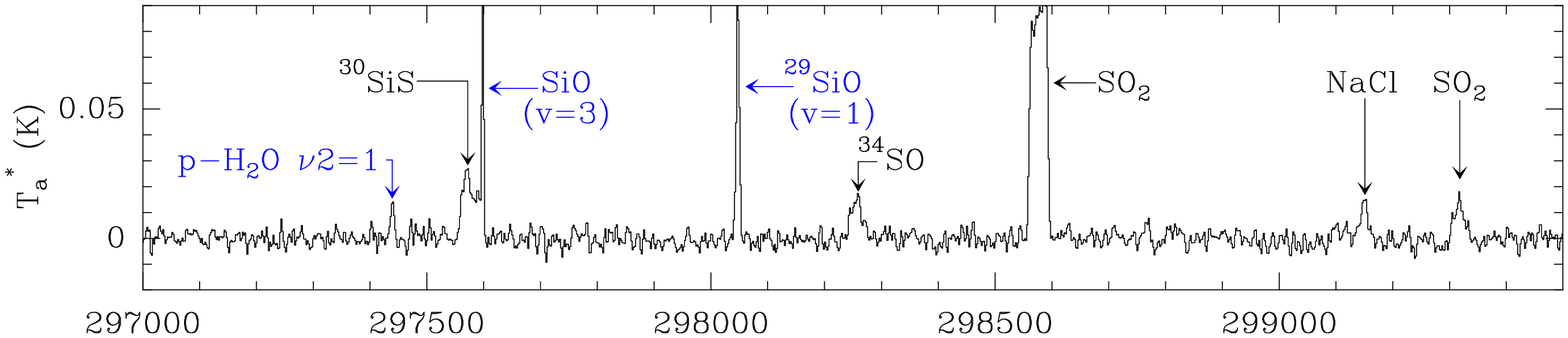}
\includegraphics[width=0.70\hsize]{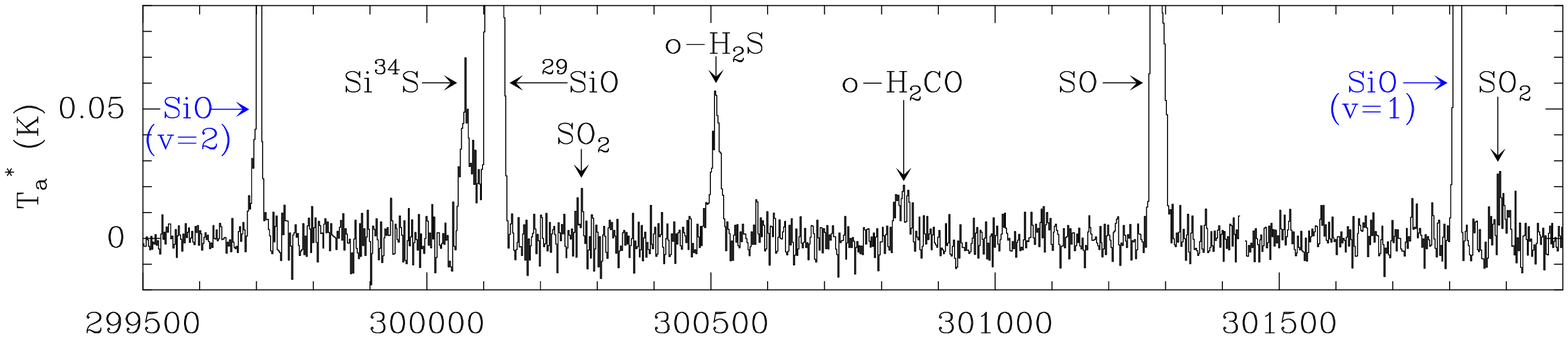}
\includegraphics[width=0.70\hsize]{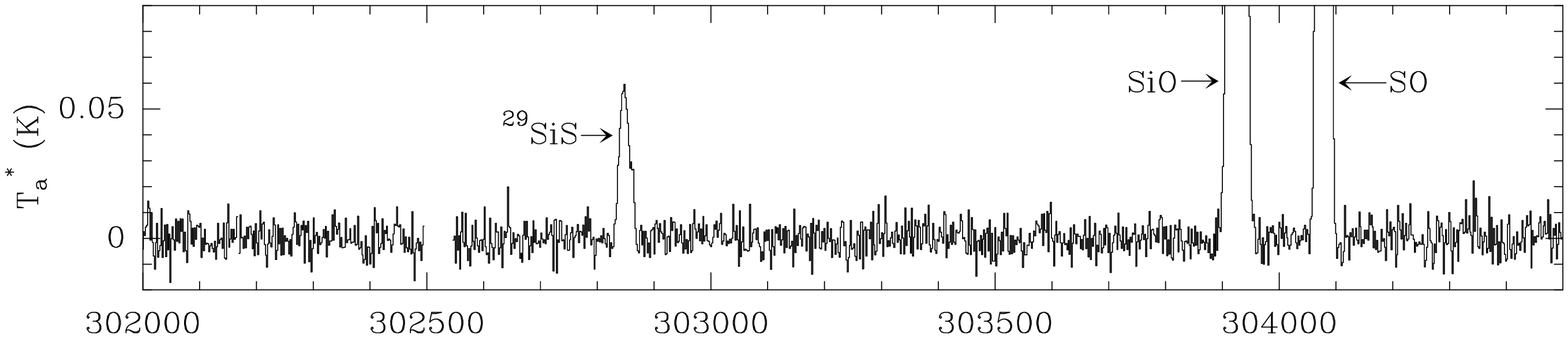}
\includegraphics[width=0.70\hsize]{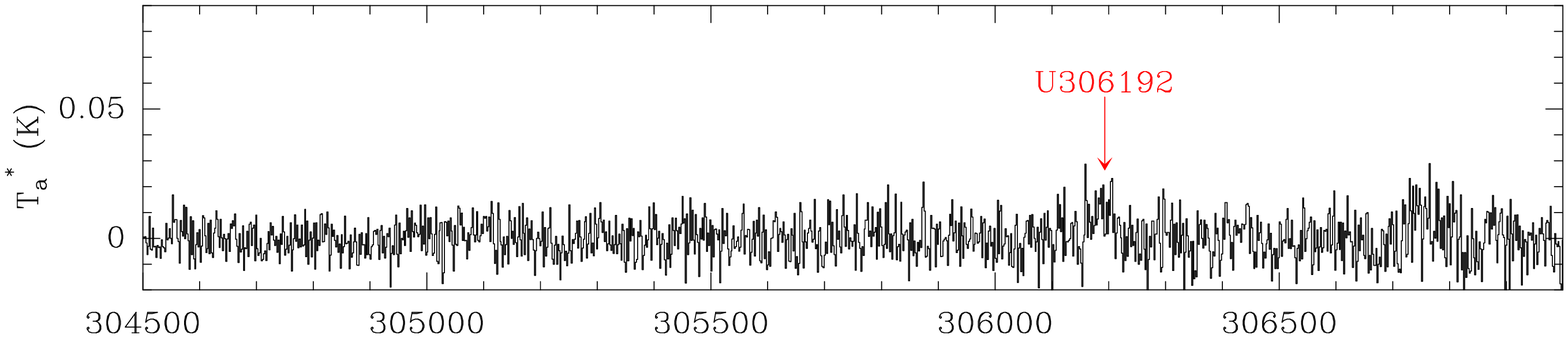}
\includegraphics[width=0.70\hsize]{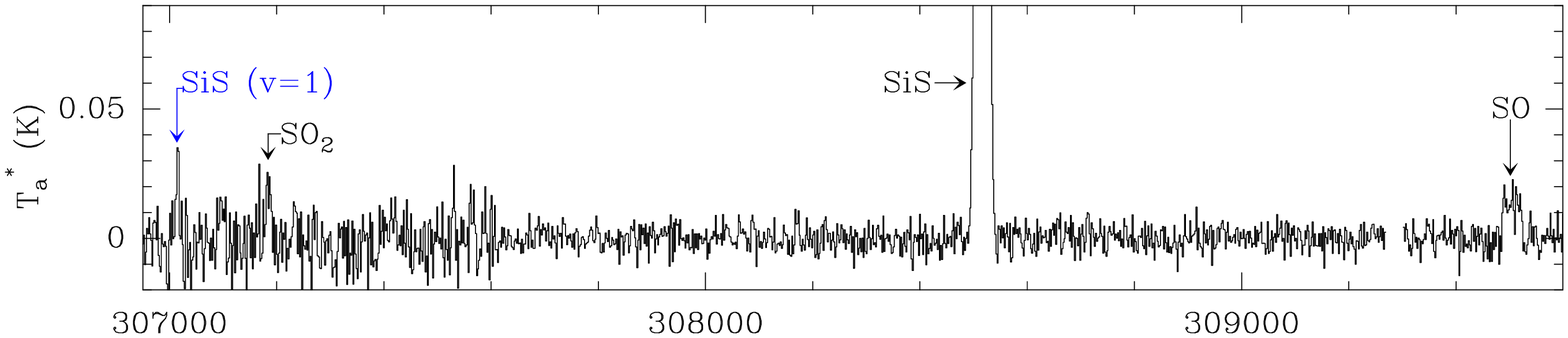}
\includegraphics[width=0.70\hsize]{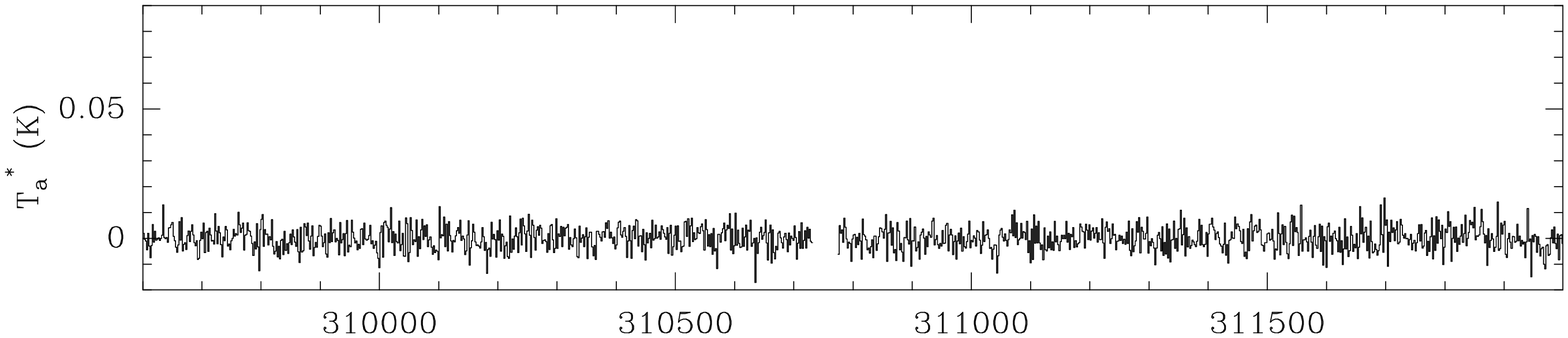}
\includegraphics[width=0.70\hsize]{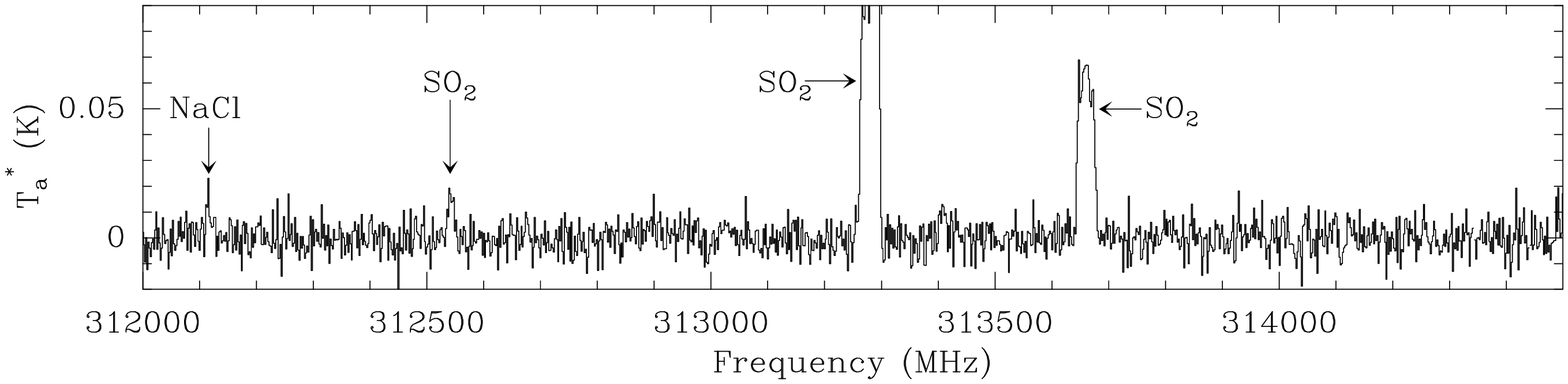}
\caption{(Continued)}
\end{figure*}

\setcounter{figure}{0}
\begin{figure*}[p] 
\centering
\includegraphics[width=0.70\hsize]{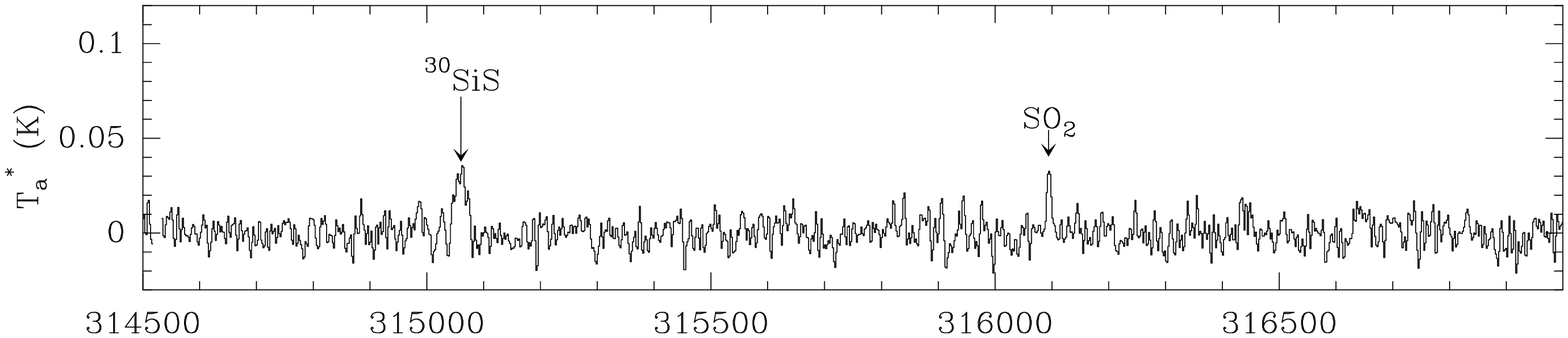}
\includegraphics[width=0.70\hsize]{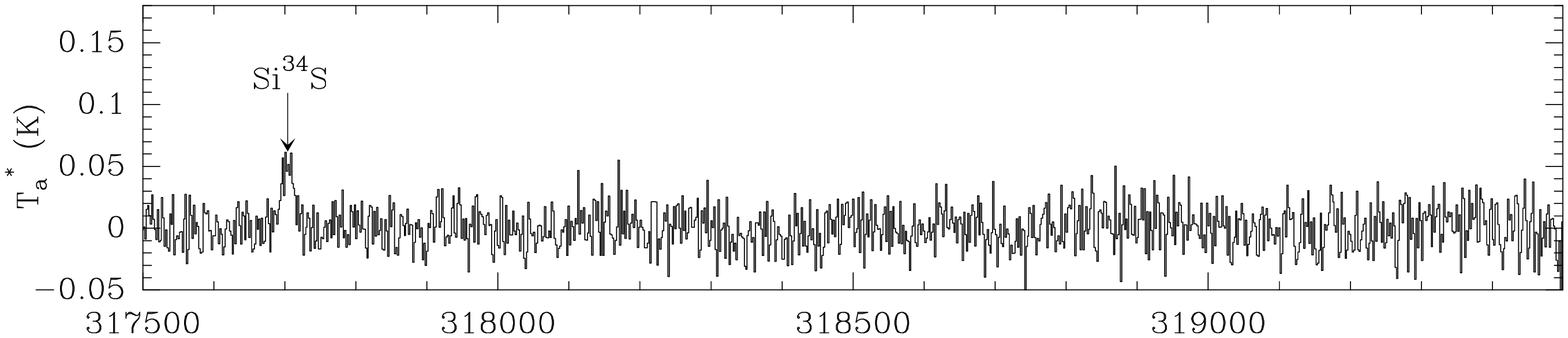}
\includegraphics[width=0.70\hsize]{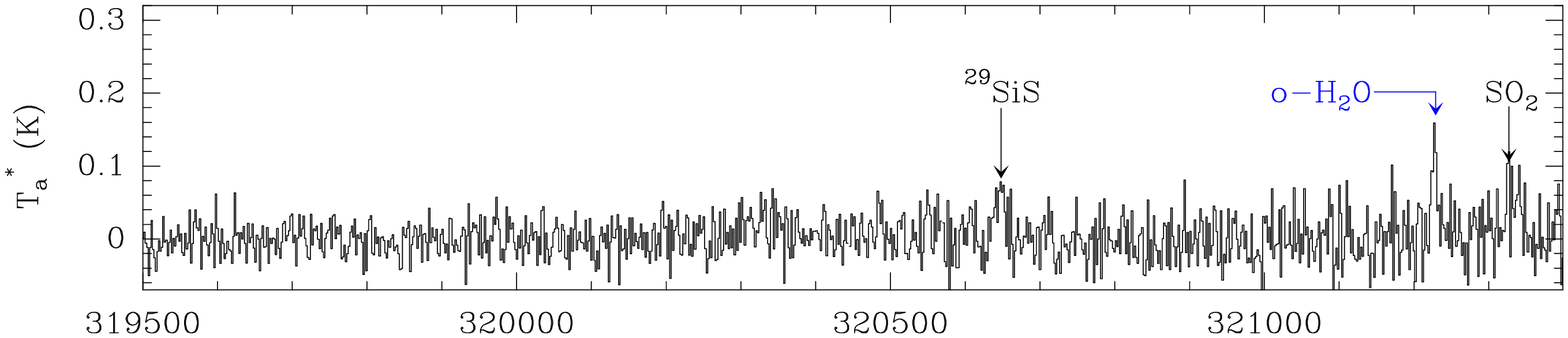}
\includegraphics[width=0.70\hsize]{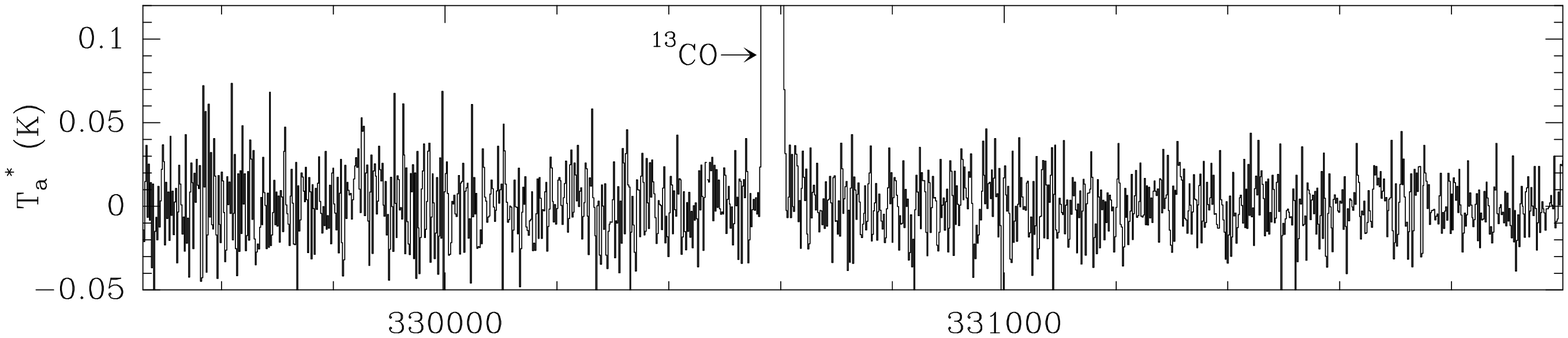}
\includegraphics[width=0.70\hsize]{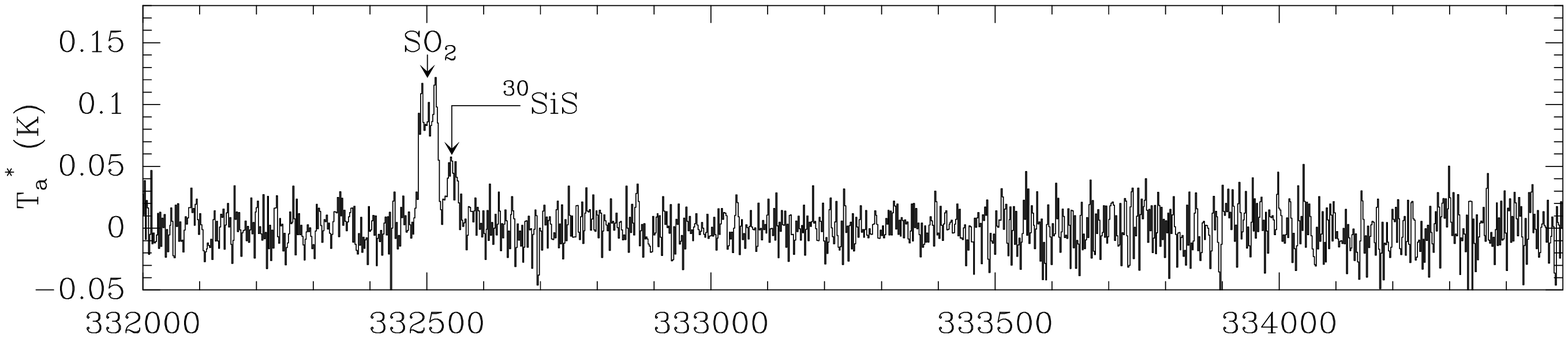}
\includegraphics[width=0.70\hsize]{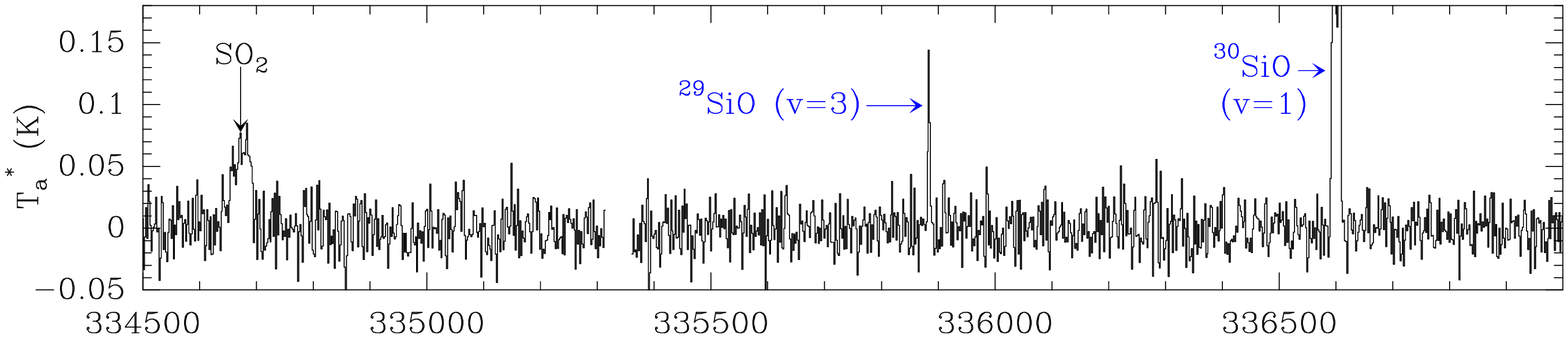}
\includegraphics[width=0.70\hsize]{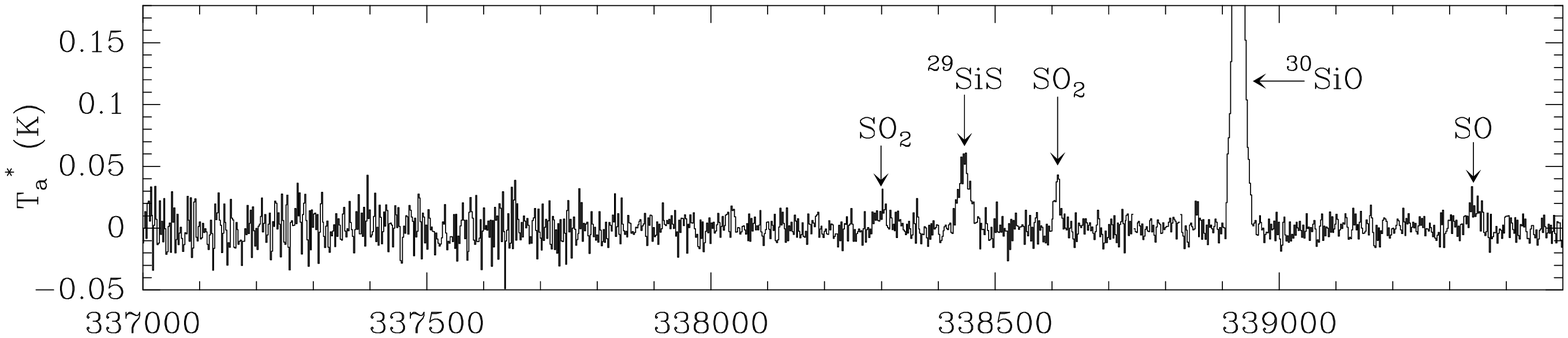}
\includegraphics[width=0.70\hsize]{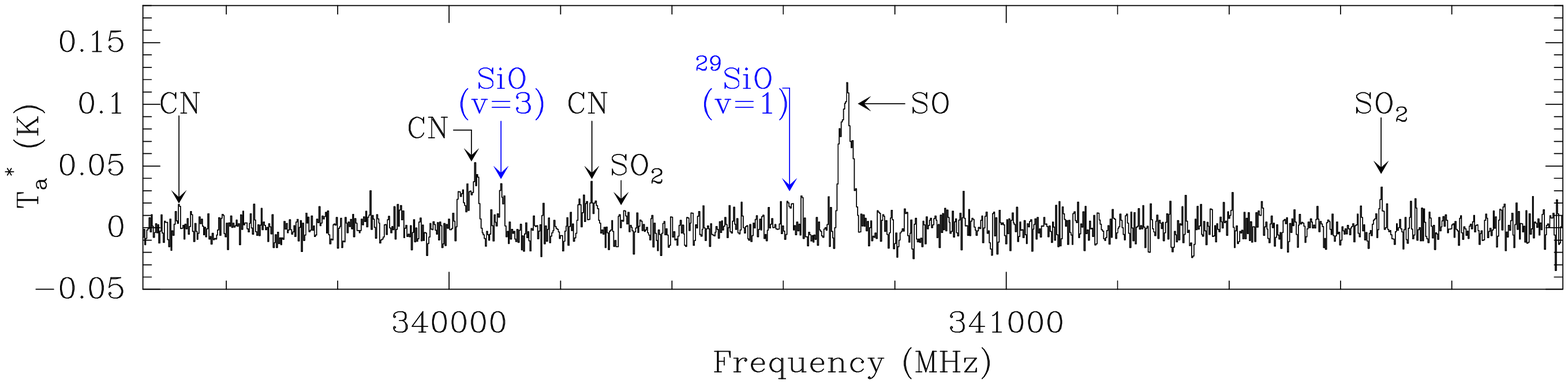}
\caption{(Continued)}
\end{figure*}

\setcounter{figure}{0}
\begin{figure*}[p] 
\centering
\includegraphics[width=0.70\hsize]{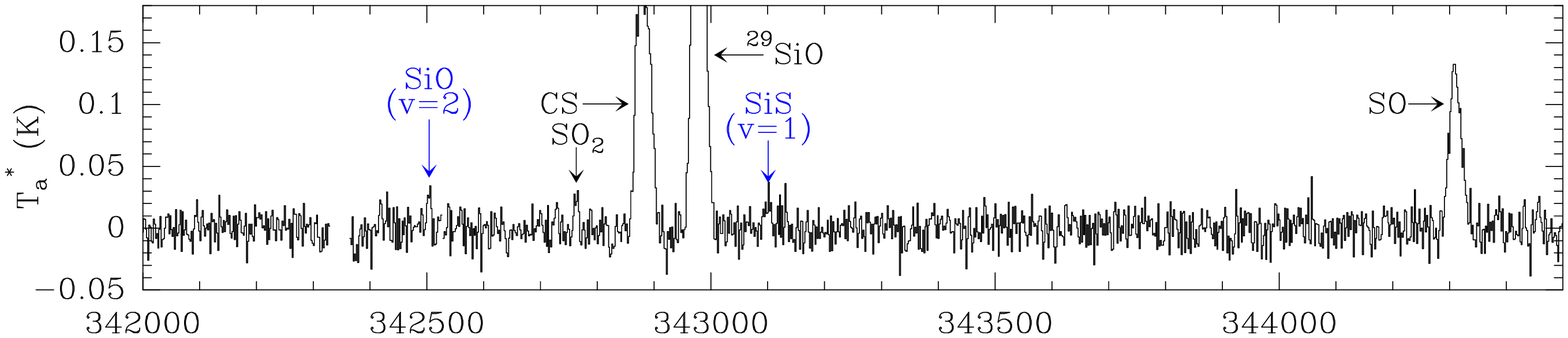}
\includegraphics[width=0.70\hsize]{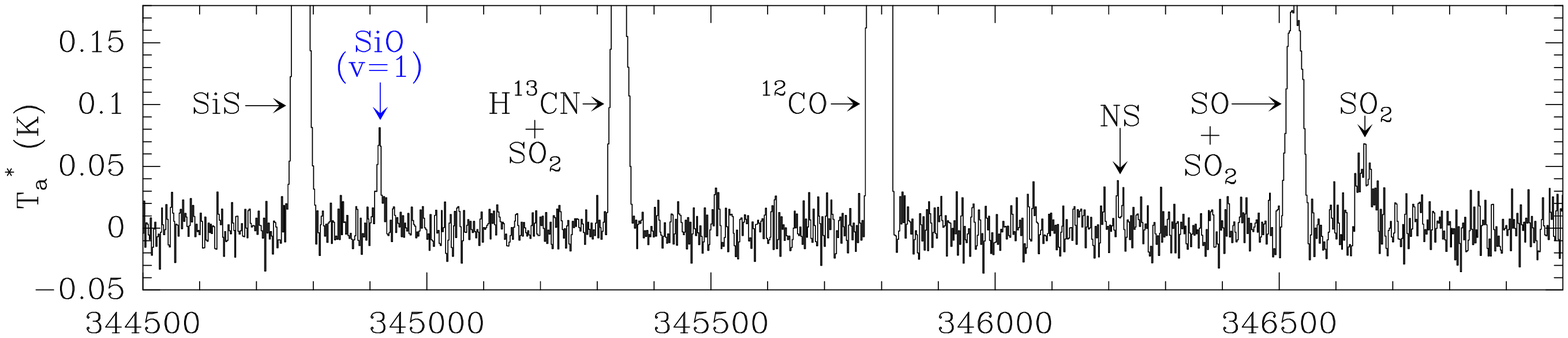}
\includegraphics[width=0.70\hsize]{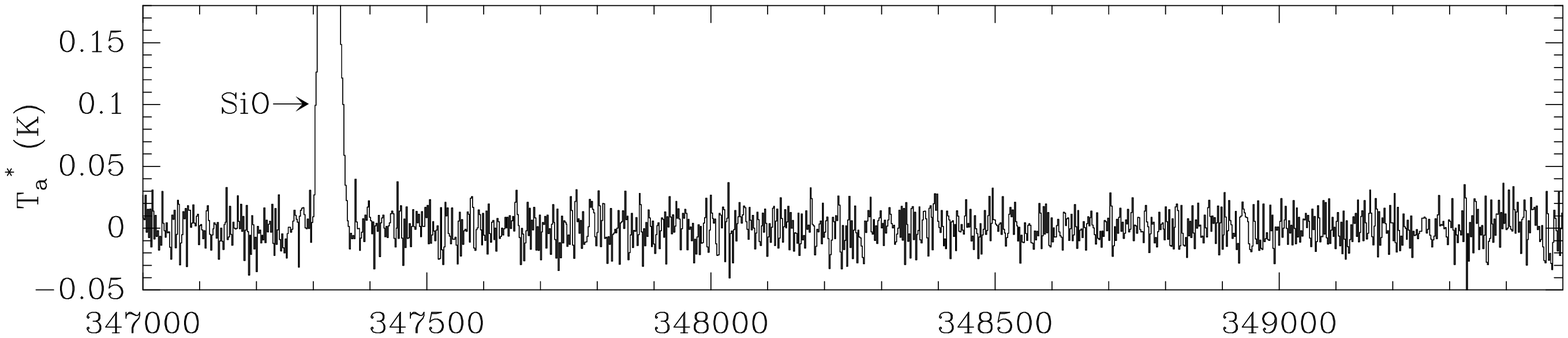}
\includegraphics[width=0.70\hsize]{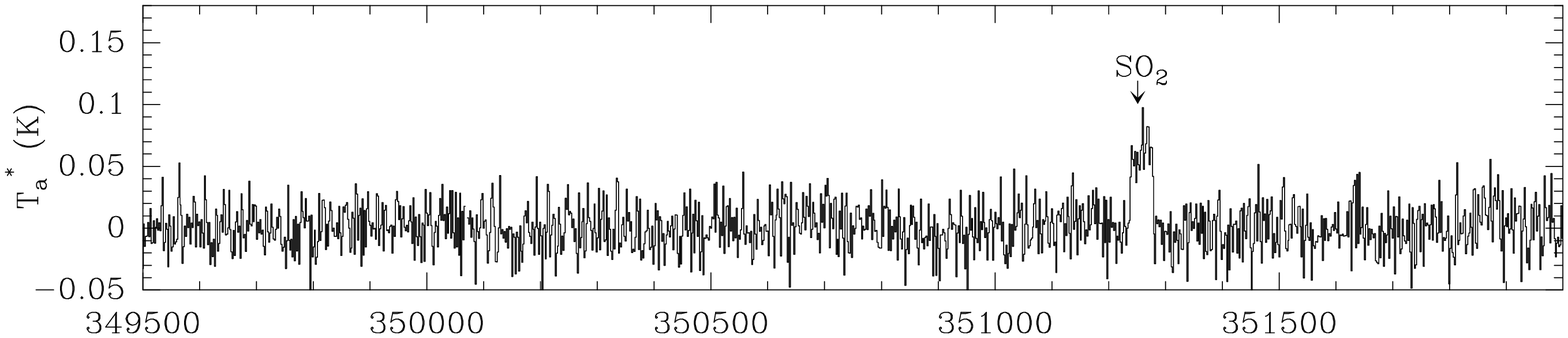}
\includegraphics[width=0.70\hsize]{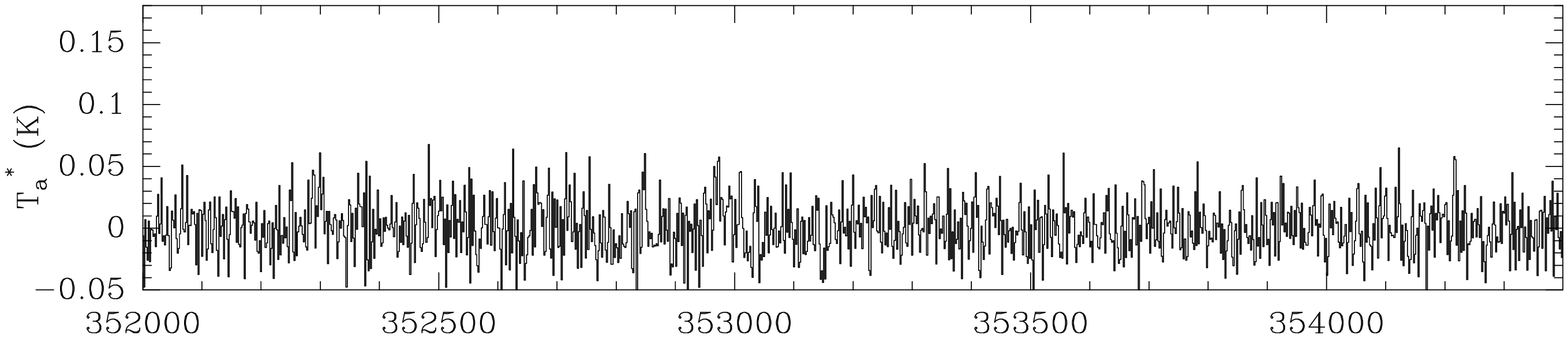}
\includegraphics[width=0.70\hsize]{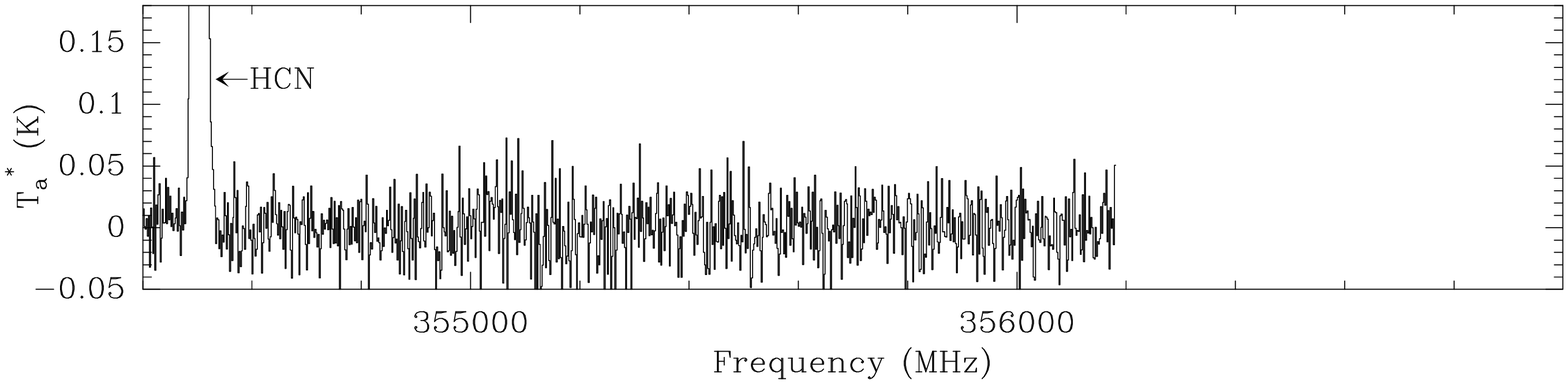}
\caption{(Continued)}
\end{figure*}

\clearpage

\section{Population diagrams}\label{sec:app_rd}

\begin{figure*}[!htp] 
\centering
\includegraphics[width=0.45\hsize]{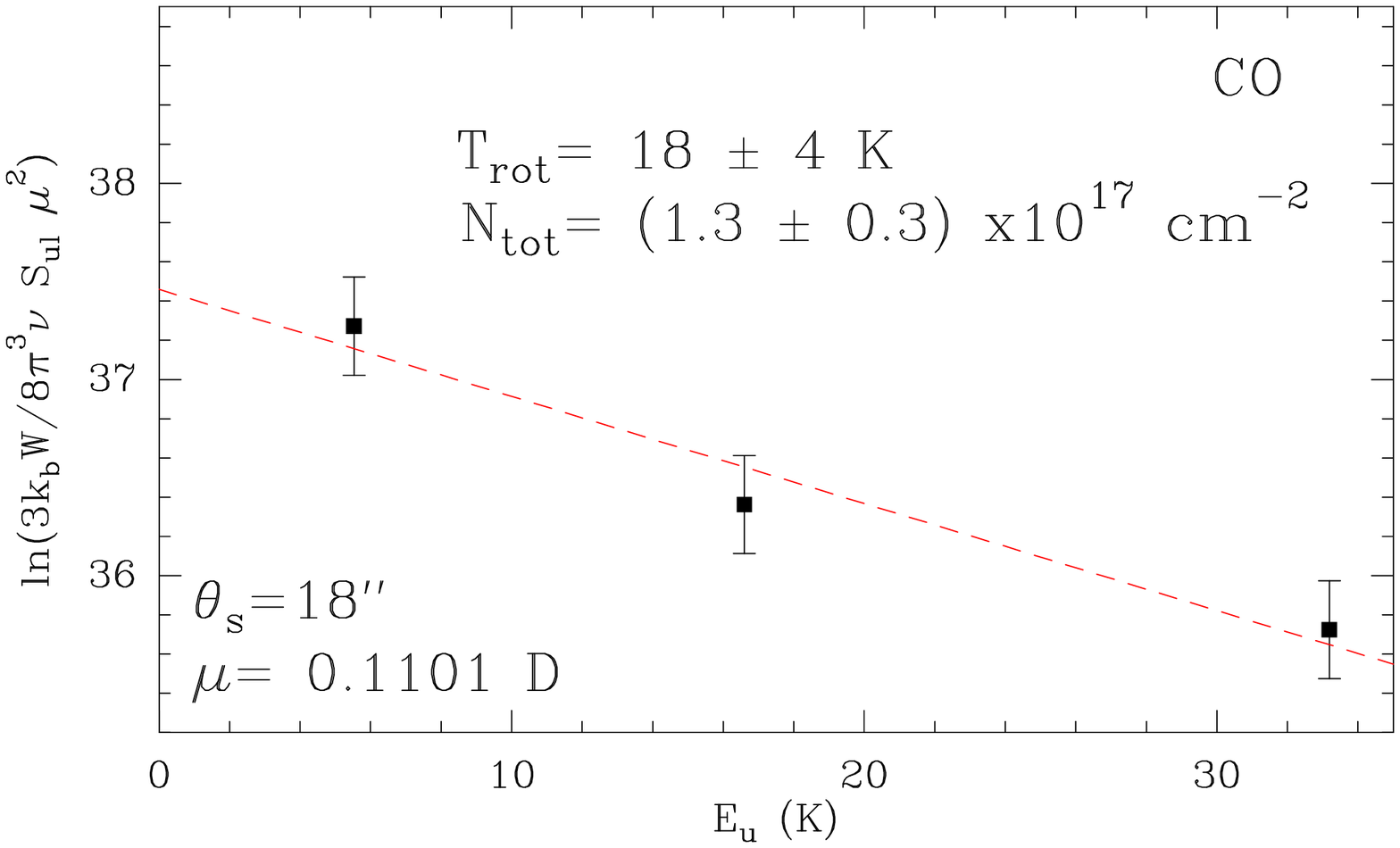}
\includegraphics[width=0.45\hsize]{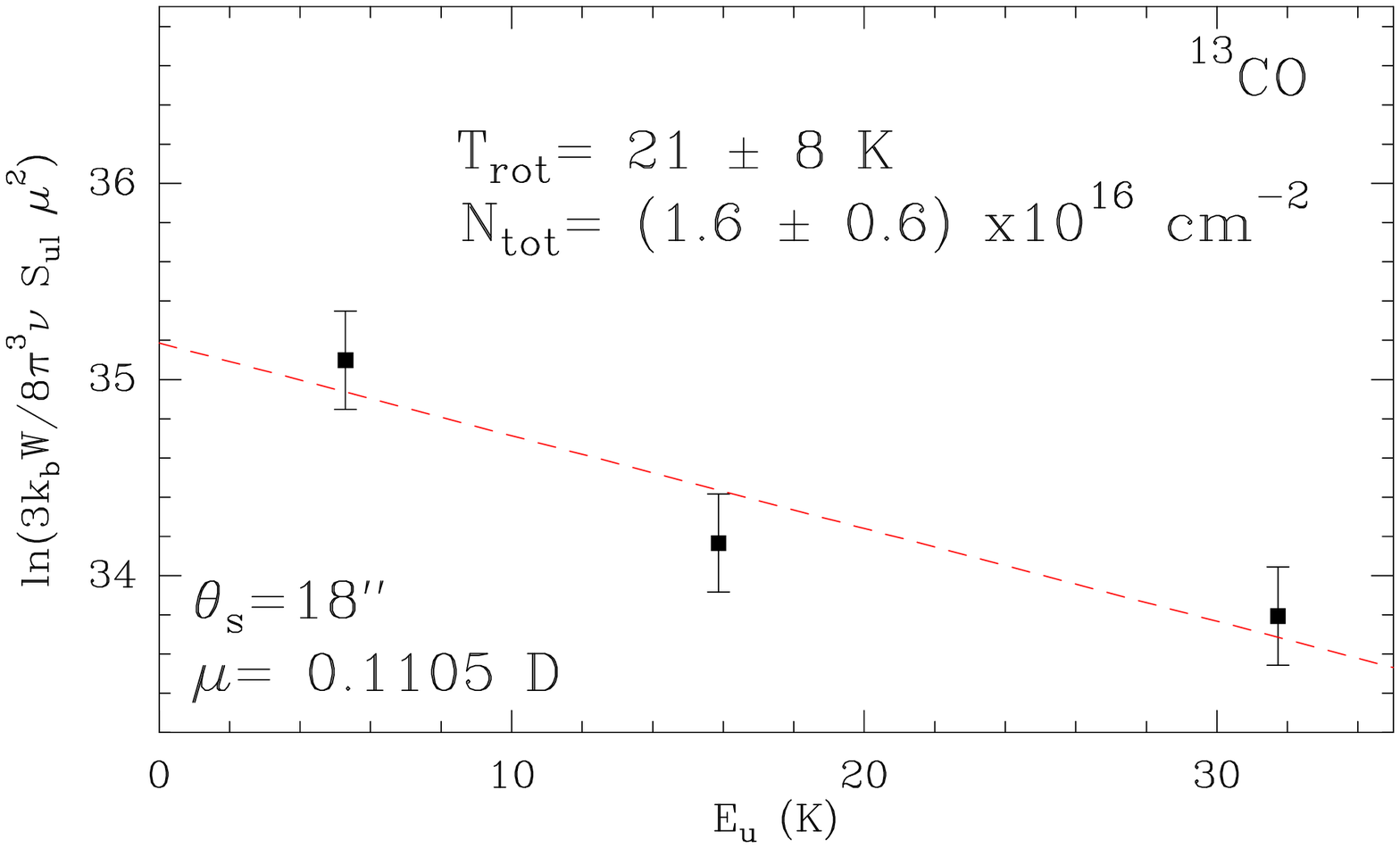}
\caption{Population diagrams for CO and $^{13}$CO. The results from the fit (red dashed line) are shown in each box. The 
emission size adopted and the dipole moment of the molecule are shown in the bottom-left corner of each box. 
Error bars include the formal uncertainty of the measurement and a 25\% uncertainty due to flux calibration or
poor baseline substraction.}
\label{fig:rdallco}
\end{figure*}

\begin{figure*}[!htp] 
\centering
\includegraphics[width=0.45\hsize]{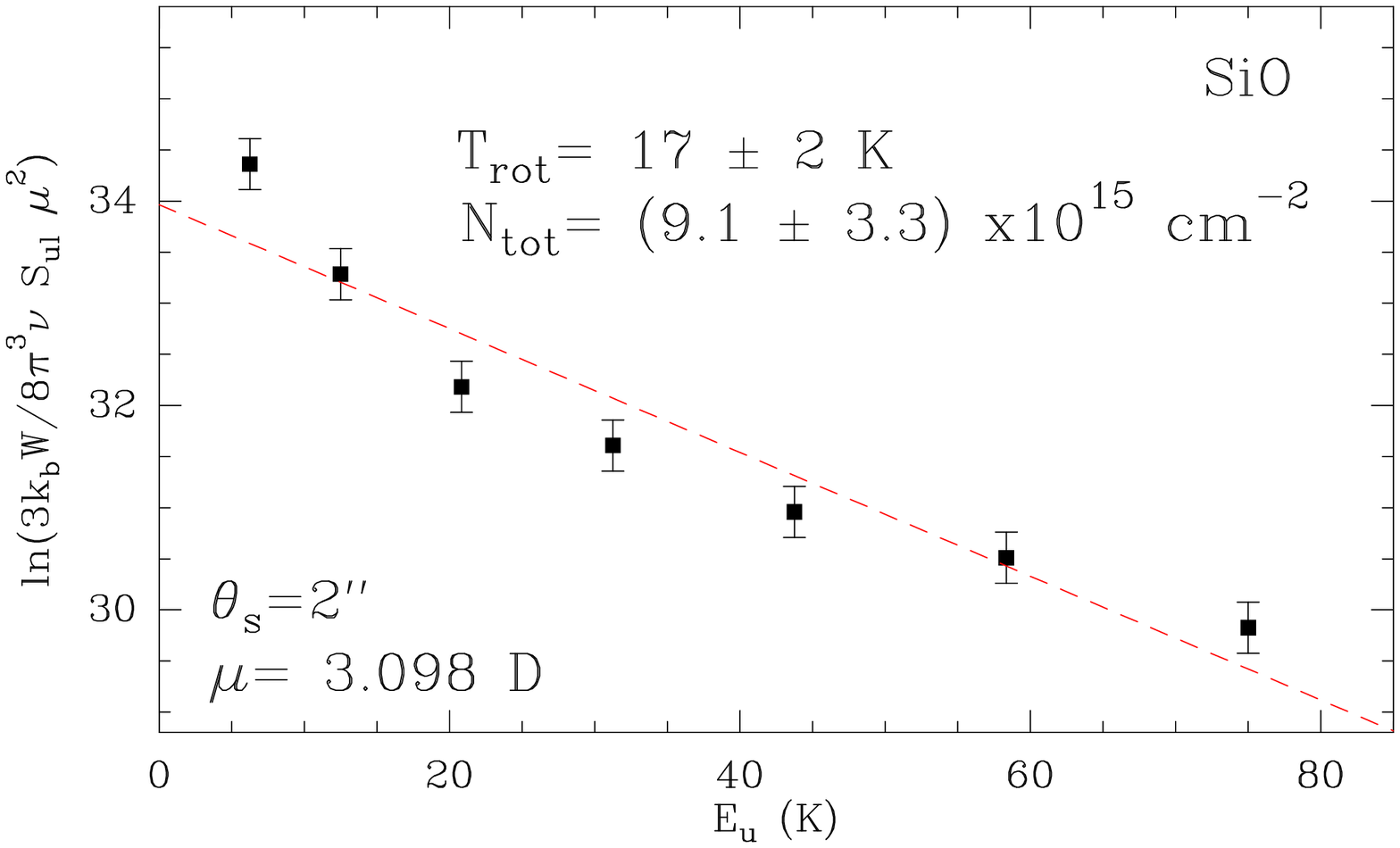}
\includegraphics[width=0.45\hsize]{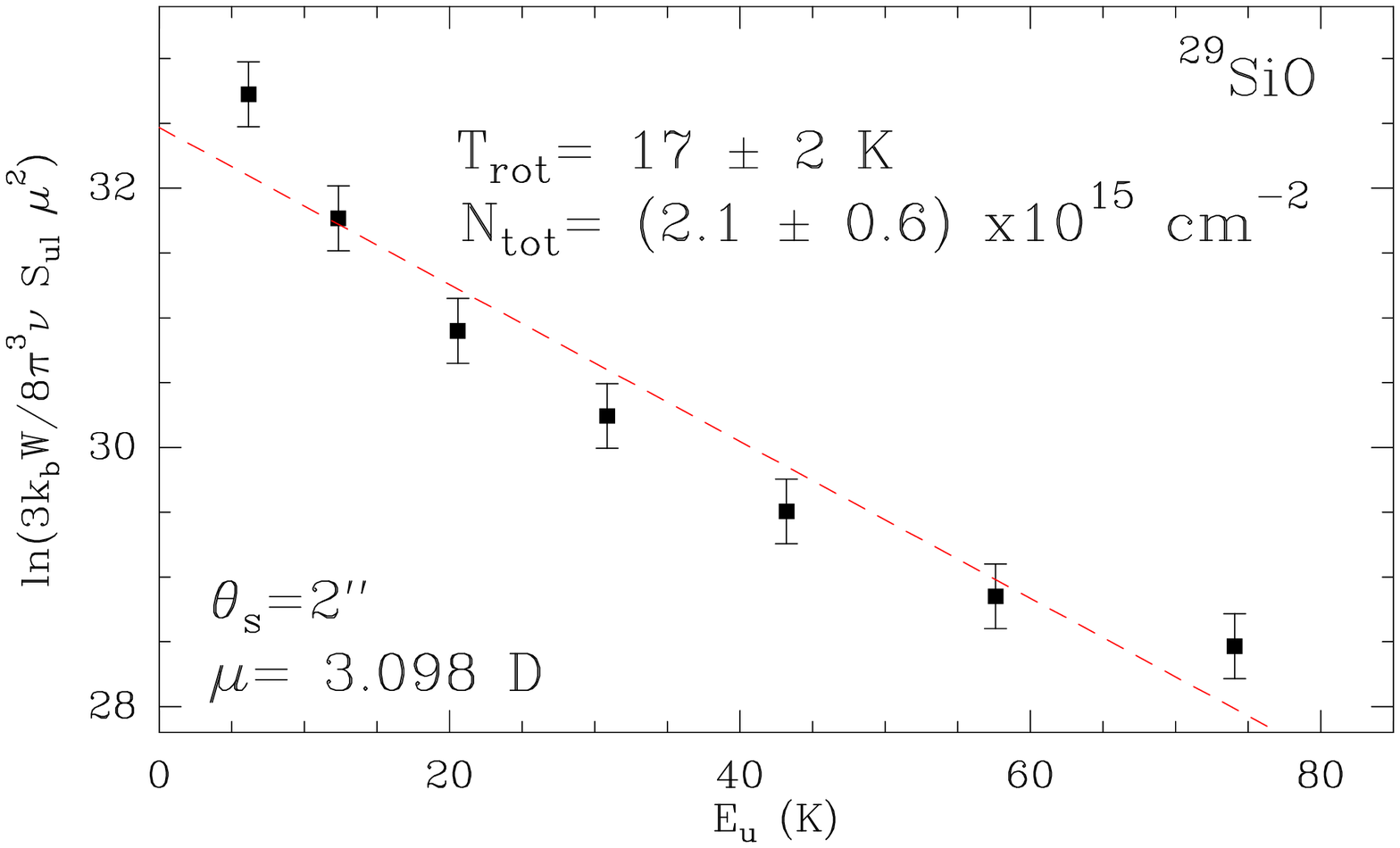}
\includegraphics[width=0.45\hsize]{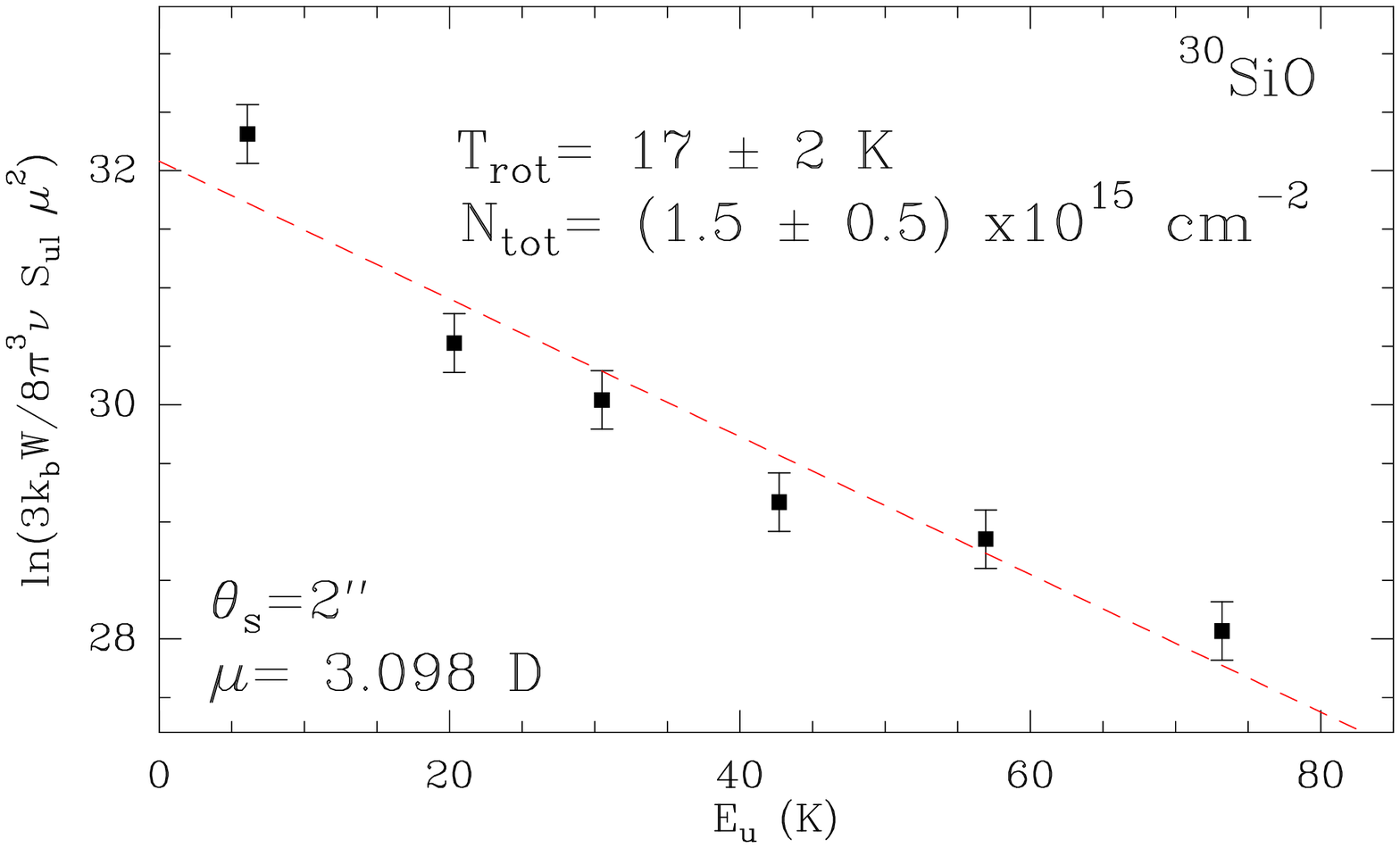}
\includegraphics[width=0.45\hsize]{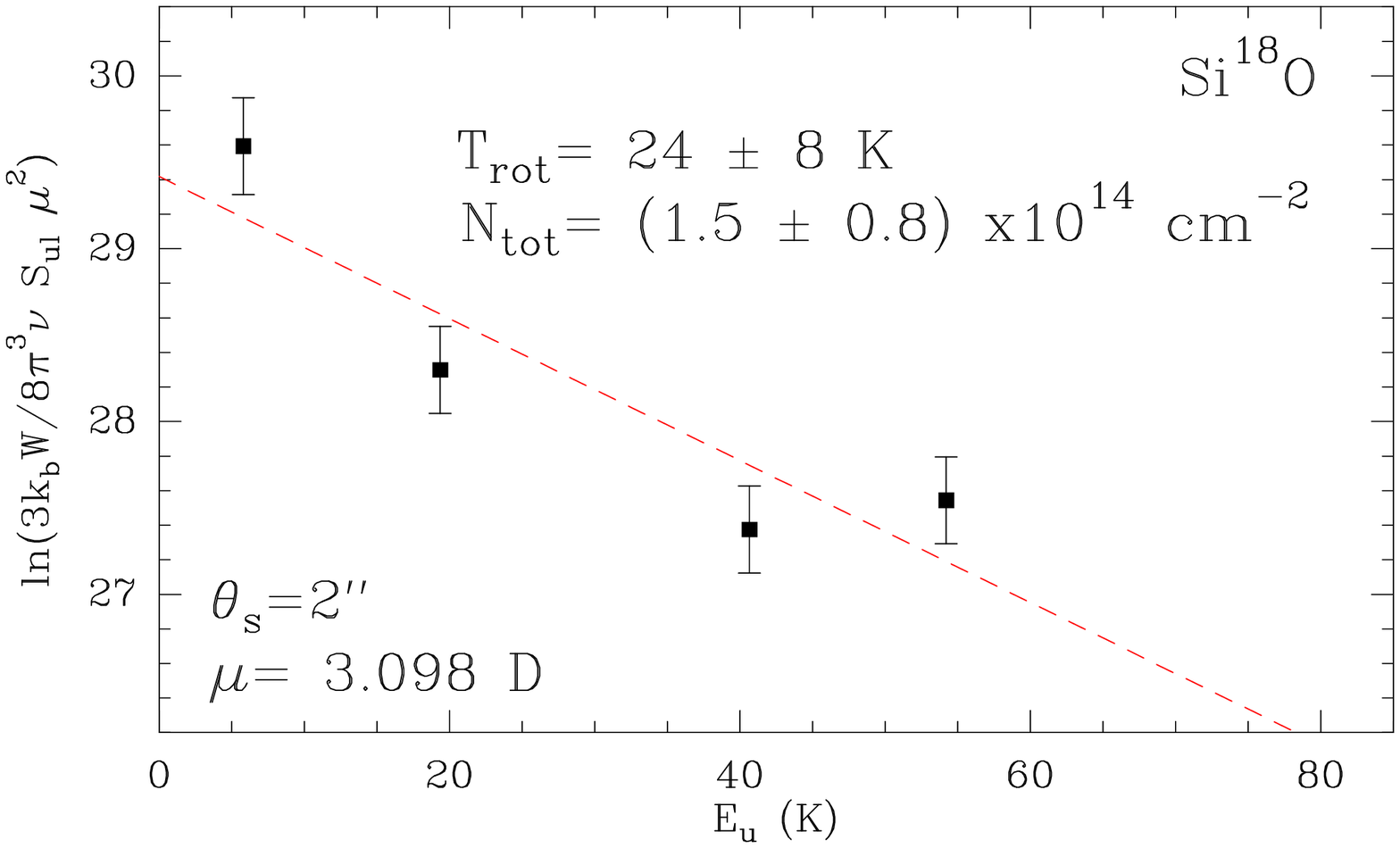}
\includegraphics[width=0.45\hsize]{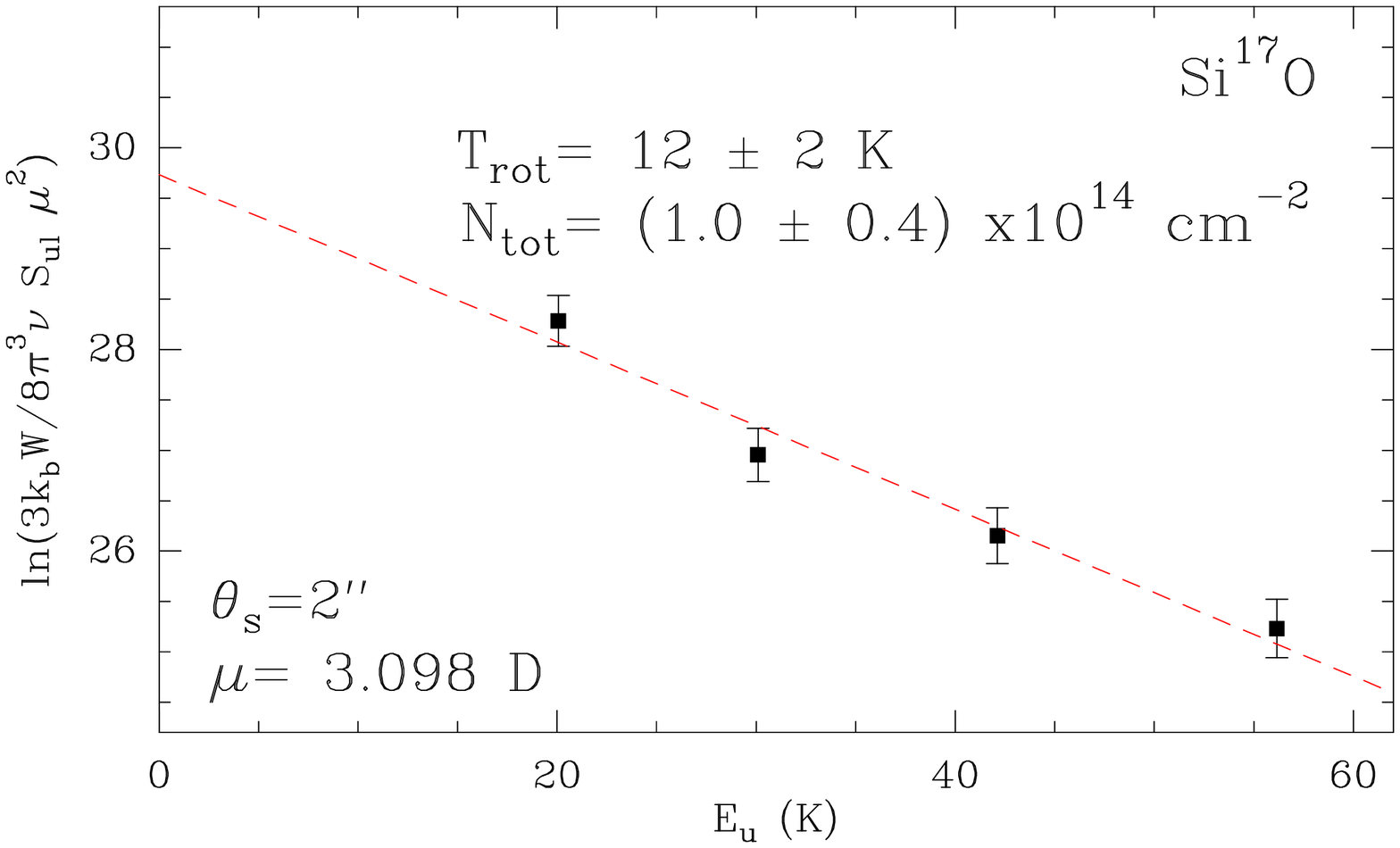}
 \vspace{-1mm}
\caption{As in Fig.\,\ref{fig:rdallco} but for SiO isotopologues.}
\label{fig:rdallsio}
\end{figure*}

\twocolumn

\begin{figure}[!hbp] 
\centering
\includegraphics[width=0.95\hsize]{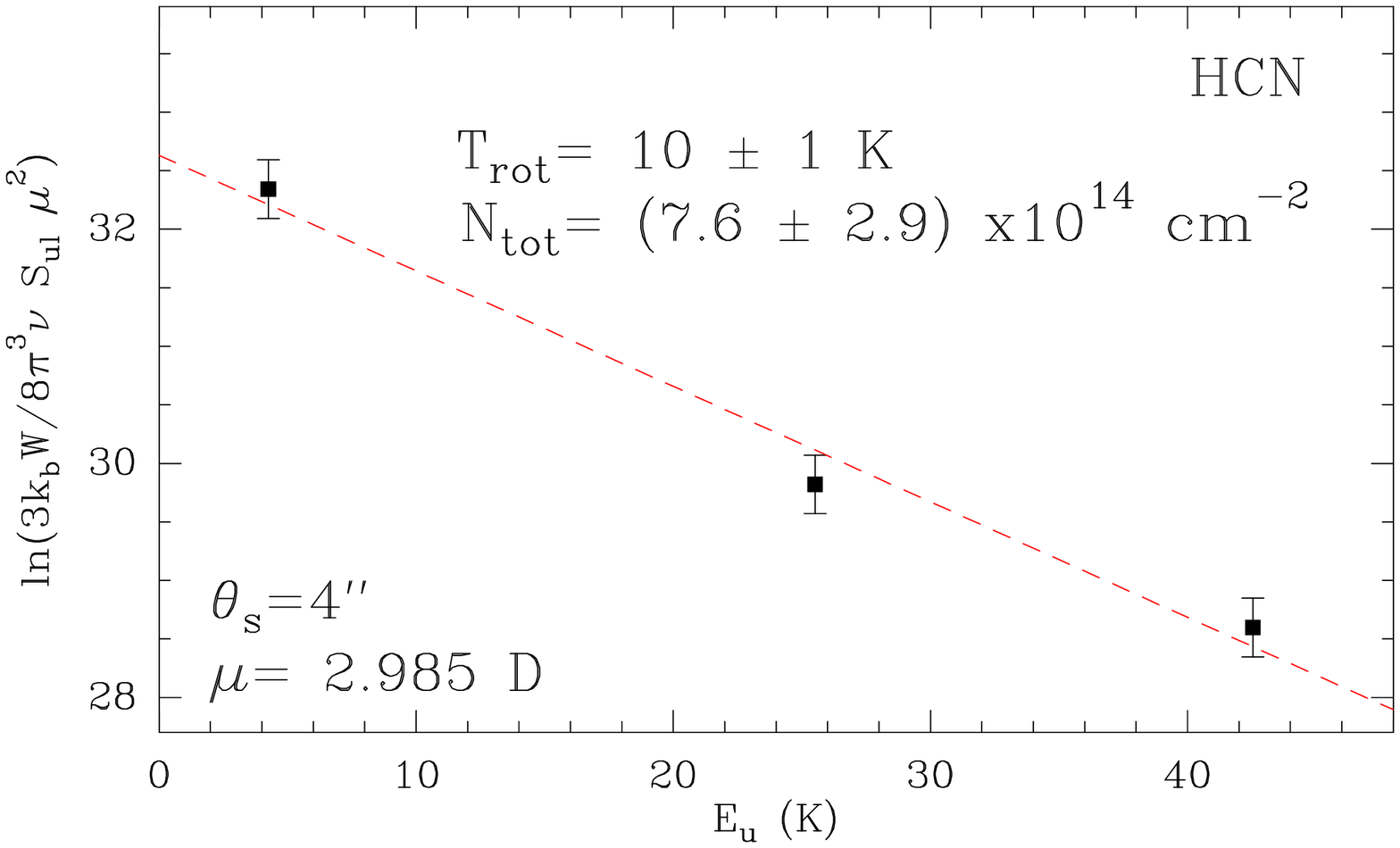}
\includegraphics[width=0.95\hsize]{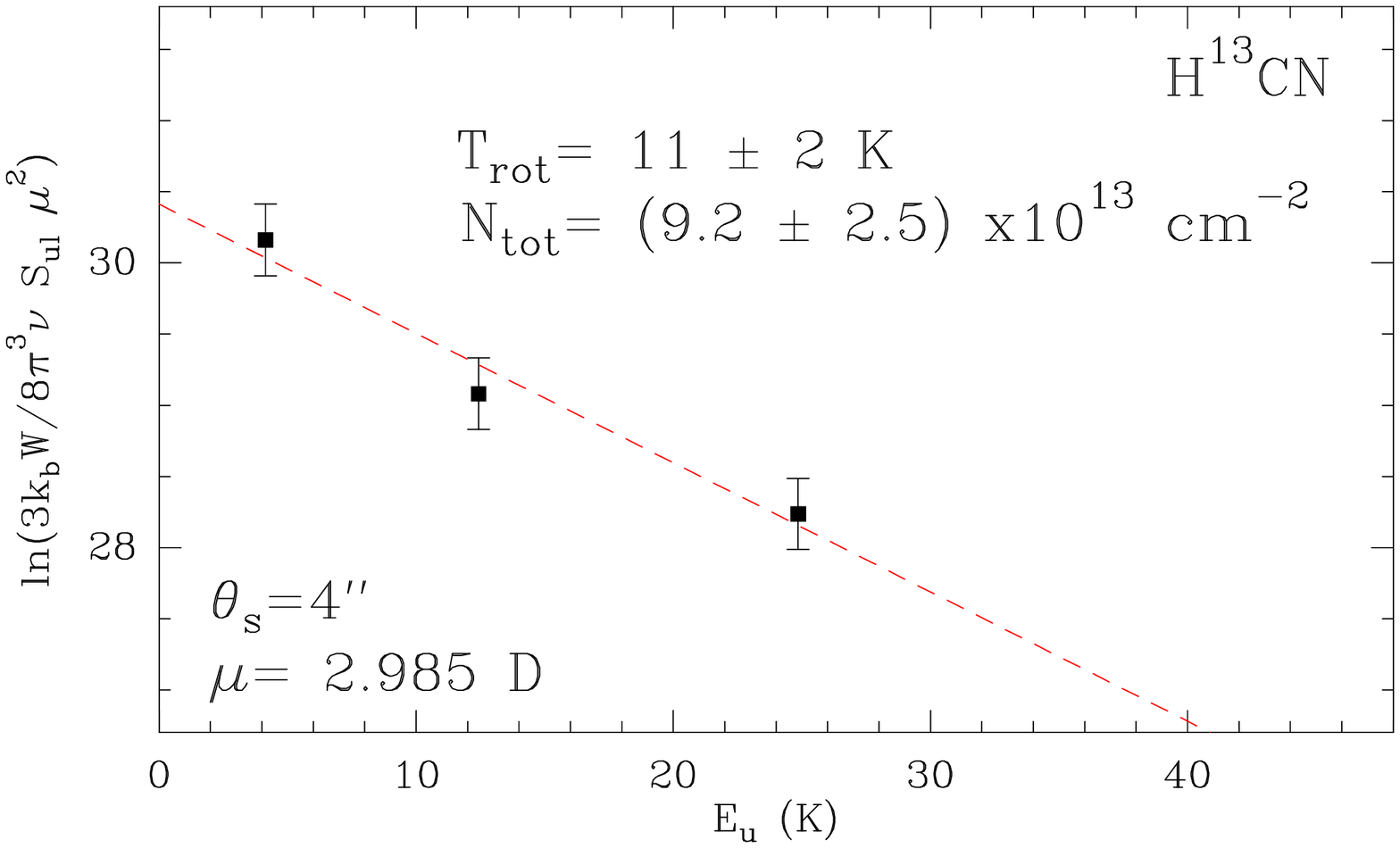}
\caption{As in Fig.\,\ref{fig:rdallco} but for HCN isotopologues.}
\label{fig:rdallhcn}
\end{figure}

\begin{figure}[!hbp] 
\centering
\includegraphics[width=0.95\hsize]{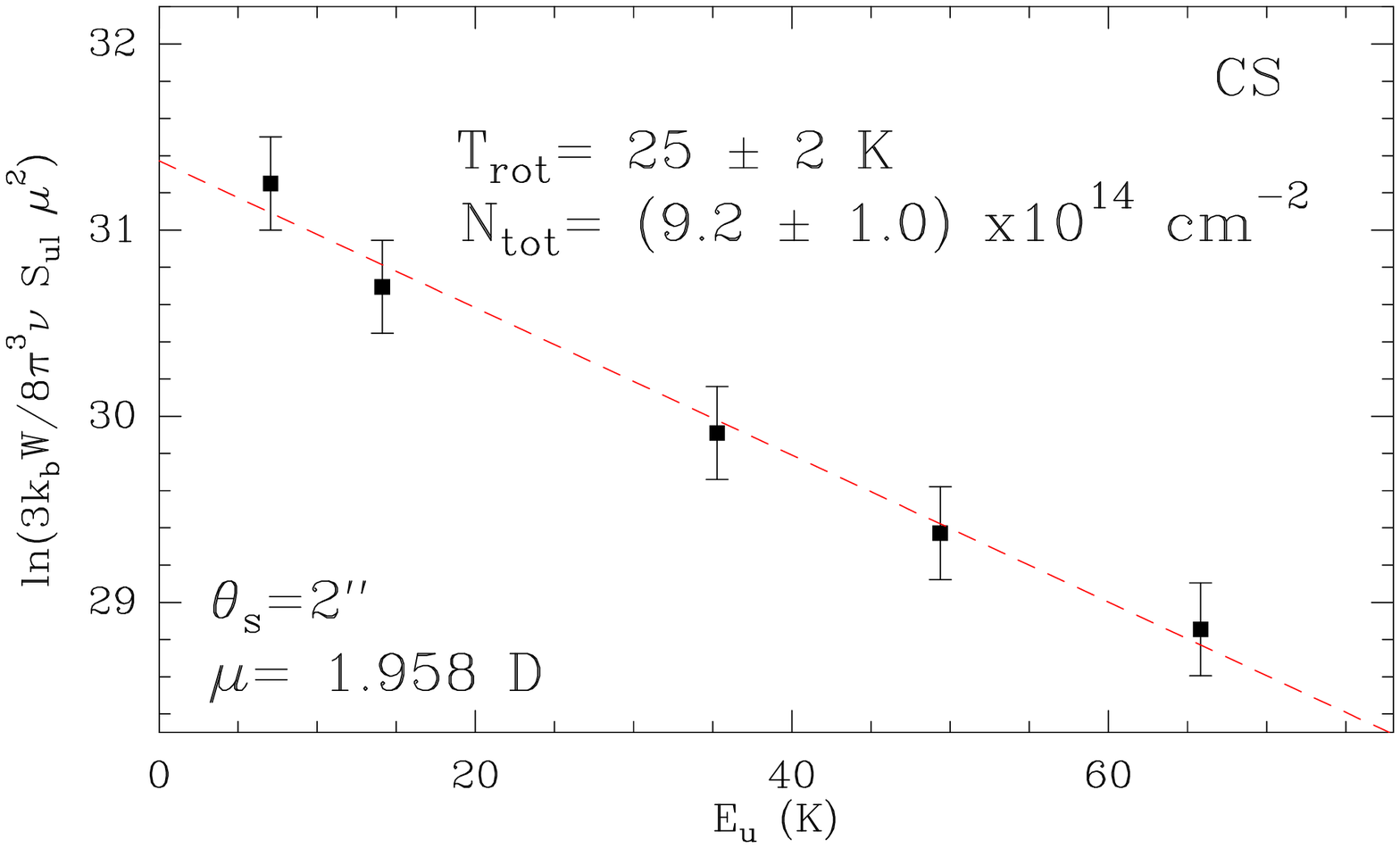}
\includegraphics[width=0.95\hsize]{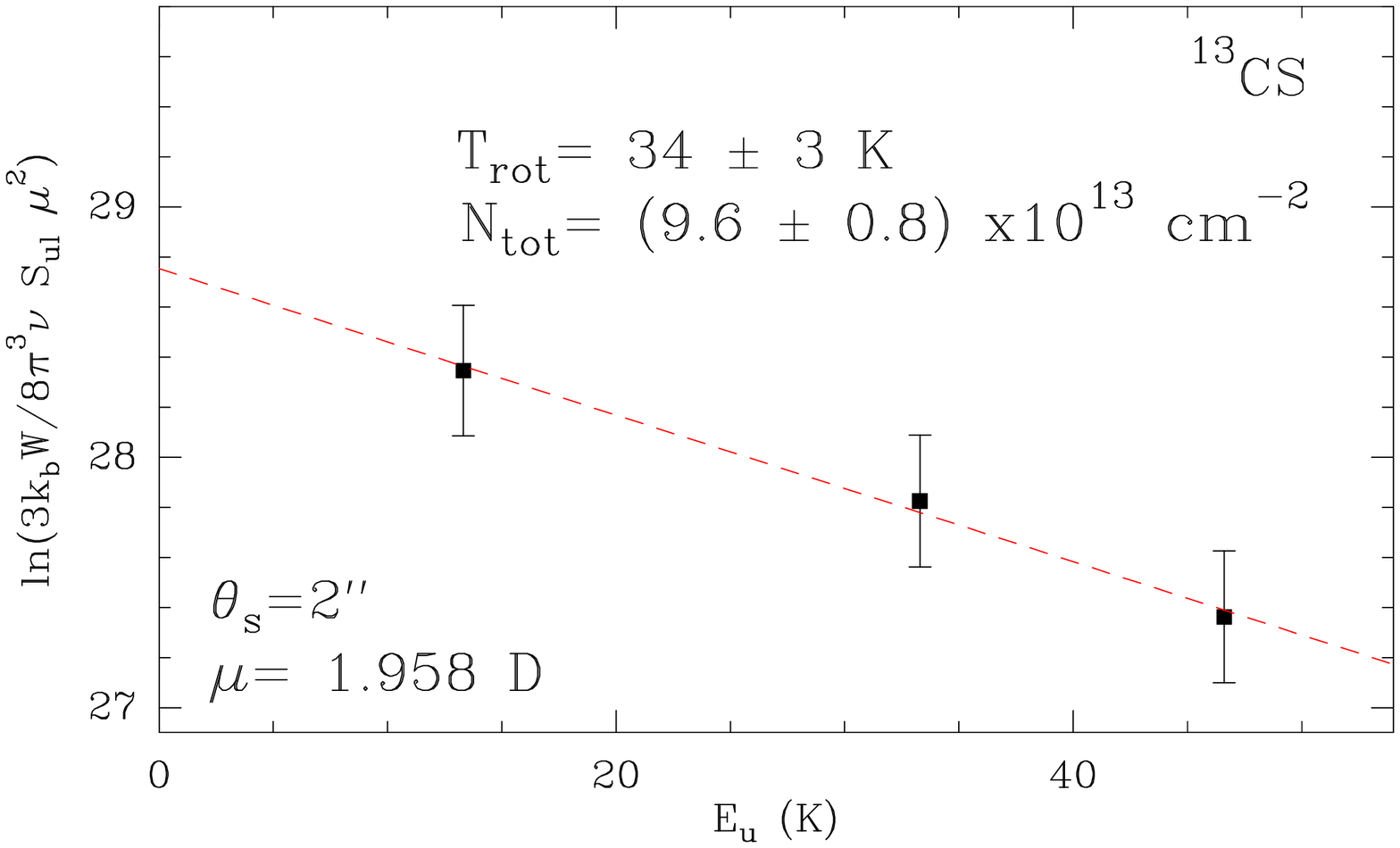}
\includegraphics[width=0.95\hsize]{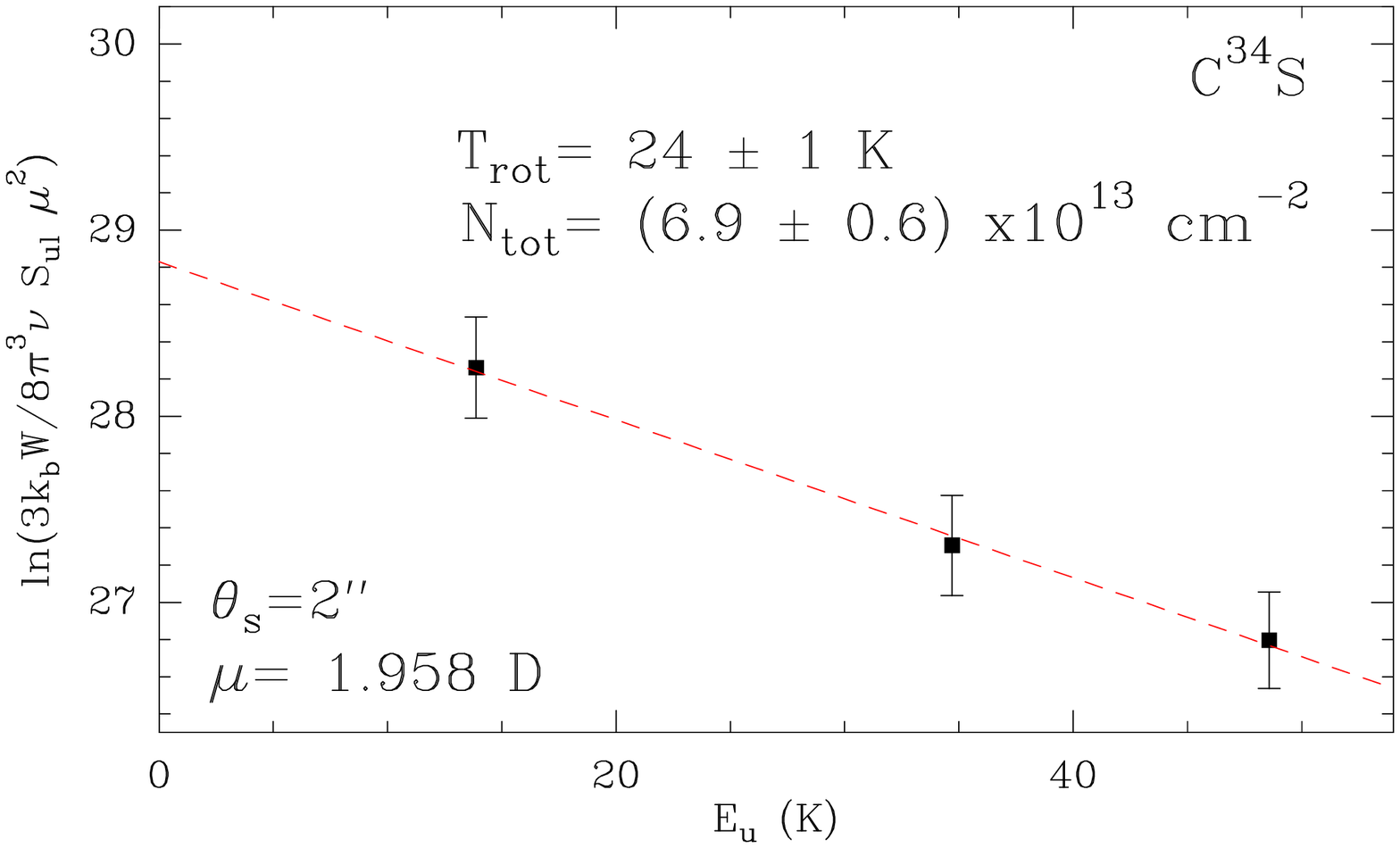}
\caption{As in Fig.\,\ref{fig:rdallco} but for CS isotopologues.}
\label{fig:rdallcs}
\end{figure}

\begin{figure}[!htp] 
\centering
\includegraphics[width=0.95\hsize]{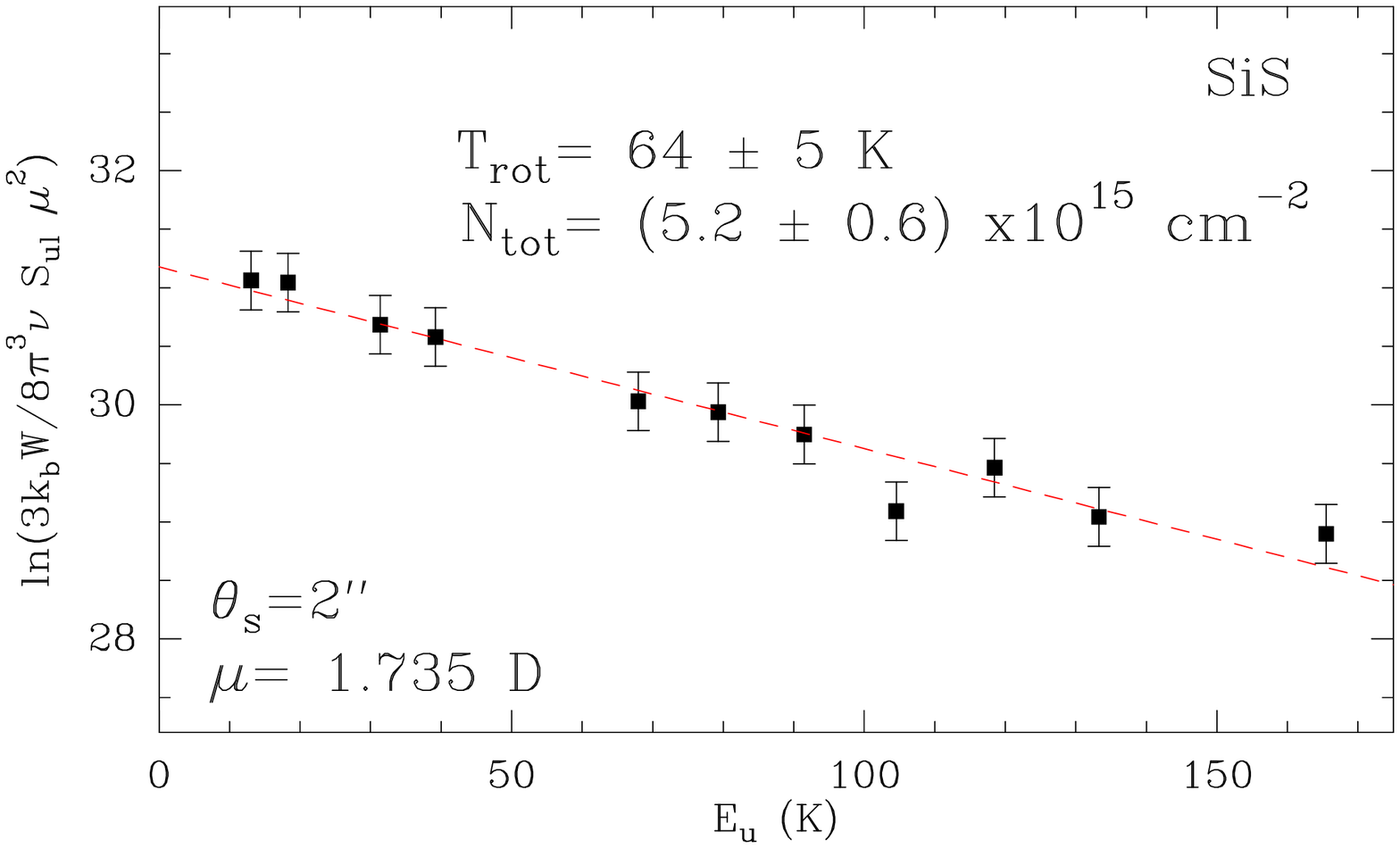}
\includegraphics[width=0.95\hsize]{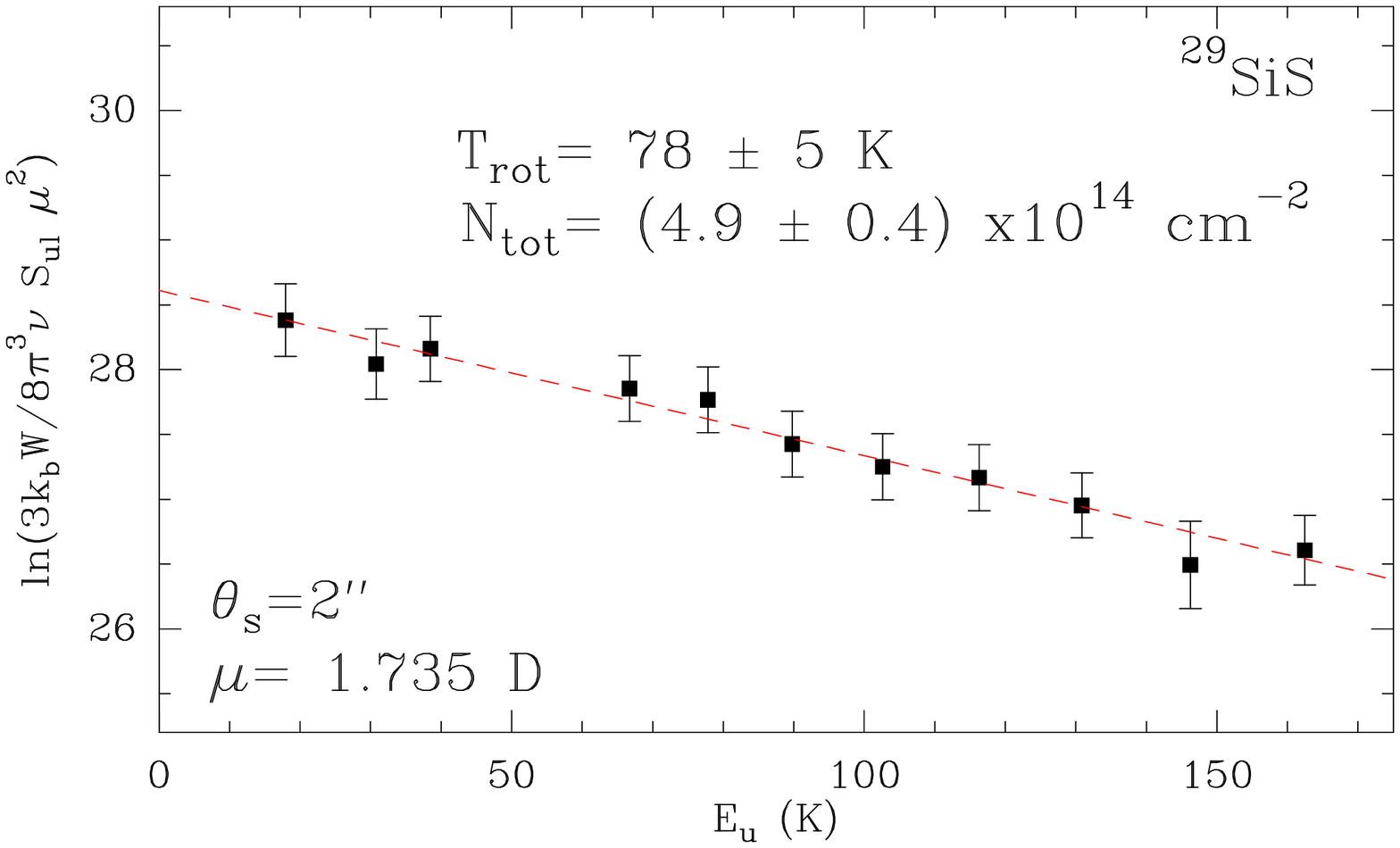}
\includegraphics[width=0.95\hsize]{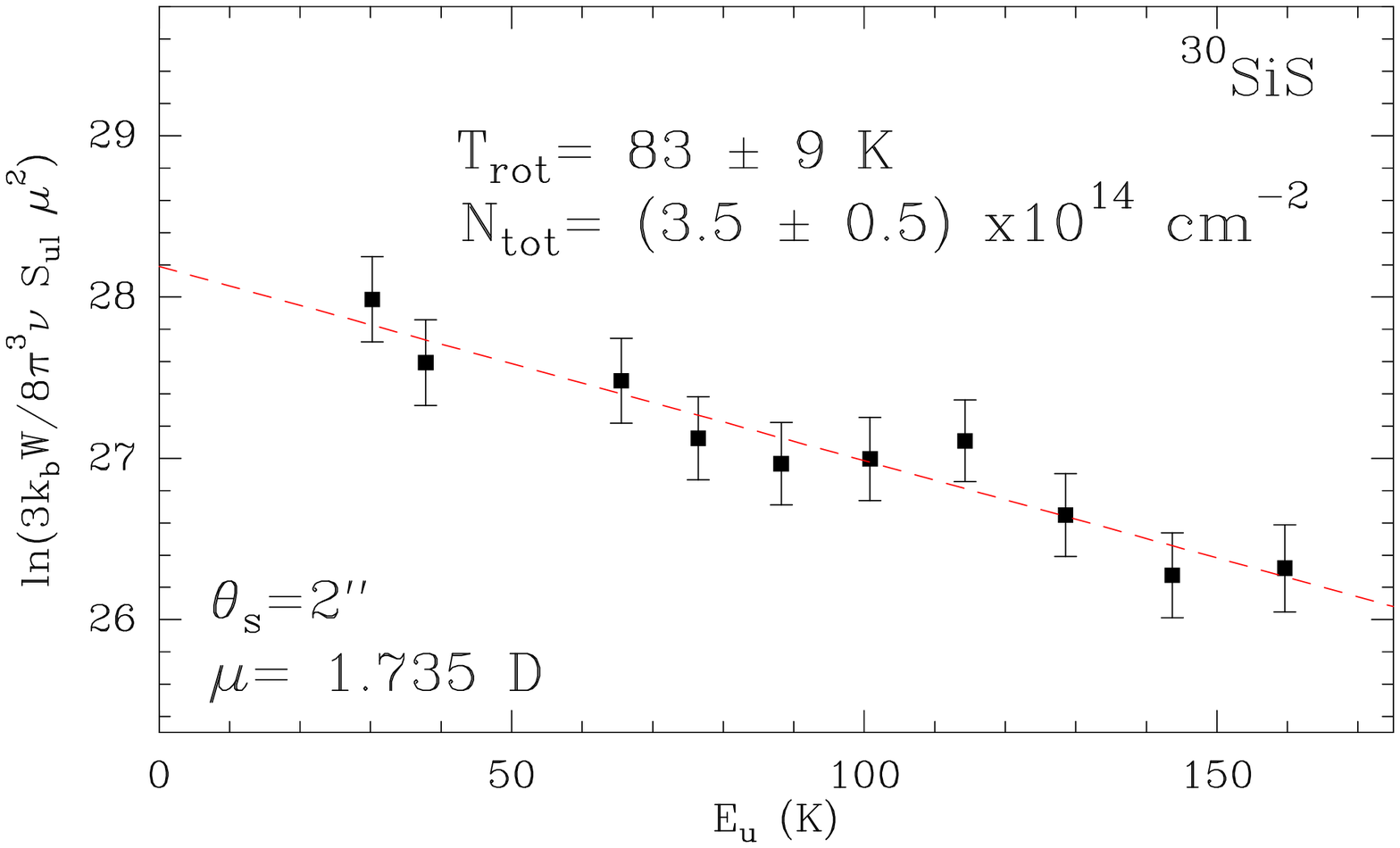}
\includegraphics[width=0.95\hsize]{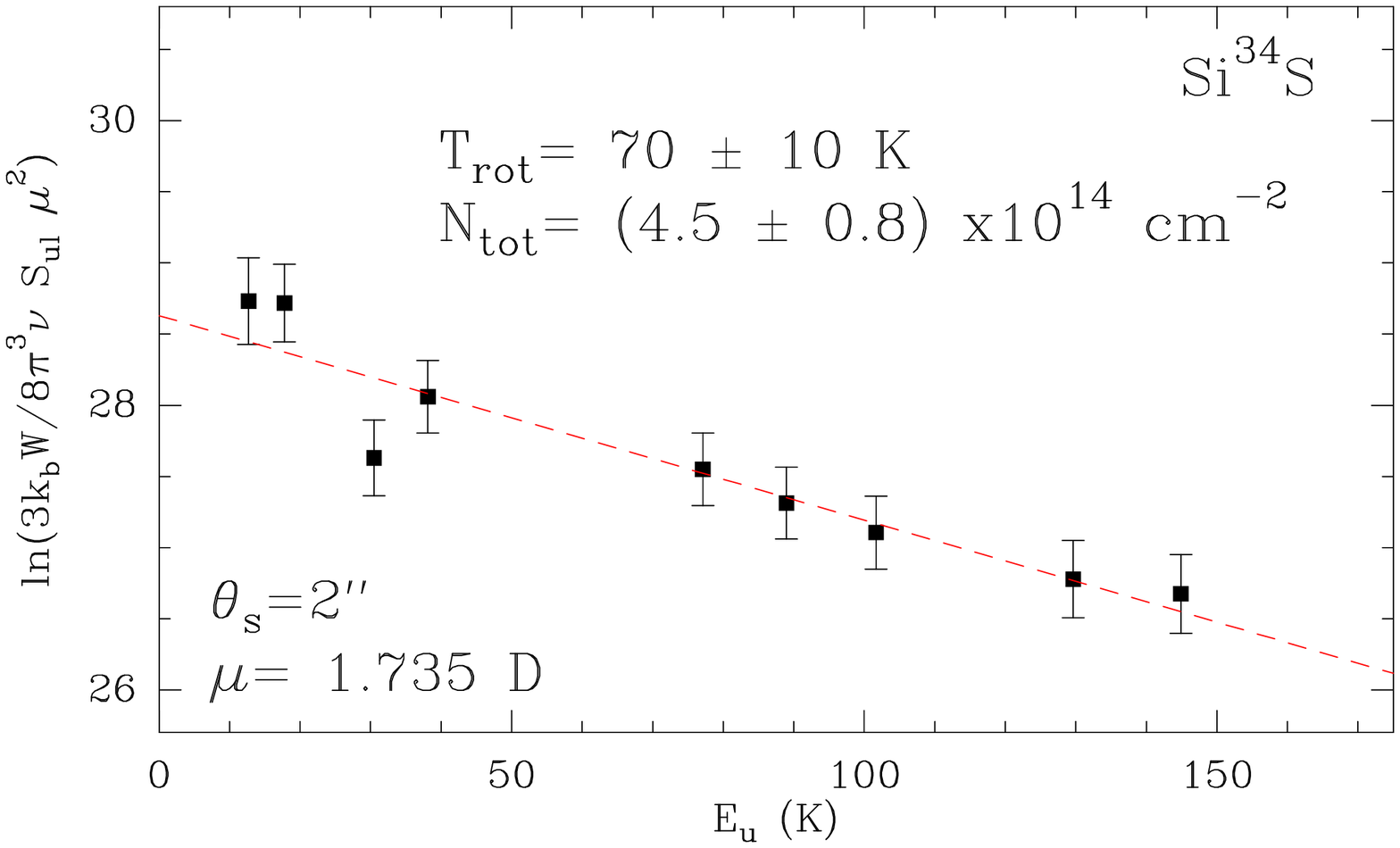}
\caption{As in Fig.\,\ref{fig:rdallco} but for SiS isotopologues.}
\label{fig:rdallsis}
\end{figure}

\begin{figure}[!htp] 
\centering
\includegraphics[width=0.95\hsize]{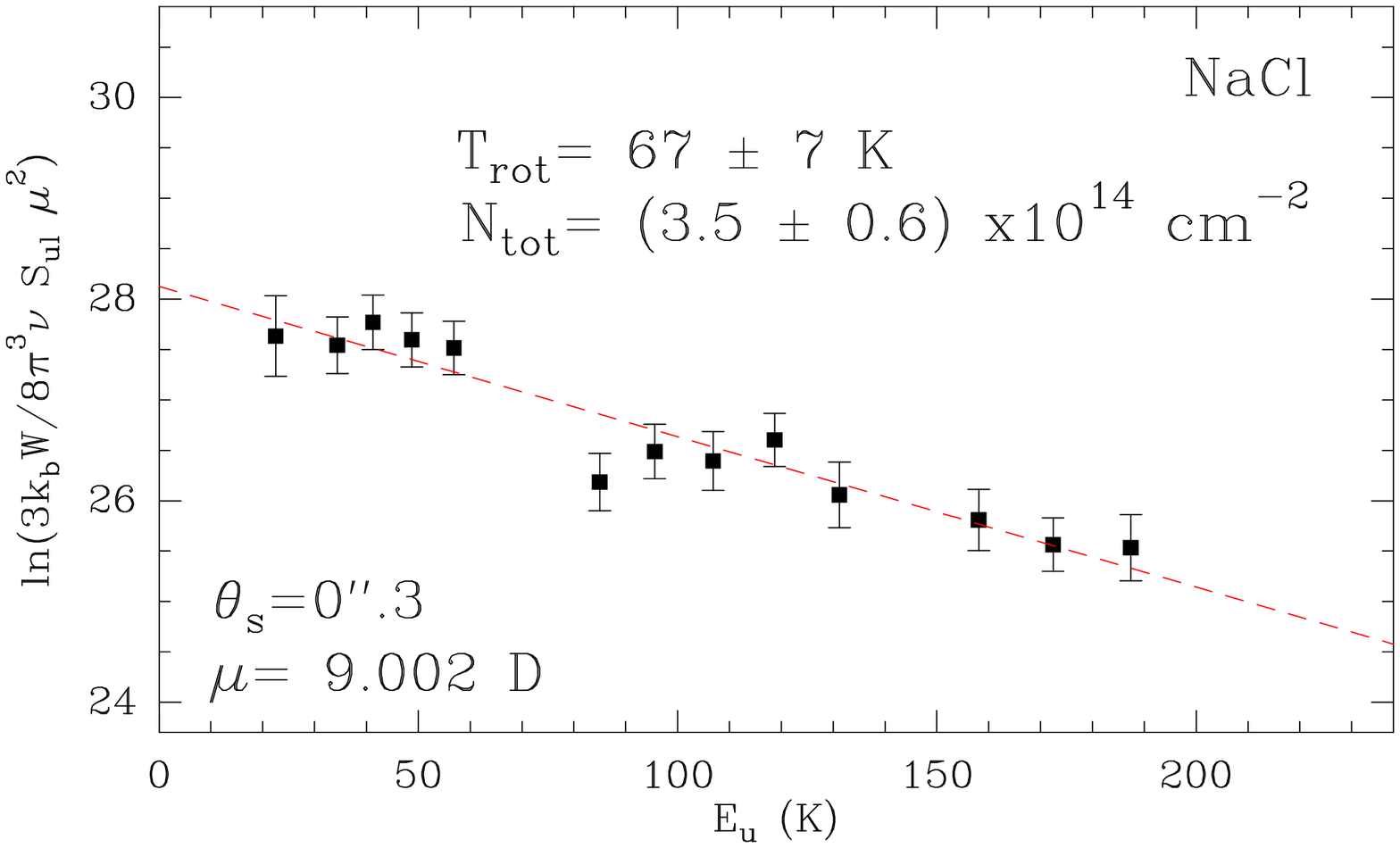}
\caption{As in Fig.\,\ref{fig:rdallco} but for NaCl.}
\label{fig:rdallnacl}
\end{figure}

\begin{figure}[!htp] 
\centering
\includegraphics[width=0.95\hsize]{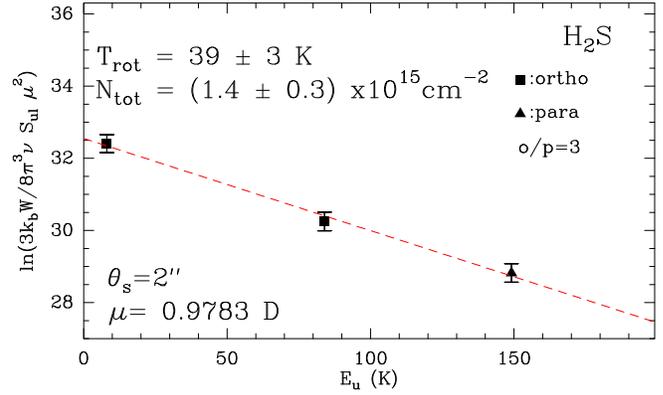}
\caption{As in Fig.\,\ref{fig:rdallco} but for H$_2$S. We fitted simultaneously the ortho and para species 
adopting an ortho-to-para ratio of 3:1 \citep{dec10a}. The line o-H$_2$S $J,K_{\mathrm{a}},k$=4$_{\mathrm{1,4}}$--3$_{\mathrm{2,1}}$  
significantly outlies the fit probably owing to its low $S/N$, therefore, it has been excluded from the diagram.}
\label{fig:rdallsh2}
\end{figure}

\begin{figure}[!htp] 
\centering
\includegraphics[width=0.95\hsize]{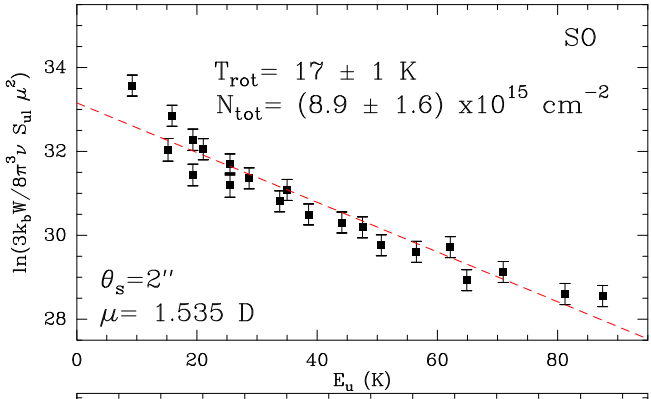}
\includegraphics[width=0.95\hsize]{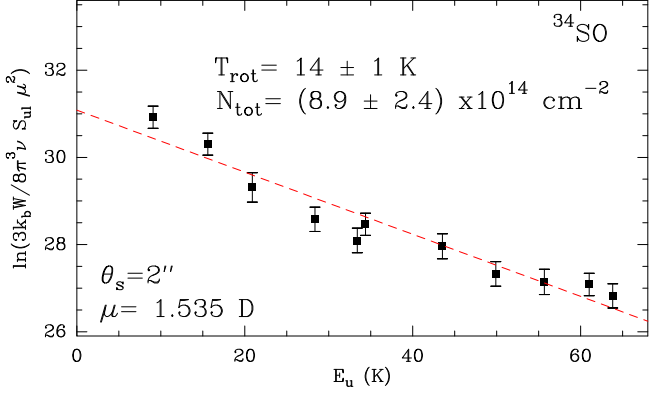}
\caption{As in Fig.\,\ref{fig:rdallco} but for SO isotopologues.}
\label{fig:rdallso}
\end{figure}

\begin{figure}[!htp] 
\centering
\includegraphics[width=0.95\hsize]{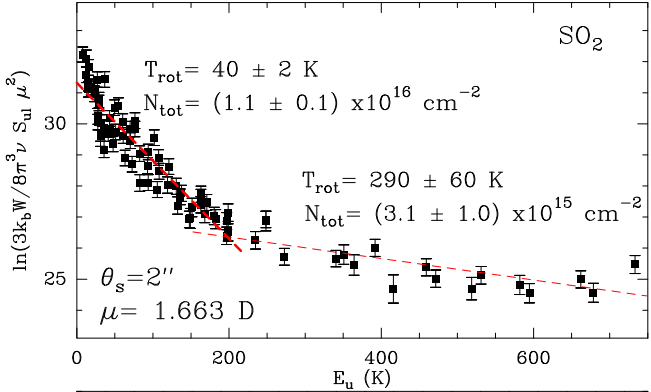}
\includegraphics[width=0.95\hsize]{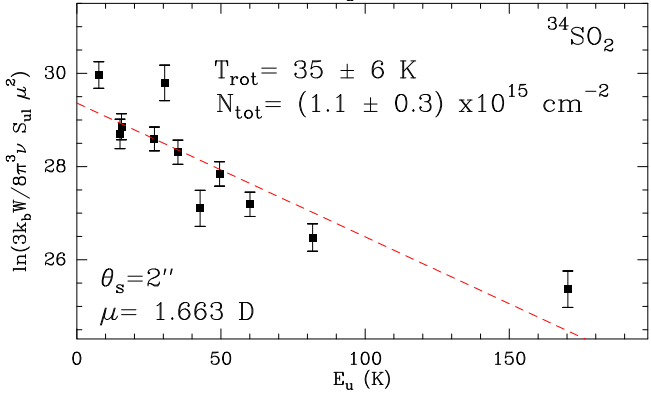}
\caption{As in Fig.\,\ref{fig:rdallco} but for SO$_2$ isotopologues.. The rotational diagram of SO$_2$ was separated
in two different trends (see Sect.\,\ref{sec:respop}).}
\label{fig:rdallso2}
\end{figure}

\begin{figure}[!htp] 
\centering
\includegraphics[width=0.95\hsize]{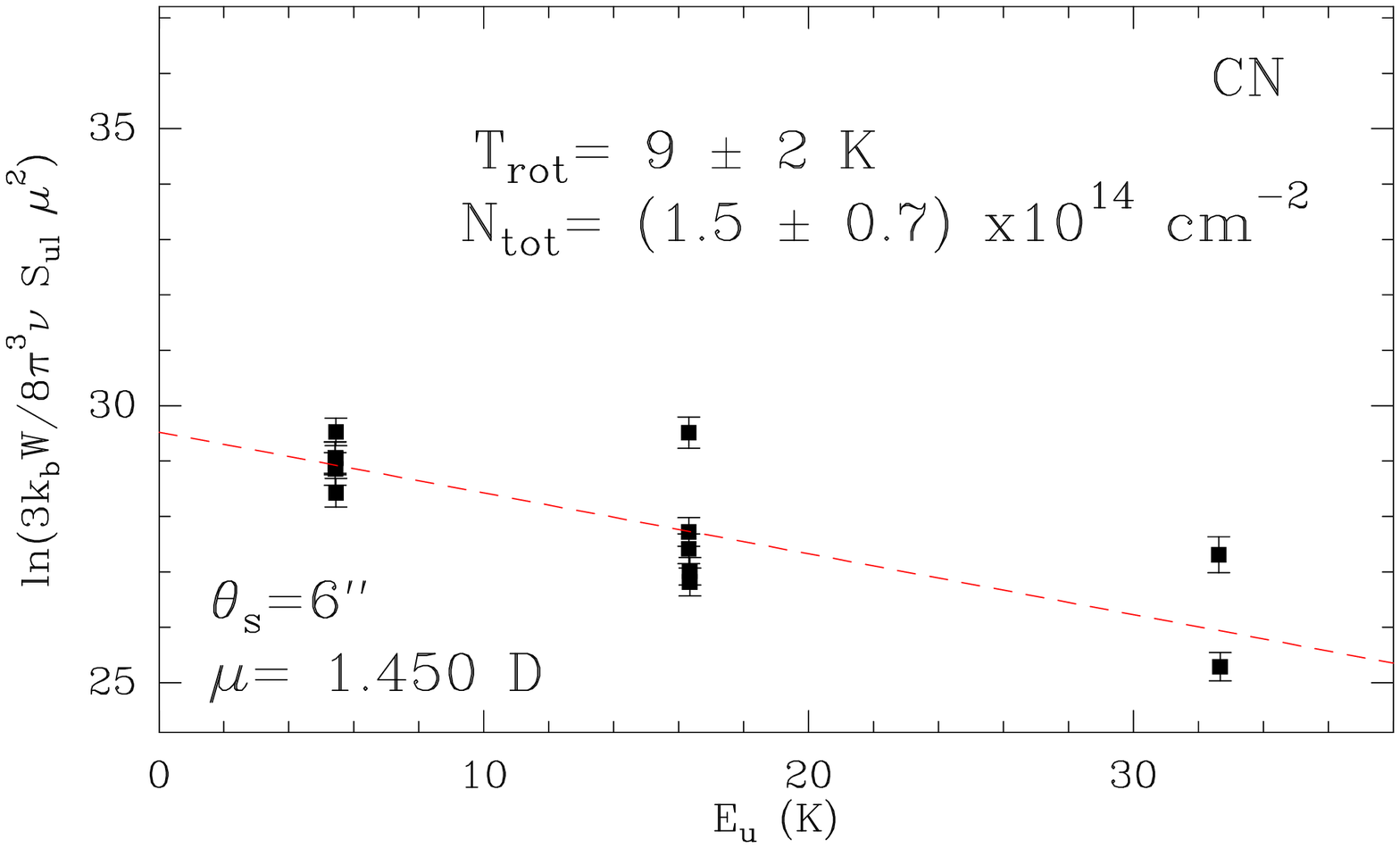}
\caption{As in Fig.\,\ref{fig:rdallco} but for CN.}
\label{fig:rdallcn}
\end{figure}

\begin{figure}[!htp] 
\centering
\includegraphics[width=0.95\hsize]{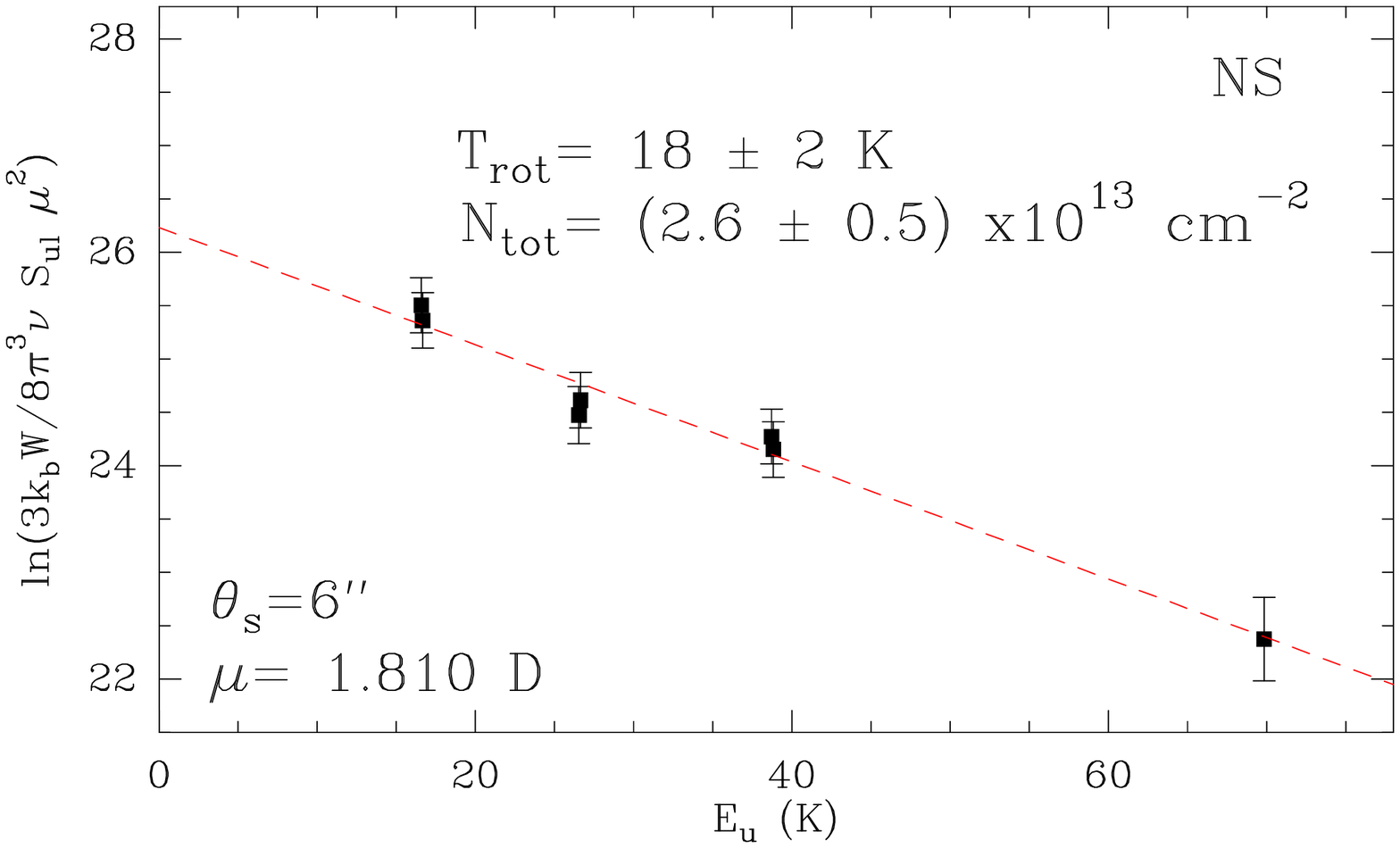}
\caption{As in Fig.\,\ref{fig:rdallco} but for NS.}
\label{fig:rdallns}
\end{figure}

\begin{figure}[!htp] 
\centering
\includegraphics[width=0.95\hsize]{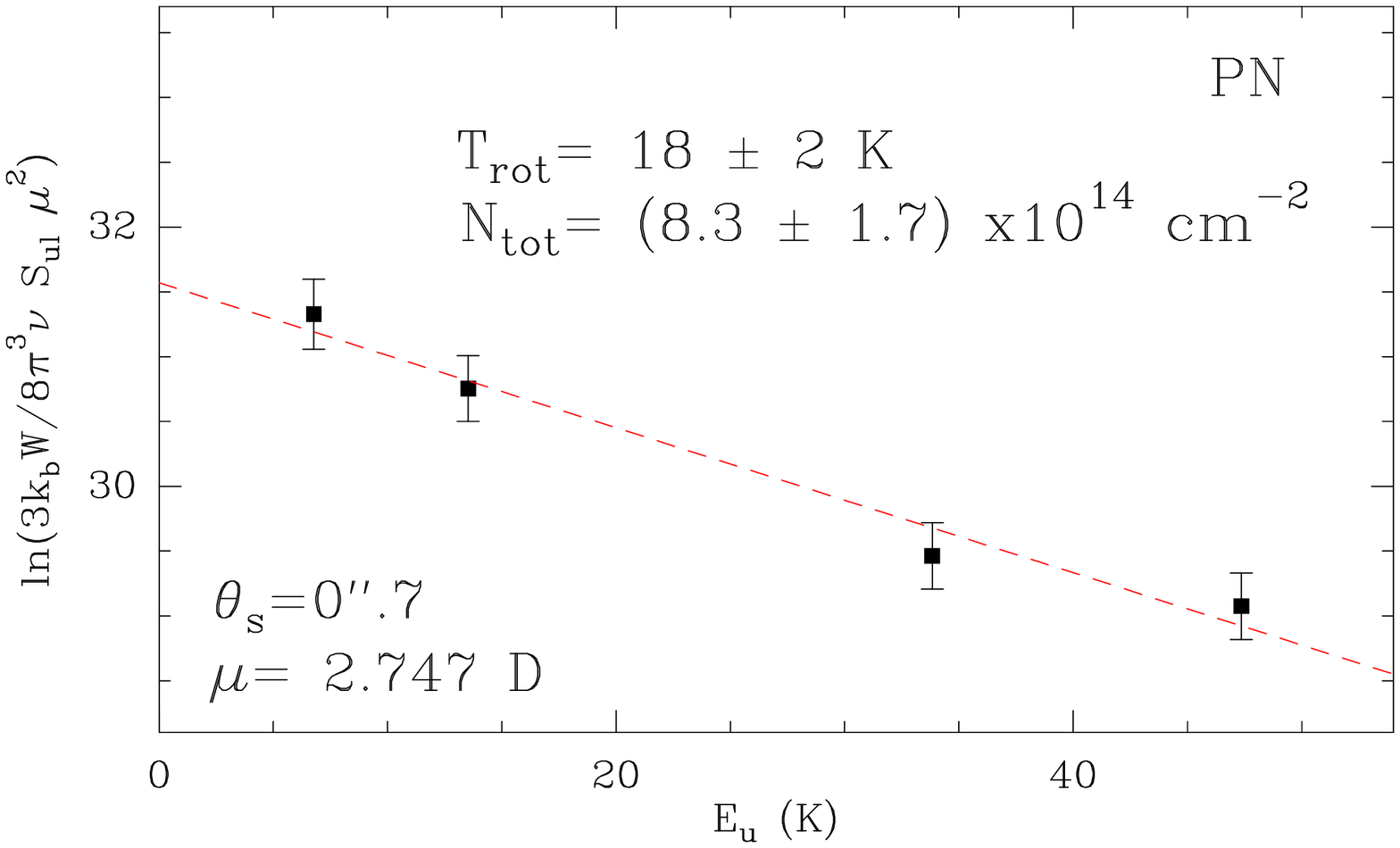}
\includegraphics[width=0.95\hsize]{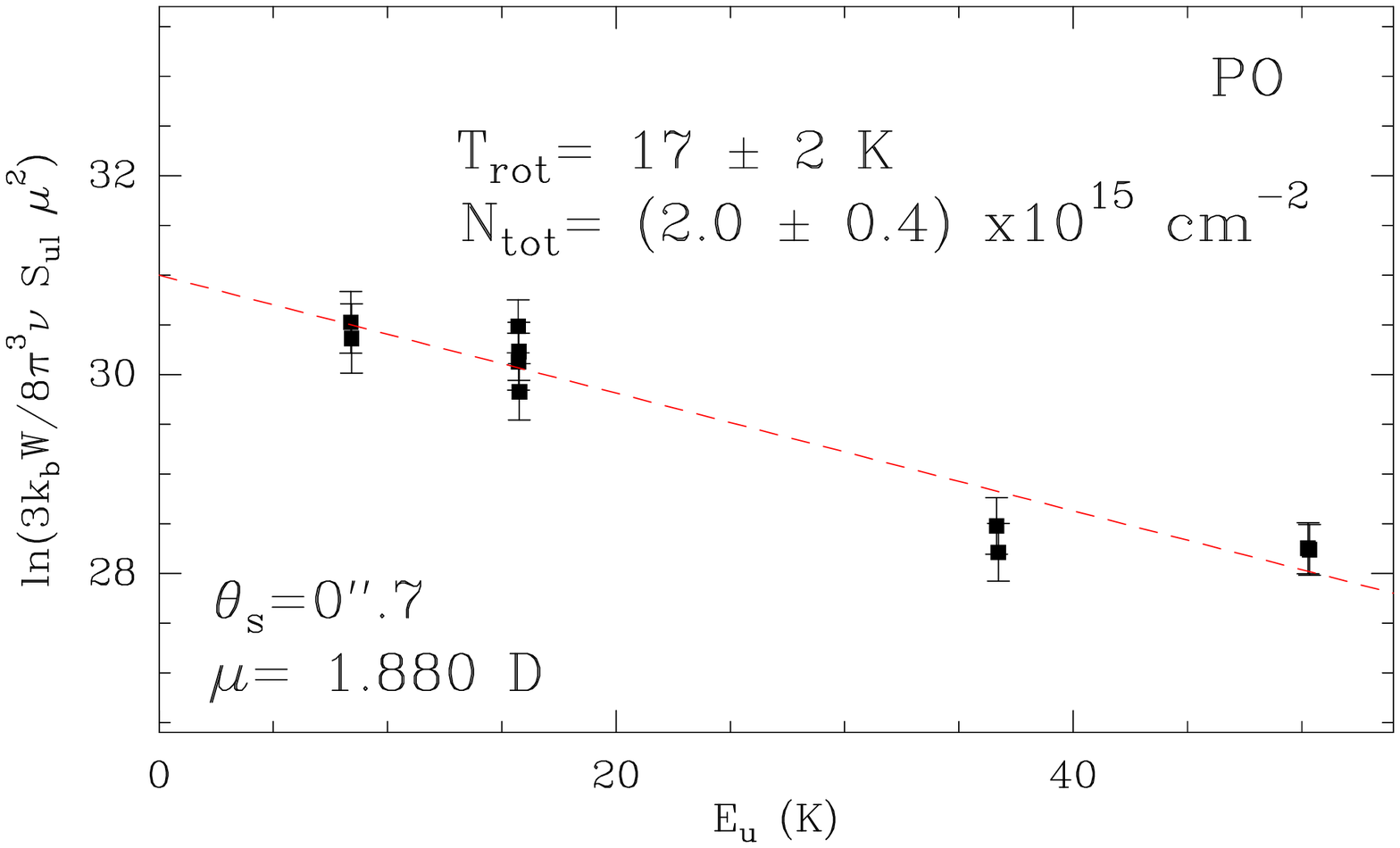}
\caption{As in Fig.\,\ref{fig:rdallco} but for P-bearing molecules. The PO line $\Omega$=1/2, e, $J,F$=(11/2,5)--(9/2,5) 
significantly outlies the fit probably owing to its low $S/N$, therefore, it has been excluded from the diagram.}
\label{fig:rdallphos}
\end{figure}

\begin{figure}[!htp] 
\centering
\includegraphics[width=0.95\hsize]{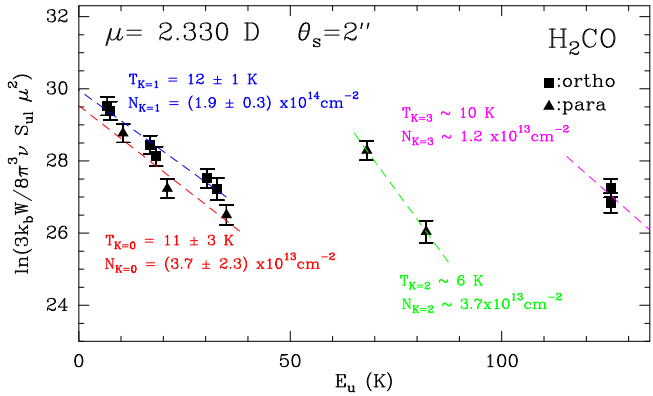}
\caption{Population diagram for H$_2$CO. 
We show the results from the fits of the rotational ladders 
$K_{\mathrm{a}}$=0 (red), $K_{\mathrm{a}}$=1 (blue), and $K_{\mathrm{a}}$=2 (green).
For the $K_{\mathrm{a}}$=3 (pink) we adopted as rotational temperature the average of the rotational 
temperatures of the $K_{\mathrm{a}}$=0, 1, and 2 ladders to estimate a rough value for the 
$K_{\mathrm{a}}$=3 column density.}
\label{fig:rdallh2co}
\end{figure}



\clearpage
\clearpage

\section{SO$_2$ IOS calculations}\label{sec:app_so2}
Close coupling (CC) calculations based on a state-of-the-art potential-energy surface (PES) \citep{spi09} have provided 
collisional para- and ortho-H$_2$ rate coefficients for temperatures ranging from 5 to 30\,K for the 31 lowest SO$_2$ rotational 
levels \citep{cer11}. However, an extension of these full CC calculations to a large number of SO$_2$ rotational levels and high 
temperatures would be prohibitive in terms of memory and CPU time, so the use of different approximate methods (coupled states (CS)
and infinite order sudden (IOS)) is needed  \citep{gre76,gre79}.  Considering the high-$J$ SO$_2$ levels observed in the 
present work, we used the IOS approximation which is expected to give reliable estimates of rate coefficients, except for 
low energies. State-to-state de-excitation rate coefficients between the 410 lowest levels (with energy up to 743.9\,K)  
were calculated for energies up to 9000 cm$^{-1}$ (value limited by the validity domain of the PES) giving data for temperatures
up to 1000\,K \citep{bal16}. The SO$_{2}$ energy levels and wave functions were obtained from spectroscopic constants 
describing the  SO$_{2}$ Hamiltonian up to the fourth order and provided by the MADEX code \citep{madex}. 
The results show a relatively flat temperature variation of the rates. \\
The accuracy of these data was studied by comparing the resulting rate coefficients with CC and CS values obtained for 
collisions with para and ortho-H$_2$ and all transitions involving the 31 lowest levels. 
Compared to CC and CS data, it is found that the IOS approach leads a systematic underestimation of para-H$_2$ rates, 
within a factor better than 50\% in average,  the agreement being better at high temperatures. However, the situation is not as 
good for collisions with ortho-H$_2$ as the IOS approximation describes collisions with para-H$_2$($j_{H_{2}}=0$) and does 
take into account all the angular couplings involved in collisions with ortho-H$_2$.  
The comparison between ortho- and para-H$_2$ CS rate coefficients shows that ortho-H$_2$ rates are systematically 
larger than para-H$_2$ rates by a factor two in average, with larger differences (up to a factor eight) for a number of transitions 
mainly identified as $\Delta K_{a}$=1 transitions. 

The rate coefficients were implemented in MADEX fitting the logarithm of the rate coefficients to a sixth order polynomial as:
\begin{equation}\label{eq:poli}
P(x)=\sum_{i=0}^{6}a_{i} x^{i},
\end{equation}
\noindent where
\begin{equation}\label{eq:x}
x=T^{3/2},
\end{equation}
\noindent 
which reproduces practically all the rates with a relative error below 30\% for temperatures between 20 and 1000\,K, 410 
energy levels and a maximum \eu=744\,K. 
We provide the energy levels involved in the calculations in Table\,\ref{tab:so2ener}, as well as 
the polynomial coefficients of the fit in Table\,\ref{tab:so2coef}.
The level identifier used in Table\,\ref{tab:so2coef} can be consulted in Table\,\ref{tab:so2ener}
to find the corresponding level.

\begin{table}[hbt!]
\caption{Parameters of the energy levels used for SO$_2$ modelling.}
\label{tab:so2ener}
\begin{center}
\begin{tabular}{llllll}
\hline\hline
Identifier & J & $K_{\mathrm{a}}$ & $K_{\mathrm{c}}$ & E & g \\
-          & - & - & - & K & - \\
\hline
    1      & 0 & 0 & 0 & 0.000000 &  1 \\
    2      & 2 & 0 & 2 & 2.751006 &  5 \\
    3      & 1 & 1 & 1 & 3.339254 &  3 \\
    4      & 2 & 1 & 1 & 5.320091 &  5 \\
    5      & 3 & 1 & 3 & 7.743834 &  7 \\
    6      & 4 & 0 & 4 & 9.151093 &  9 \\
\multicolumn{6}{l}{\ldots}             \\
\hline
\hline
\end{tabular}
\end{center}
\tablefoot{Table\,\ref{tab:so2ener}, available at the CDS, contains the following information:
(Col. 1) identifier for the energy level used in Table\,\ref{tab:so2coef}; 
(Col. 2) quantum number $J$; 
(Col. 3) quantum number $K_{\mathrm{a}}$;
(Col. 4) quantum number $K_{\mathrm{c}}$;
(Col. 5) energy of the level in Kelvin;
(Col. 6) degeneracy of the level.
}
\end{table}

\begin{table*}[hbt!]
\caption{Polynomial coefficients of the fit to the rate coefficients.}
\label{tab:so2coef}
\begin{center}
\begin{tabular}{lllllllll}
\hline\hline
Upper level id & Lower level id & $a_0$ & $a_1$ & $a_2$ & $a_3$ & $a_4$ & $a_5$ & $a_6$ \\
-              & -              & -     & -     & -     & -     & -     & -     & -     \\
\hline
   2 &  1 & -9.48415 & -29.3454 &  350.453 & -2071.21 & 6590.82 & -11012.6 &  7610.91 \\
   3 &  1 & -9.58292 & -35.9159 &  406.891 & -2385.77 & 7513.84 & -12378.6 &  8410.64 \\
   3 &  2 & -9.99698 & -14.3352 &  120.467 & -442.266 & 635.155 &  52.7172 & -657.401 \\
   4 &  1 &  0.00000 &  0.00000 &  0.00000 &  0.00000 & 0.00000 &  0.00000 &  0.00000 \\
   4 &  2 & -9.53669 & -26.9597 &  289.292 & -1598.36 & 4761.52 & -7413.75 &  4772.86 \\
   4 &  3 & -9.09926 & -29.6351 &  346.972 & -2025.33 & 6401.05 & -10614.5 &  7267.38 \\
\multicolumn{9}{l}{\ldots}             \\
\end{tabular}
\end{center}
\tablefoot{Table\,\ref{tab:so2coef}, available at the CDS, contains the following information:
(Col. 1) identifier for the upper energy level described in Table\,\ref{tab:so2ener}; 
(Col. 2) identifier for the lower energy level described in Table\,\ref{tab:so2ener}; 
(Col. 3) to (Col. 9) coefficients of the 6th order polynomial fit to the rate coefficients. 
}
\end{table*}

\end{document}